%% file: main.tex
\PassOptionsToPackage{dvipsnames}{xcolor}
\documentclass[acmsmall,nonacm]{acmart}

\acmJournal{PACMPL}
\acmVolume{1}
\acmNumber{CONF} 
\acmArticle{1}
\acmYear{2018}
\acmMonth{1}
\acmDOI{} 
\startPage{1}

\setcopyright{none}

\usepackage{booktabs}   
\usepackage{subcaption} 
\usepackage{cleveref}
\usepackage{xcolor}
\usepackage{colortbl}
\usepackage{bm}
\usepackage{enumitem}
\usepackage{colortbl}
\usepackage{graphicx}
\usepackage{wrapfig}
\usepackage{pifont} 
\usepackage{quiver}
\usepackage{multirow}

\input{listings.tex}
\input{macros.tex}

\usetikzlibrary{calc}

\begin{document}

\title{Sketch-Guided Equality Saturation}         
\subtitle{Scaling Equality Saturation to Complex Optimizations of Functional Programs}                     

\newcommand{\ourcomments}[4]{\todo[color=#1]{\color{#2}#4}}

\newcommand{\tkcomment}[1]{\ourcomments{purple}{white}{TK}{#1}}
\newcommand{\pwtcomment}[1]{\ourcomments{blue}{white}{PWT}{#1}}
\newcommand{\michelcomment}[1]{\ourcomments{green}{black}{MS}{#1}}

\newcommand{\tkcommentforphil}[1]{\tkcomment{{\color{blue}Phil,} #1}}

\makeatletter
\global\let\tikz@ensure@dollar@catcode=\relax
\makeatother


\author{Thomas K{\oe}hler}
\orcid{0000-0001-8461-8075}             
\affiliation{
  \institution{University of Glasgow}              
  \state{Scotland}
  \country{UK}                    
}
\email{thomas.koehler@thok.eu}          

\author{Phil Trinder}
\affiliation{
  \institution{University of Glasgow}             
  \state{Scotland}
  \country{UK}                   
}
\email{Phil.Trinder@glasgow.ac.uk}         

\author{Michel Steuwer}
\orcid{0000-0001-5048-0741}             
\affiliation{
  \institution{University of Edinburgh}             
  \state{Scotland}
  \country{UK}                   
}
\email{michel.steuwer@ed.ac.uk}         

\begin{abstract}

Generating high-performance code for diverse hardware and application domains like image processing, physics simulation, and machine learning is challenging.
Functional array programming languages with patterns like map and reduce have been successfully combined with semantics-preserving term rewriting to define and explore optimization spaces.
However, deciding what sequence of rewrites to apply is hard and has a huge impact on the performance of the rewritten program. 
Equality saturation avoids the issue by automatically exploring many possible ways to apply rewrites.
It is made feasible by an efficient representation of many equivalent programs in an e-graph data structure.

Equality saturation has some limitations for compiler optimizations that rewrite functional language terms.
Currently, there are only naive encodings of the lambda calculus for equality saturation.
We present new techniques for encoding polymorphically typed lambda calculi, and show that the efficient encoding reduces the runtime and memory consumption of equality saturation by orders of magnitude.

Moreover, equality saturation does not yet scale to complex compiler optimizations.
These emerge from long rewrite sequences of thousands of rewrite steps, and may use pathological combinations of rewrite rules that cause the e-graph to quickly grow too large. This paper introduces \emph{sketch-guided equality saturation}, a semi-automatic technique that allows programmers to provide program sketches to guide rewriting.
Sketch-guided equality saturation is evaluated for seven complex matrix multiplication optimizations, including loop blocking, vectorization, and multi-threading.
Even with efficient lambda calculus encoding, unguided equality saturation can locate only the two simplest of these optimizations, the remaining five are undiscovered even with an hour of compilation time and 60GB of RAM.
By specifying three or fewer sketch guides all seven optimizations are found in seconds of compilation time, using under 1GB of RAM, and generating high performance code.

\end{abstract}



\maketitle

\input{section/introduction.tex}
\input{section/background.tex}
\input{section/sketching.tex}
\input{section/eqsat-bindings.tex}
\input{section/evaluation.tex}
\input{section/related-work.tex}
\input{section/conclusion.tex}


\bibliography{reference.bib}



\end{document}

%% file: listings.tex
\usepackage{listings}

\crefname{lstlisting}{listing}{listings}
\Crefname{lstlisting}{Listing}{Listings}

\definecolor{rise}{HTML}{117733}
\definecolor{sketch}{HTML}{1c4ca6}
\definecolor{highlight}{HTML}{c7143e}

\lstdefinestyle{rise}{
  morekeywords = {def},
  morecomment = [l][\itshape\color{gray}]{|},
  commentstyle=\itshape\color{gray},
  morekeywords= [6]{ 
    app, dot,
    map, mapSeq, mapSeqUnroll, mapPar,
    mapGlobal, mapVec, mapStream,
    toMem,
    zip, fst, snd, generate, unzip,
    reduce, reduceSeq, reduceSeqUnroll,
    asVector, asScalar, vectorFromScalar,
    padEmpty, padClamp, take,
    slide, split, join, transpose,
    circularBuffer, rotateValues, iterateStream,
    global, local, private 
  },
  keywordstyle= [6]{\bfseries\color{rise}},
  basicstyle=\ttfamily\scriptsize,
  commentstyle=\itshape, 
  xleftmargin=2.5em,
  numbers=left, 
  numberstyle=\scriptsize, 
  tabsize = 2,
  numbersep=8pt, 
  breaklines=true, 
  frame=tb, 
  mathescape=true,
  moredelim=[is][\bfseries\color{highlight}]{*}{*},
	captionpos=b,
}
\lstnewenvironment{rise}[1][]{\lstset{style=rise, #1}}{}
\newcommand{\inlineRise}{\lstinline[style=rise]}

\lstdefinestyle{rise-sketch}{
  morekeywords = {def},
  morecomment = [l][\itshape\color{gray}]{|},
  commentstyle=\itshape\color{gray},
  otherkeywords={?, ::},
  morekeywords= [5]{ 
    ?, contains, ::,
    containsMap, containsMapPar,
    containsReduceSeq, containsReduceSeqUnroll,
    containsAddMul, containsAddMulVec,
    isSlide, isCircularBuffer,
    containsGrayLine, containsSobelLine, containsCoarsityLine,
  },
  keywordstyle= [5]{\bfseries\color{sketch}},
  morekeywords= [6]{ 
    app, dot,
    map, mapSeq, mapSeqUnroll,
    mapGlobal, mapVec, mapStream,
    toMem,
    zip, fst, snd, generate, unzip,
    reduce, reduceSeq, reduceSeqUnroll,
    asVector, asScalar, vectorFromScalar,
    padEmpty, padClamp, take,
    split, join, transpose,
    rotateValues, iterateStream,
    global, local, private 
  },
  keywordstyle= [6]{\bfseries\color{rise}},
  basicstyle=\ttfamily\scriptsize,
  commentstyle=\itshape, 
  xleftmargin=2.5em,
  numbers=left, 
  numberstyle=\scriptsize, 
  tabsize = 2,
  numbersep=8pt, 
  breaklines=true, 
  frame=tb, 
  mathescape=true,
  moredelim=[is][\bfseries\color{highlight}]{*}{*},
	captionpos=b,
}
\lstnewenvironment{rise-sketch}[1][]{\lstset{style=rise-sketch, #1}}{}
\newcommand{\inlineRiseSketch}{\lstinline[style=rise-sketch]}

\lstnewenvironment{c-code}[1][]{\lstset{
  language=C,
#1}}{}

\lstdefinestyle{elevate-rise}{
  language=scala,
  morekeywords = {apply},
  deletekeywords={try},
  otherkeywords = {;, |>, =>, >>, <<, <+, @},
  keywordstyle=\bfseries\color{black!75}, 
  morekeywords= [2]{ 
  ;,seq,
  <+,lChoice,
  try,
  repeat,
  all,one,some,
  topDown,
  allTopDown,
  tryAll,
  allBottomUp,
  bottomUp,
  normalize,
  Strategy,
  RewriteResult,
  Success,
  Failure,
  @
  },
  keywordstyle= [2]{\bfseries\color{MidnightBlue!75}},
  morekeywords = [3]{ 
    separateDot,isTransposedB,
    lowerToSeqC,
    mapFusion,
    fuseReduceMap,
    isReduce,
    isMap,isTranspose,isToMem,
    lowerToC,inLambda,isLambda,
    blocking,
    baseline,
    loopPerm,inToMem,par,
    arrayPacking,
    packB,
    parallelizeCopy,
    outermost,mapNest,innermost,
    body,isTransposeB, extractInToMem, permuteB,
    lowerStore,storeInMemory,isTranspose,transposedB,
    function,
    argument,appliedMap,
    fmap,appliedReduce,isApplied,fissionReduceMap,
    reorder,
    interchange,
    argOf,
    not,isFun,isApp,isApplication,
    betaReduction,
    etaAbstraction,etaReduction,
    id,fail,
    buildGet, lenBuild, letPartialEval, letApp, funToLet,
    parallel,vectorize,unroll,tile,tileND,DFNF,BENF,
  },
  keywordstyle = [3]{\color{RoyalPurple}},
  morekeywords = [4]{ 
    p, e, s
  },
  keywordstyle = [4]{\color{Plum}},
  morekeywords = [5]{
    bf,
    mm,
    dot,
    threemaps,
    mt
  },
  keywordstyle = [5]{\bfseries\color{RedViolet}},
  morekeywords= [6]{ 
    |>, >>, <<,
    fun,
    app,
    map,
    zip,Identifier,
    toMem,
    fst,snd,
    reduce,reduceSeq,reduceSeqUnroll,
    mapSeq,mapSeqUnroll,
    mapSeq,
    mapVec,
    asVector,
    asScalar,
    map2D,
    pad,
    pad2D,
    slide,
    slide2D,
    join,
    transpose,
    split,
    add,mult
  },
  keywordstyle= [6]{\bfseries\color{OliveGreen!95}},
  morekeywords = [7]{ 
    dot,
    mm
  },
  keywordstyle = [7]{\color{RawSienna}},
  morekeywords = [8]{ 
    weights2d,weightsH,weightsV,
    nbh,
    x,y,
    f,nf,
    g,
    h,
    xs,
    img,
    a, na,
    b, nb,
    M,K,N,as,bs,ab,
    ak, bk,arow,bcol,
    float,
  },
  keywordstyle = [8]{\color{RedOrange}},
	basicstyle=\ttfamily\scriptsize, 
	commentstyle=\itshape\color{gray}, 
	stringstyle=\itshape, 
	xleftmargin=2.5em,
	numbers=left, 
	numberstyle=\scriptsize, 
	stepnumber=1, 
  tabsize = 2,
	numbersep=8pt, 
	showstringspaces=false, 
	breaklines=true, 
	backgroundcolor = \color{black!05},
	frame=lines, 
  escapechar=\#,
	captionpos=b,
  mathescape=true,
  literate={`}{\lq}1 
}

\newcommand{\elevateConstruct}{\lstinline[basicstyle=\ttfamily\scriptsize\color{RoyalPurple}]}

%% file: macros.tex
\newcommand{\Lift}{\textsc{Lift}}
\newcommand{\Rise}{\textsc{Rise}}
\newcommand{\Elevate}{\textsc{Elevate}}
\newcommand{\kles}{\textsc{Risegg}}

\newcommand{\rewritesTo}{\operatorname{\longmapsto}}

\newcommand{\goalStyle}[1]{\textit{\textsf{#1}}}

%% file: section/introduction.tex

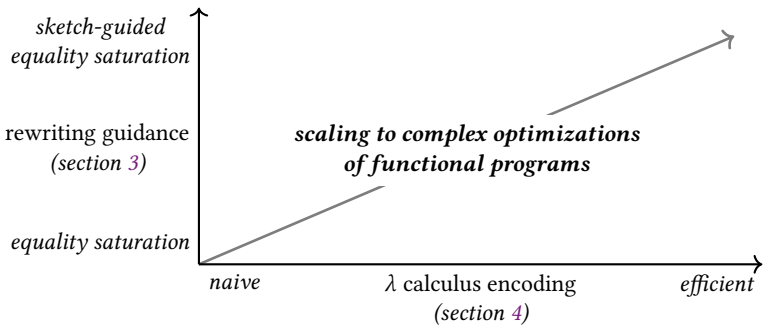
\begin{figure}[b]
    \centering
\begin{tikzpicture}[scale=1.5]
    \tikzstyle{every node}=[font=\small\itshape, align=center]
    \coordinate (origin)        at (0,0);
    \coordinate (top-left)      at (0,2.25);
    \coordinate (bottom-right)  at (5,0);
    \coordinate (top-right)     at (4.75,2);
    \draw [->, thick, color={rgb,255:red,125;green,125;blue,125},line width=1pt] (origin) -- (top-right);
    \draw [->, thick] (origin) -- (top-left);
    \draw [->, thick] (origin) -- (bottom-right);
    \node[anchor=north east] at (top-left) {sketch-guided\\ equality saturation};
    \node[anchor=east,yshift=-.5em] at ($(origin)!0.5!(top-left)$) {\normalfont{rewriting guidance}\\(\cref{sketching})};
    \node[anchor=south east] at (origin) {equality saturation};
    \node[anchor=north west] at (origin) {naive};
    \node[anchor=north] at ($(origin)!0.5!(bottom-right)$) {$\lambda$~\normalfont{calculus encoding}\\(\cref{eqsat-bindings})};
    \node[anchor=north east] at (bottom-right) {efficient};
    \node[fill=white] at ($(origin)!0.5!(top-right)$) {\bfseries scaling to complex optimizations\\\bfseries of functional programs};
\end{tikzpicture}
\caption{This paper is about scaling equality saturation to complex optimizations of functional programs by combining an efficient $\lambda$ calculus encoding with sketch-guided equality saturation.}
    \label{fig:overview}
\end{figure}

\section{Introduction}

Term rewriting has been effective in optimizing compilers for decades \cite{dershowitz1993-rewrite-systems}.
More recently, functional array languages like \Lift{}~\cite{lift-rewrite-2015} and \Rise{}~\cite{hagedorn2020-elevate} produce high-performance code for diverse hardware by using rewrite rules to define and explore optimization spaces.
However, deciding when to apply each rewrite rule, the so-called \emph{phase ordering problem}, is hard and has a huge impact on the performance of the rewritten program.
The challenge is that the global benefit of applying a rewrite rule depends on future rewrites.
Maximizing local benefit in a greedy fashion is not sufficient in the absence of a convergence property, i.e. confluence and termination, as local optima may be far away from global optima.

Rewriting strategies \cite{visser1998-strategies} allow programmers to control when to apply each rewrite rule, step-by-step.
Prior work on \Elevate{}~\cite{hagedorn2020-elevate} has shown that rewriting strategies can achieve \emph{complex optimizations} of \Rise{} programs in less than a second of compilation time.
However, these complex optimizations emerge from thousands of rewrite steps making the rewriting strategies challenging to write, as the phase ordering problem is passed on to the programmer.

\vspace{8em}

Equality saturation \cite{tate2009-equality-saturation, willsey2021-egg} mitigates the phase ordering problem by exploring many possible ways to apply rewrite rules.
Starting from an input program, equality saturation grows an equality graph (\emph{e-graph}) by applying all possible rewrites iteratively until reaching a fixed point (saturation), achieving a performance goal, or timing out.
An e-graph efficiently represents a large set of equivalent programs, and is grown by repeatedly applying rewrite rules in a purely additive way.
Instead of replacing the matched left-hand side of a rewrite rule by its right-hand side, the equality between the left-hand side and the right-hand side is recorded in the e-graph.
After growing the e-graph, the best program found is extracted from it using a cost model, e.g. one that selects the fastest program.

Unfortunately, equality saturation does not scale to complex optimizations such as the ones applied to \Rise{} with \Elevate{}, producing huge e-graphs that exceed the memory available in most machines.
\textbf{To scale equality saturation to complex optimizations of functional programs}, this paper makes advances in two directions, as shown by the two axes in~\cref{fig:overview}.

\textbf{Rewriting Guidance.}
On each equality saturation iteration, the e-graph tends to grow larger as every possible rewrite rule is applied in a purely additive way. 
The growth rate is extremely rapid for some combinations of useful rewrite rules, like associativity and commutativity.
This makes exploring long rewrite sequences requiring many iterations unfeasible.
One way to address this issue is to limit the number of rules applied \cite{wang2020-spores, willsey2021-egg}, but this risks not finding optimizations that require an omitted rule.
An alternative is to use an external solver to speculatively add equivalences \cite{nandi2020-synthesizing-CAD}, but this requires the identification of sub-tasks that can benefit from being delegated.

This paper proposes \emph{sketch-guided equality saturation} that factors complex optimizations into a sequence of smaller optimizations, each sufficiently simple to be found by equality saturation.
The programmer guides rewriting by describing how a program
should evolve during optimization, through a sequence of \emph{sketches}: program patterns that leave details unspecified.
While sketches have previously been used as a starting point for program synthesis \cite{solar2008-synthesis-sketching},
our work uses sketches in a novel way as checkpoints to guide program optimization.

\textbf{Lambda Calculus Encoding.}
As almost all programming languages use variables, and hence name binding, this paper explores practical ways of efficiently encoding languages with name bindings for the purposes of equality saturation.
We focus on encoding the lambda calculus as it is the standard formalism for functional languages such as \Rise{}. Previously, equality saturation has used a naive lambda calculus encoding that is simple but inefficient \cite{willsey2021-egg}: the size of the e-graph quickly blows up.
The inefficiency can be avoided by explicitly avoid name bindings in the rewritten language~\cite{smith2021-access-patterns}. 
We show that lambda calculus name bindings can be managed by using De Bruijn indices to avoid overloading the e-graph with $\alpha$-equivalent terms, and that an approximate implementation of substitution avoids overloading the e-graph with intermediate substitution steps.

\smallskip

Our evaluation demonstrates that combining sketch-guided equality saturation with an efficient lambda calculus encoding enables complex optimizations of \Rise{} functional programs, that require tens of thousands of rewrite steps in term rewriting.
We first evaluate the effectiveness of our lambda calculus encoding techniques for three rewrite goals. Thereafter, we show that sketch guiding enables seven complex optimizations of matrix multiplication to be applied.
\Elevate{} rewriting strategies apply these complex  optimizations in less than a second with low memory consumption, but require ordering thousands of rewrite rules.
Unguided equality saturation abstracts over rewrite ordering, but can only locate the simplest of the seven optimizations before the e-graph exceeds the available memory.
Sketch-guided equality saturation applies all seven optimizations in seconds with low memory consumption, and only requires ordering up to three sketch guides.

\smallskip

To summarize, the main contributions of this paper are:
\begin{itemize}
  \item Proposing \emph{sketch-guided equality saturation}, a semi-automated process offering a novel, practical trade-off between rewriting strategies and equality saturation.
  The programmer guides multiple equality saturations by specifying a sequence of sketches describing how the program should evolve during optimization (\cref{sketching}).
  
  %
  \item Exploring new techniques to efficiently encode a polymorphically typed lambda calculus such as \Rise{} for the purpose of equality saturation.
  In particular, De Bruijn indices are used to avoid overloading the e-graph with $\alpha$-equivalent terms, and an approximate (\emph{extraction-based}) substitution is used to avoid overloading the e-graph with intermediate substitution steps (\cref{eqsat-bindings}).

  \item Two systematic evaluations of \Rise{} applications.
  The effectiveness of our lambda calculus encoding is demonstrated by optimizing a binomial filter application (\cref{lambda-eval}).
  Sketch-guided equality saturation is evaluated by exploring seven complex optimizations of a matrix multiplication application, including loop blocking, vectorization, and multi-threading.
 These optimizations are infeasible with unguided equality saturation due to excessive runtime and memory consumption, but sketch-guided equality saturation performs the optimizations in seconds and with low memory consumption.
  No more than three sketch guides need to be written for any of the seven optimizations (\cref{mm-eval}).



\end{itemize}

%% file: section/background.tex
\section{Motivation and Background}
\label{background}

This section introduces \Rise{} programs and their optimization, as well as prior work on \Elevate{} rewriting strategies and equality saturation.
We highlight limitations and motivate our work.

\subsection{The \Rise{} functional language}
\label{rise}

\Rise{}~\cite{hagedorn2020-elevate} is a functional array programming language.
It is a spiritual successor of \Lift{}~\cite{lift-rewrite-2015,lift-ir-2017} that demonstrated performance portability across hardware by automatically applying semantics-preserving rewrite rules to optimize programs from various domains, including scientific code~\cite{lift-stencil-2018} and convolutions~\cite{mogers2020-convolution}.

As a typed lambda calculus, \Rise{} provides standard lambda abstraction (\inlineRise{$\lambda$x. b}), function application (\inlineRise{f x}), identifiers and literals.
\Rise{} expresses data-parallel computations as compositions of high-level computational patterns over multi-dimensional arrays, such as
\inlineRise{map} which applies a function to each element of an array.
\inlineRise{reduce} combines all elements of an array to a single value given a binary reduction operator.
\inlineRise{split}, \inlineRise{join}, \inlineRise{transpose}, \inlineRise{zip}, \inlineRise{unzip} and \inlineRise{slide} reshape arrays in various ways.

High-level programs, such as a matrix multiplication in the top left of \cref{fig:mm-rewrite}, express their computations without committing to a particular implementation strategy.

Implementation choices are explicitly encoded in \Rise{} programs by applying rewrite rules that introduce low-level patterns which directly correspond to a particular implementation strategy.
For example, \inlineRise{reduceSeq} is a sequential implementation of \inlineRise{reduce}.
For \inlineRise{map}, there exist multiple low-level patterns corresponding to different sequential and parallel implementations.
After rewriting, \Rise{} programs are translated to low-level imperative code such as C or OpenCL for execution.

The \Rise{} program on the right of \cref{fig:mm-rewrite} shows an optimized version of matrix multiplication.
A common loop blocking optimization that improves data locality, and therefore memory usage, has been introduced by rewriting.
Its impact is visualized in the bottom-left of \cref{fig:mm-rewrite}.


\newcommand{\mydim}[1]{\color{red}\tikz[baseline=(char.base)]{
    \node[shape=circle,draw,inner sep=1pt,minimum size=0pt] (char) {\tiny #1};}}

\begin{figure}
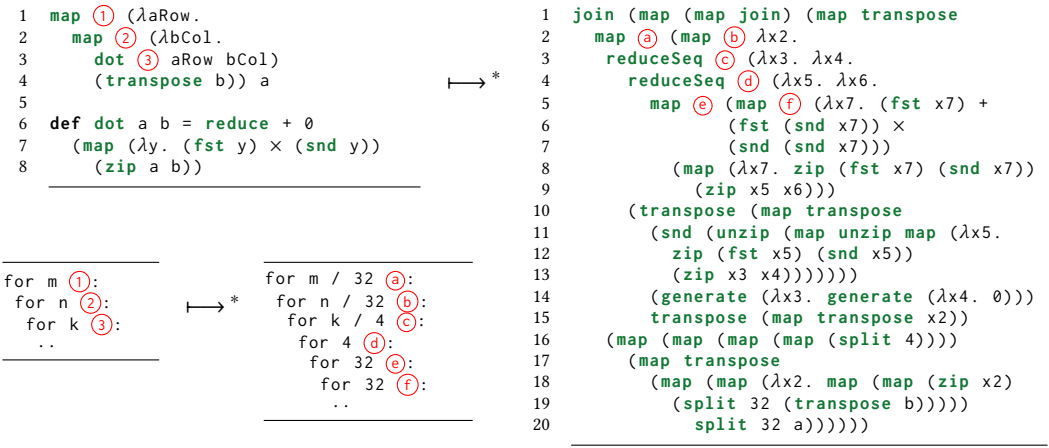

  \begin{minipage}[t]{0.5\linewidth}\vspace{0pt}%
  \begin{minipage}{0.8\linewidth}
  \begin{rise}
map $\mydim{1}$ ($\lambda$aRow.
  map $\mydim{2}$ ($\lambda$bCol.
    dot $\mydim{3}$ aRow bCol)
    (transpose b)) a
    
def dot a b = reduce + 0
  (map ($\lambda$y. (fst y) $\times$ (snd y))
    (zip a b))
  \end{rise}%
  \end{minipage}%
  \begin{minipage}{0.2\linewidth}
  \hspace{1em}$ \rewritesTo{}^* $\hspace{1em}
  \end{minipage}%
\vspace{2em}
  \begin{minipage}{\linewidth}
  \begin{minipage}[t]{0.3\linewidth}\vspace{0pt}%
  \begin{lstlisting}[mathescape=true]
for m $\mydim{1}$:
 for n $\mydim{2}$:
  for k $\mydim{3}$:
   ..
  \end{lstlisting}
  \end{minipage}%
  \begin{minipage}[t]{0.2\linewidth}
  \vspace{1.5em}
  \hspace{1em}$ \rewritesTo{}^* $\hspace{1em}
  \end{minipage}%
  \begin{minipage}[t]{0.4\linewidth}\vspace{0pt}%
  \begin{lstlisting}[mathescape=true]
for m / 32 $\mydim{a}$: 
 for n / 32 $\mydim{b}$:
  for k / 4 $\mydim{c}$:
   for 4 $\mydim{d}$:
    for 32 $\mydim{e}$:
     for 32 $\mydim{f}$:
      ..
  \end{lstlisting}
  \end{minipage}
\end{minipage}
  \end{minipage}%
  \begin{minipage}[t]{0.5\linewidth}\vspace{0pt}%
  \begin{rise}
join (map (map join) (map transpose
  map $\mydim{a}$ (map $\mydim{b}$ $\lambda$x2.
   reduceSeq $\mydim{c}$ ($\lambda$x3. $\lambda$x4.
     reduceSeq $\mydim{d}$ ($\lambda$x5. $\lambda$x6.
       map $\mydim{e}$ (map $\mydim{f}$ ($\lambda$x7. (fst x7) +
              (fst (snd x7)) $\times$
              (snd (snd x7)))
         (map ($\lambda$x7. zip (fst x7) (snd x7))
           (zip x5 x6)))
     (transpose (map transpose
       (snd (unzip (map unzip map ($\lambda$x5.
         zip (fst x5) (snd x5))
         (zip x3 x4)))))))
       (generate ($\lambda$x3. generate ($\lambda$x4. 0)))
       transpose (map transpose x2))
   (map (map (map (map (split 4))))
     (map transpose
       (map (map ($\lambda$x2. map (map (zip x2)
         (split 32 (transpose b)))))
           split 32 a))))))
\end{rise}
\end{minipage}
\caption{Applying a blocking optimization to matrix multiplication via rewriting in \Rise{}.
In the initial program (top-left), a \inlineRise{dot} product is computed between each row of \inlineRise{a} (\inlineRise{aRow}) and column of \inlineRise{b} (\inlineRise{bCol}).
To define the \inlineRise{dot} product, the \inlineRise{zip} pattern combines two arrays \inlineRise{a} and \inlineRise{b} whose elements are multiplied pairwise using \inlineRise{map} before they are summed using \inlineRise{reduce}.
In the final program (right), a blocking optimization has been applied.
Intuitively, each red circle identifies a loop characteristic of the optimization (bottom-left).
Understanding the remaining program details is not required, as they are only shown to visualize program complexity.}
\label{fig:mm-rewrite}
\end{figure}

\begin{lstlisting}[float,style=elevate-rise,escapechar={\#}, caption={An \Elevate{} strategy that applies the blocking optimization to matrix multiplication \cite{hagedorn2020-elevate}},label=lst:elevate-blocking]
def blocking = ( baseline `;`
  tile(32,32)      `@` outermost(mapNest(2))   `;;` #\label{line:elevate-tile}#
  fissionReduceMap `@` outermost(appliedReduce)`;;` #\label{line:elevate-fission}#
  #\text{\color{RoyalPurple}{split}}#(4)          `@` innermost(appliedReduce)`;;` #\label{line:elevate-split}#
  reorder(List(1,2,5,6,3,4))) #\label{line:elevate-reorder}#
\end{lstlisting}

\subsection{\Elevate{} rewriting strategies}
\label{elevaterewrite}

To specify optimizations \Rise{} is complemented by a second language: \Elevate{}~\cite{hagedorn2020-elevate} that describes complex optimizations as compositions of rewrite rules, called \emph{rewriting strategies}.
The performance of the code generated by \Rise{} and \Elevate{} is comparable with state-of-the-art compilers, e.g. with the TVM  deep learning compiler~\cite{chen2018-tvm} for matrix multiplication~\cite{hagedorn2020-elevate}; and with the Halide image processing compiler~\cite{halide-2012} for the Harris corner detection~\cite{koehler2021-elevate-imgproc}.

For example, the \Elevate{} matrix multiplication blocking strategy is shown in~\cref{lst:elevate-blocking}.
This strategy describes the sequence of rewrite steps that are required to rewrite the \Rise{} program on the left of \cref{fig:mm-rewrite} to the one on the right.
32$\times$32 blocks (or tiles) are created in line \ref{line:elevate-tile} and another block of 4 is created in line \ref{line:elevate-split}.
Finally, loops are reordered in line \ref{line:elevate-reorder} to create 4$\times$32$\times$32 blocks and hence produce the loop nest in \cref{fig:mm-rewrite}.
All abstractions like \elevateConstruct{tile}, \elevateConstruct{split}, \elevateConstruct{reorder}, \elevateConstruct{outermost} and \elevateConstruct{innermost} are not built-in but programmer-defined.

\paragraph{Limitations of manual rewriting with strategies}
Although \Elevate{} enables the development of abstractions that help write concise strategies, strategies remain challenging to write.
The authors of \cite{hagedorn2020-elevate} and \cite{koehler2021-elevate-imgproc} estimate\footnote{in private communication with us} that they spent between two and five person-weeks developing the \Elevate{} strategies for their matrix multiplication and image processing case studies.
These case studies implement complex optimizations by applying tens of thousands of rewrite steps.

A particular limitation is that strategies are often program-specific and complex to implement.
For example, the implementation of the \elevateConstruct{reorder} strategy is 43 lines long, involves the definition of 8 internal strategies, and carefully and recursively composes them together with some generic strategies\footnote{\tiny\url{https://github.com/rise-lang/shine/blob/sges/src/main/scala/rise/elevate/strategies/algorithmic.scala}}.
Despite its name, this \elevateConstruct{reorder} strategy is not capable of reordering arbitrary nestings of \inlineRise{map} and \inlineRise{reduce} patterns, but only works for the matrix multiplication example.
Developing generic strategies is difficult because small program differences require adjustments to the rewrite sequence.
The fundamental problem is that while \Elevate{} empowers programmers to manually control the rewrite process, this requires the programmer to order all rewrite steps deterministically, i.e.\  to deal with fine-grained phase ordering.

To reduce the programmer effort required to optimize \Rise{} programs, we look at equality saturation as an automation technique that promises to mitigate the phase ordering problem.

\subsection{Equality saturation}
\label{eqsat}

Equality saturation \cite{tate2009-equality-saturation, willsey2021-egg} is a technique for efficiently implementing rewrite-driven compiler optimizations without committing to a single rewrite choice.
We demonstrate how equality saturation mitigates the phase ordering problem with a rewriting example where greedily reducing a cost function is not sufficient to find the optimal program.

Rewriting is often used to fuse operators and avoid writing results to memory, for example:
\begin{align*}
\tag{A} \label{overview-start}
(map~(map~f)) &\circ (transpose \circ (map~(map~g))) \\
&\rewritesTo{}^* \\
\tag{B} \label{overview-goal}
(map~(map~(f &\circ g))) \circ transpose \\
\end{align*}
The initial term (\ref{overview-start}) applies function $g$ to each element of a two-dimensional matrix (using two nested $map$s), transposes the result, and then applies function $f$ to each element.
The optimized term (\ref{overview-goal}) avoids storing an intermediate matrix in memory and transposes the input before applying $g$ and $f$ to each element.
Applying the following rewrite rules in the correct order is sufficient to perform this optimization:
\begin{align}
\label{overview-move-transpose}
transpose \circ map~(map~a) \longleftrightarrow{}& map~(map~a) \circ transpose\\
\label{overview-comp-assoc}
a \circ (b \circ c) \longleftrightarrow{}& (a \circ b) \circ c\\
\label{overview-map-fusion}
map~a \circ map~b \longleftrightarrow{}& map~(a \circ b)
\end{align}

Rule (\ref{overview-move-transpose}) states that transposing a two-dimensional array before or after applying a function to the elements is equivalent.
Rule (\ref{overview-comp-assoc}) states that function composition is associative.
Finally, rule (\ref{overview-map-fusion}) is the rewrite rule for map fusion.
In this example, minimizing the term size results in maximizing fused maps and is, therefore, a good cost model. 

If we greedily apply rewrite rules that lower term size, we will only apply rule (\ref{overview-map-fusion}) as this is the only rule that reduces term size.
However, rule (\ref{overview-map-fusion}) cannot be directly applied to term (\ref{overview-start}): we are in a local optimum.
The only way to reduce term size further is to first apply the other rewrite rules, which may or may not pay off depending on future rewrites.

\begin{figure}
  \begin{subfigure}[b]{0.23\linewidth}
  \includegraphics[width=\linewidth]{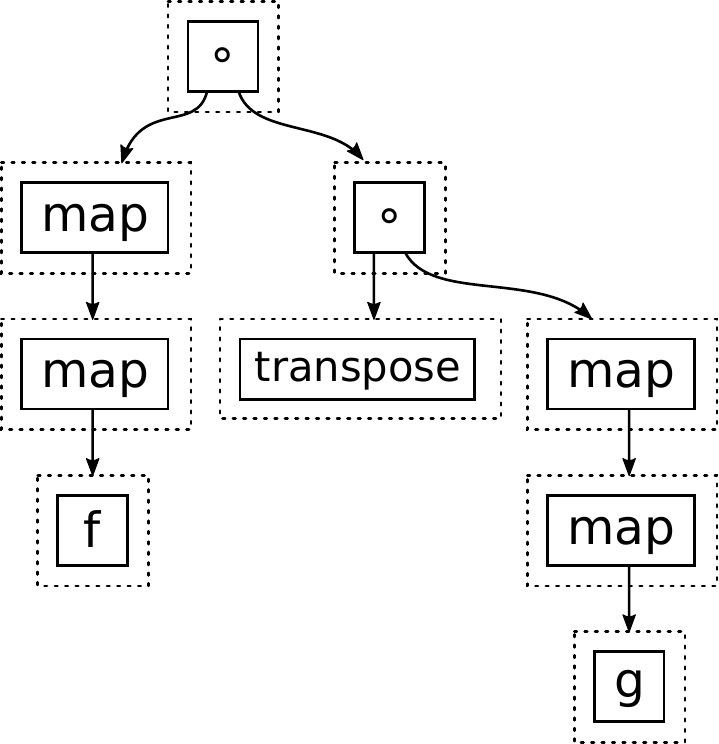}
  \caption{initialization}
  \label{fig:eqsat-ex-0}
  \end{subfigure}\hfill%
  \begin{subfigure}[b]{0.23\linewidth}
  \includegraphics[width=\linewidth]{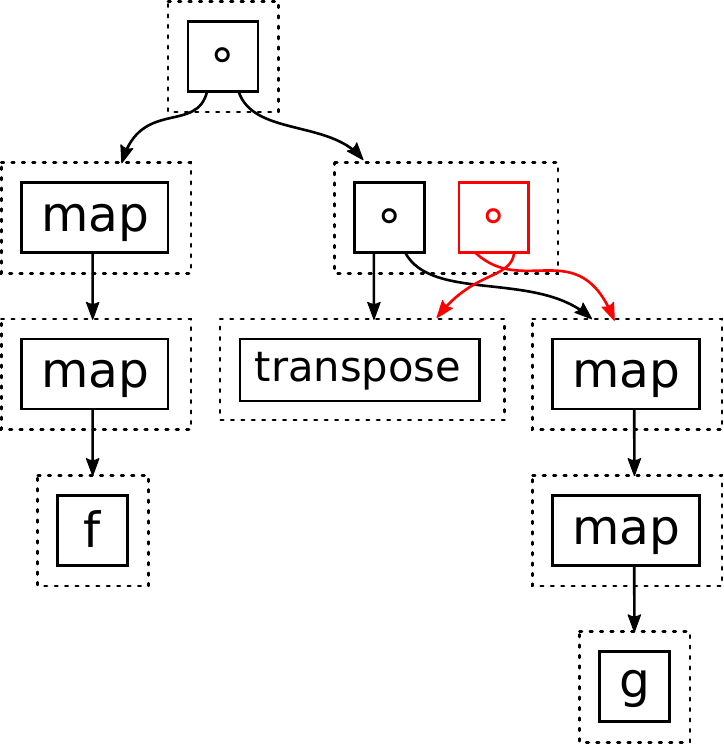}
  \caption{after rewrite (\ref{overview-move-transpose})}
  \label{fig:eqsat-ex-1}
  \end{subfigure}\hfill%
  \begin{subfigure}[b]{0.23\linewidth}
  \includegraphics[width=\linewidth]{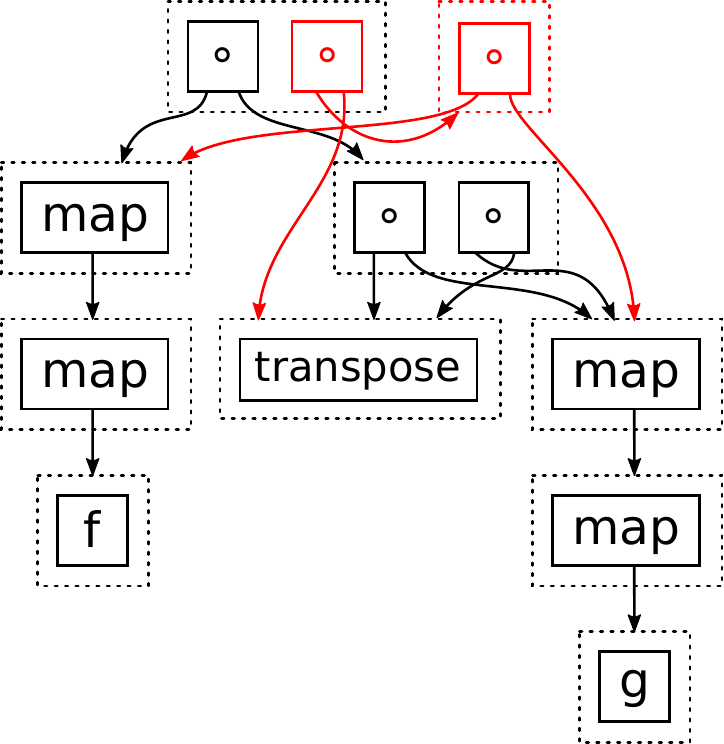}
  \caption{after rewrite (\ref{overview-comp-assoc})}
  \label{fig:eqsat-ex-2}
  \end{subfigure}\hfill%
  \begin{subfigure}[b]{0.29\linewidth}
  \includegraphics[width=\linewidth]{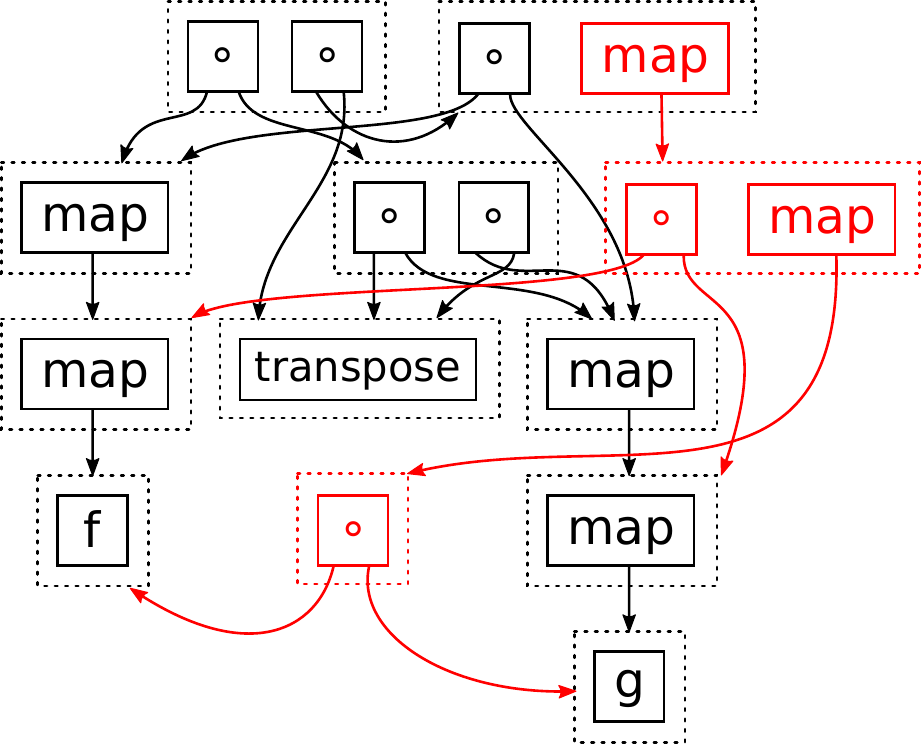}
  \caption{after two rewrite (\ref{overview-map-fusion})}
  \label{fig:eqsat-ex-3}
  \end{subfigure}
  \caption{Growing an e-graph for the term $(map~(map~f)) \circ (transpose \circ (map~(map~g)))$.
  An e-graph is a set of e-classes themselves containing equivalent e-nodes.
  The dashed boxes are e-classes, and the solid boxes are e-nodes.
  New e-nodes and e-classes are shown in red.}
  \label{fig:eqsat-ex}
\end{figure}
\smallskip

Equality saturation minimizes term size by exploring many rewrites while avoiding local minima.
First, an equality graph (e-graph) representing the initial term is constructed (\cref{fig:eqsat-ex-0}).
An \emph{e-graph} is a set of equivalence classes (e-classes).
An \emph{e-class} is a set of equivalent nodes (e-nodes).
An \emph{e-node} $F(e_1, .., e_n)$ is an $n$-ary function symbol ($F$) from the term language, associated with $n$ child e-classes ($e_i$).
Examples of symbols are $map$, $transpose$, and $\circ$.
The e-graph data structure is used during equality saturation to efficiently represent and rewrite a set of equivalent programs.

The initial e-graph is iteratively grown by applying rules non-destructively (\cref{fig:eqsat-ex-1,fig:eqsat-ex-2,fig:eqsat-ex-3}).
While standard term rewriting picks a single possible rewrite in a depth-first manner, equality saturation explores all possible rewrites in a breadth-first manner.
In an equality saturation iteration, rewrites are applied independently: they only depend on rewrites from previous iterations.
For the sake of simplicity, we only apply a handful of rewrite rules in \cref{fig:eqsat-ex}.
When applying a rewrite rule, the equality between its left-hand side and its right-hand side is recorded in the e-graph.
Rewrite rules stop being applied when a fixed point is reached (saturation), or when another stopping criteria is reached (e.g. timeout).
If saturation is reached, it means that all possible rewrites have been explored.

\begin{figure}
  \begin{minipage}[b]{0.45\linewidth}
    \centering
    \includegraphics[width=0.6\linewidth]{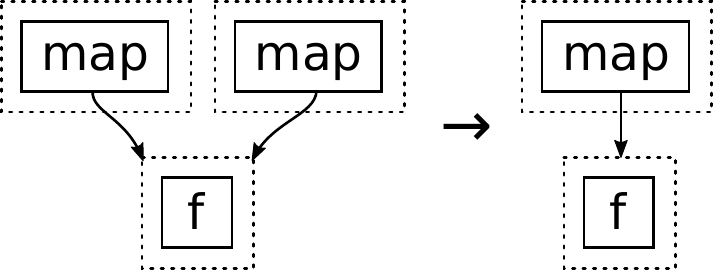}
    \caption{The congruence invariant simplifies the e-graph on the left by merging two identical e-nodes for $map~f$ into a single e-node as shown on the right.}
    \label{fig:congruence-invariant}
  \end{minipage}%
  \hfill
  \begin{minipage}[b]{0.5\linewidth}
    \centering
    \includegraphics[width=0.8\linewidth]{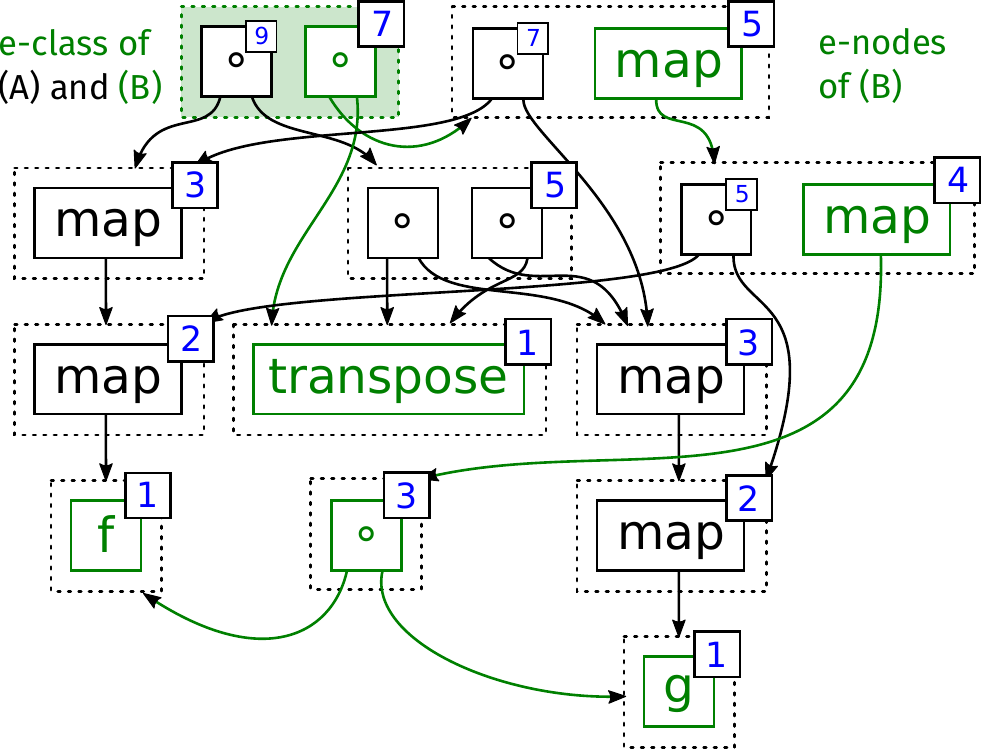}
    \caption{Smallest term size computed using an e-class analysis, shown for each e-class in the top-right corner in blue. Where it differs from its e-class value, the smallest term size of e-nodes is also shown. The e-class and e-nodes of the smallest term (\ref{overview-goal}) are shown in green.}
    \label{fig:eqsat-ex-extract}
  \end{minipage}
\end{figure}

An e-graph represents many equivalent terms and is far more compact than some set of terms as equivalent sub-terms are shared.
E-graphs can represent exponentially many terms in polynomial space, and even infinitely many terms in the presence of cycles \cite{willsey2021-egg}.
To maximize sharing, a \emph{congruence invariant} is maintained: intuitively identical e-nodes should not be in different e-classes (\cref{fig:congruence-invariant}).
Later we will see that even extensive sharing does not necessarily prevent e-graph sizes from exploding.

After rewriting completes, an \emph{extraction} procedure selects the best term from the e-graph using a cost function. Extraction can use a relatively simple bottom-up e-graph traversal if a \emph{local cost function} $c$  can be defined~\cite{panchekha2015-herbie}. With costs of type $K$, $c$ has signature $c(F(k_1: K, .., k_n: K)) : K$.
More complex cost functions require more complex extraction procedures \cite{wang2020-spores, wu2019-carpentry}.

An \emph{e-class analysis} \cite{willsey2021-egg} propagates analysis data of type $D$ in a bottom-up fashion, and can be used for extraction when the cost function is local.
An e-class analysis is defined by providing two functions: one to $make$ the analysis data from an $n$-ary symbol $F$ combined with the data $d_i$ of its child e-classes; and one to $merge$ the analysis data of e-nodes in the same e-class.
$$ make(F(d_1 : D, .., d_n : D)) : D $$
$$ merge(d_1 : D, d_2 : D) : D $$

To demonstrate e-class analysis, we compute the smallest term in each e-class of our example using an e-class analysis with analysis data $D = (\text{term size}, \text{term})$ as well as $make$ and $merge$ functions below.
The term sizes from the analysis are shown in \cref{fig:eqsat-ex-extract}.
The analysis reveals that there is a smaller term (\ref{overview-goal}) of size 7 in the same e-class as the original term (\ref{overview-start}) which has a size of 9 (top left in \cref{fig:eqsat-ex-extract}).

$$ make(F(d_1, .., d_n)) = (1 + \sum_{i} fst(d_i), F(snd(d_1), .., snd(d_n))) $$
$$ merge(d_1, d_2) = \text{if } fst(d_1) \leq fst(d_2) \text{ then } d_1 \text{ else } d_2 $$

\noindent

\paragraph{Limitations of automatic rewriting with equality saturation}
In practice, the applicability of equality saturation is limited by scaling issues.
As the e-graph grows, iterations become slower and require more memory.
The growth rate is aggravated by some combinations of rewrite rules, such as associativity and commutativity, that generate an exponential number of equivalent permutations \cite{wang2020-spores, nandi2020-synthesizing-CAD, willsey2021-egg}.
This makes exploring long rewrite sequences inherently hard, as the breadth-first exploration of all possible rewrites leads to an exponential increase of the search space.
We encounter these issues when attempting to optimize matrix multiplication in \Rise{} using equality saturation, as we discuss in \Cref{evaluation}.

Our contributions ameliorate these limitations by minimizing the size of the e-graphs required.
Sketch-guiding factors unfeasible equality saturations into a sequence of smaller, and hence feasible, equality saturations (\cref{sketching}).
Efficient encoding of the $\lambda$ calculus reduces the sizes of the e-graphs produced when optimizing functional programs, like \Rise{} programs (\cref{eqsat-bindings}).

%% file: section/sketching.tex
\section{Sketch-Guided Equality Saturation} 
\label{sketching}

This section introduces \emph{sketch-guided equality saturation}, a novel semi-automated process offering a trade-off between manually written rewriting strategies and fully automated equality saturation.
The programmer guides multiple equality saturations by specifying a sequence of sketches that describe how a program should evolve during optimization.
While rewriting strategies require fine-grained phase ordering, sketch-guided equality saturation enables coarse-grained phase ordering.
For example, instead of specifying how to sequence thousands of rewrite rules to achieve matrix multiplication blocking, specifying just two sketches suffices.

\subsection{The Intuition for Sketches}

When designing optimizations, programmers often visualize the desired \emph{shape} of the optimized program. An optimization, e.g.~with rewriting strategies or schedules, is often illustrated with program snippets that explain its effect \cite{koehler2021-elevate-imgproc, ragan2013-halide, chen2018-tvm, adams2019-learning, sioutas2020-schedule, ikarashi2021-guided, anderson2021-efficient}. Indeed, this is exactly how we have explained the loop nest blocking optimization in \cref{fig:mm-rewrite}.

Our \emph{key new insight} is that explanatory program snippets can be formalized as sketches and used to guide an optimization search, instead of manually writing strategies or schedules that perform the optimization directly.
\emph{Sketches} are program patterns that capture intuitions about the program shape while leaving details unspecified.
The guided optimization search still performs rewriting with semantic preserving rules, allowing sketches to leave out details without sacrificing correctness.
We define sketches formally in \cref{sketch-def}.

\begin{figure}
\begin{minipage}{0.75\linewidth}
\begin{center}
\begin{rise-sketch}[caption={Sketch for the \goalStyle{baseline} matrix multiplication goal}, label={mm-baseline-sketch}]
containsMap(m,                | for m:
 containsMap(n,               |  for n:
  containsReduceSeq(k,        |   for k:
   containsAddMul)))          |     .. + .. * ..
\end{rise-sketch}
\begin{rise-sketch}[caption={Sketch for the \goalStyle{blocking} matrix multiplication goal}, label={mm-blocking-sketch}]
containsMap(m / 32,           | for m / 32:
 containsMap(n / 32,          |  for n / 32:
  containsReduceSeq(k / 4,    |   for k / 4:
   containsReduceSeq(4,       |    for 4:
    containsMap(32,           |     for 32:
     containsMap(32,          |      for 32:
      containsAddMul))))))    |        .. + .. * ..
\end{rise-sketch}
\begin{rise-sketch}[caption={Sketch guide specifying how to split loops for \goalStyle{blocking}}, label={mm-blocking-split-sketch}]
containsMap(m / 32,           | for m / 32:
 containsMap(32,              |  for 32:
  containsMap(n / 32,         |   for n / 32:
   containsMap(32,            |    for 32:
    containsReduceSeq(k / 4,  |     for k / 4:
     containsReduceSeq(4,     |      for 4:
      containsAddMul))))))    |        .. + .. * ..
\end{rise-sketch}
\end{center}
\end{minipage}
\end{figure}

We illustrate by presenting the sketches for matrix multiplication blocking.
\Cref{mm-baseline-sketch} shows the sketch for the unoptimized \goalStyle{baseline} goal, specifying the desired program structure as two nested \inlineRise{map} patterns and a nested \inlineRise{reduce}, with innermost addition and multiplication operations. The formal definitions of \inlineRiseSketch{containsMap}, \inlineRiseSketch{containsReduceSeq} and \inlineRiseSketch{containsAddMul} are in \cref{sketch-def}.
The comments on the right show the equivalent nested \inlineRise{for} loops, using the same intuition as in \cref{fig:mm-rewrite}.

The sketch describing the \goalStyle{blocking} goal in \cref{mm-blocking-sketch} corresponds to the optimized program where the "loop nests" have been split and reordered so that the iteration space is chunked into blocks of $4 \times 32 \times 32$, each processed by the three innermost \inlineRise{for} loops.

Searching for the \goalStyle{blocking} goal can be made more tractable by specifying intermediate \emph{sketch guides}, and \cref{mm-blocking-split-sketch} is an example.
This sketch guide describes a program shape where the \inlineRise{map} and \inlineRise{reduce} patterns have been split but not yet reordered.

\Elevate{} rewriting strategies, as seen before in \cref{lst:elevate-blocking}, are detailed \emph{imperative} specifications of how to rewrite the program. In contrast, a sketch is a \emph{declarative} specification of the optimization goal, and equality saturation is used to search for a sequence of rewrites to achieve that goal.
A sequence of sketches (e.g., \cref{mm-blocking-split-sketch} followed by \cref{mm-blocking-sketch}) may be used to achieve a desired optimization when the equality saturation search with a single sketch as a goal does not succeed.



\subsection{Sketch Definition}
\label{sketch-def}


\newcommand{\sLang}{\textsc{SketchBasic}}

Sketches are specified in a  \sLang{} language with just four constructors.
The  syntax of \sLang{} and the set of terms that the constructors represent are defined in \cref{fig:sketch-lang-rep}.
A sketch $s$ represents a set of terms $R(s)$, such that $R(s) \subset T$ where $T$ denotes all terms in the language we rewrite.
We say that any $t \in R(s)$ satisfies sketch $s$.

The $?$ sketch is the least precise, representing all terms in the language.
The $F(s_1, .., s_n)$ sketch represents all terms that match a specific $n$-ary function symbol $F$ from the term language, and whose $n$ children satisfy sketches $s_i$.
The $contains(s)$ sketch represents all terms containing a term that satisfies sketch $s$.
Finally, the $s_1 \lor s_2$ sketch represents terms satisfying either $s_1$ or $s_2$.

\begin{figure}
  \begin{lstlisting}[mathescape, basicstyle={\linespread{.75}\ttfamily\normalsize}, frame=none, xleftmargin=0.25\linewidth]
  $S$ ::= $?$ | $F(S, .., S)$ | $contains(S)$ | $S \lor S$
  \end{lstlisting}
    \begin{align*}
      R(?) &= T = \{ F(t_1, .., t_n) \} \\
      R(F(s_1, .., s_n)) &= \{ F(t_1, .., t_n) \mid t_i \in R(s_i) \} \\
      R(contains(s)) &= R(s) \cup \{ F(t_1, .., t_n) \mid \exists t_i \in R(contains(s)) \} \\
      R(s_1 \lor s_2) &= R(s_1) \cup R(s_2)\\[-2em]
    \end{align*}
    \caption{Grammar of \sLang{} (top) and terms represented by \sLang{} (bottom).}
    \label{fig:sketch-lang-rep}
\end{figure}

When rewriting terms in a typed language, sketches may be annotated with a type sketch ($s :: pt$) constraining the type of terms.
If $R(pt)$ denotes the set of terms satisfying the type sketch $pt$, then $R(s :: pt) = R(s) \cap R(pt)$.
The grammar of type sketches depends on the language we rewrite.
We elide type sketches from our definition of \sLang{} for simplicity, but use them for \Rise{}.

Writing a useful sketch to guide an optimization search requires striking a balance between being too precise and too vague.
An overly precise sketch may exclude valid optimized programs with a slightly different structure.
An overly vague sketch may lead to finding undesirable programs.
This balance also interacts with the rewrite rules involved, since programs that may be found by the search are $R(s) \cap EQ(t, rules)$ where $EQ(t, rules)$ represents the set of terms that can be discovered to be equivalent to the initial term $t$ according to the given $rules$.
This means that using a more restricted set of rules generally enables specifying less precise sketches.
How to best select rules and write effective sketches are topics for future work.

\textit{Sketch Abstractions}.
Sketch abstractions are defined by combining generic constructs from \sLang{} with type annotations from the term language.
To illustrate, \cref{rise-sketch-abs} shows the sketch abstractions for our \Rise{} matrix multiplication case study.
Here \inlineRise{$\to$} is a function type, and \inlineRise{n.dt} an array type of \inlineRise{n} elements of domain type \inlineRise{dt}.
The type annotations restrict the iteration domains of patterns like \inlineRise{map} and \inlineRise{reduceSeq}.
We use similar definitions for other language constructs.


\begin{figure}
\begin{minipage}{0.75\linewidth}
\begin{center}
\begin{rise-sketch}[label={rise-sketch-abs}, caption={Some sketch abstractions used in the paper.}]
def containsMap(n: NatSketch, f: Sketch): Sketch =
  contains(app(map :: ?t $\to$ n.?dt $\to$ ?t, f))
def containsReduceSeq(n: NatSketch, f: Sketch): Sketch =
  contains(app(reduceSeq :: ?t $\to$ ?t $\to$ n.?dt $\to$ ?t, f))
def containsAddMul: Sketch =
  contains(app(app($+$, ?), contains($\times$)))
\end{rise-sketch}
\end{center}
\end{minipage}
\end{figure}

\begin{figure}[b]
  \centering
  \includegraphics[width=0.9\linewidth]{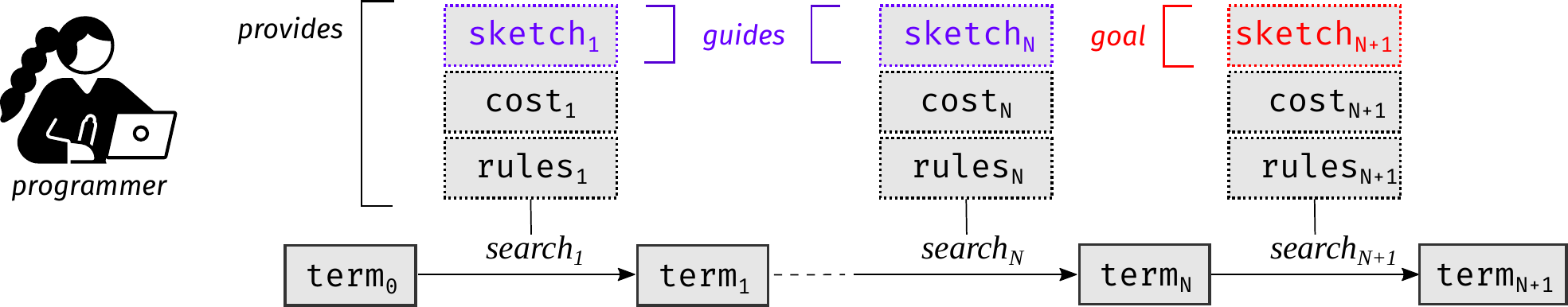}
  \caption{Sketch-Guided Equality Saturation.
  The programmer provides $N$ intermediate sketch guides and $1$ final goal sketch.
  Starting from the input term, $N+1$ consecutive equality saturation searches attempt to find a term satisfying each sketch, using the associated cost models and sets of rules (\cref{fig:search-diagram}).}
  \label{fig:sketching}
\end{figure}

\subsection{Sketch-Guided Equality Saturation}
\label{sketching-sub}

\begin{figure}[b]
\begin{minipage}{0.52\linewidth}
\begin{center}
\begin{lstlisting}[numbers=left, xleftmargin=2em, label={fig:sketching-alg}, caption={Sketch-Guided Equality Saturation Algorithm}, escapechar=|, mathescape=true, morekeywords={def, if, then, else},captionpos=b,abovecaptionskip={.5\baselineskip},moredelim={*[is][\rmfamily\itshape\color{black!50}]{~}{~}}]
def SGES(term, params): Option[Term] = |\label{line:sges}|
 if params.isEmpty
 then Some(term)
 else search(term, params.head) |\label{line:sges-search}|
      .and_then($\lambda$t. SGES(t, params.tail)) |\label{line:sges-rec}|
   
def search(term, param): Option[Term] =
 (sketch, cost, rules) = param
 g = ~create empty e-graph~
 normTerm = normalize(term) |\label{line:normalize}|
  ~using a configurable normal form~
 e = g.add(normTerm)
 ~grow~ g ~using~ rules ~until~ found(g, e, sketch) 
 if found(g, e, sketch) then |\label{line:found}|
  Some(extract(g, e, sketch, cost)) |\label{line:beam-extraction}|
 else
  None
\end{lstlisting}
\end{center}
\end{minipage}\hfill%
\begin{minipage}{0.45\linewidth}
\begin{center}
\includegraphics[width=\linewidth]{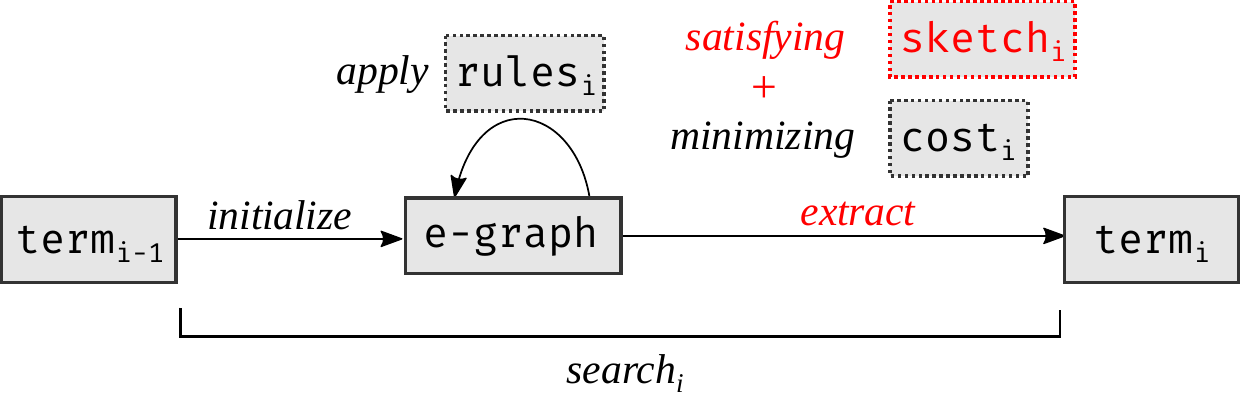}
\captionof{figure}{Sketch-Satisfying Equality Saturation implementing each $search_i$.  Changes made to standard equality saturation are highlighted in red.}\label{fig:search-diagram}%
\end{center}
\end{minipage}
\end{figure}

The process of guiding equality saturation with a sequence of sketches is illustrated in \cref{fig:sketching}.
The programmer provides a sequence of $N$ intermediate sketch guides and a final goal sketch: $sketch_1,\,..,\,sketch_{N+1}$.
Successive equality saturation searches are performed to find equivalent terms that satisfy each sketch in the sequence. As each sketch may be satisfied by many terms, the programmer must also provide a sequence of cost models $cost_1, .., cost_{N+1}$ to select the term to be used as the starting point for the next search. Sets of rewrite rules ($rules_1, .., rules_{N+1}$) are provided to grow the e-graph in each search. The cost model and set of rules may be identical for many or all of the searches, but we show examples in \cref{evaluation} where selectively restricting the set of rules reduces search runtime. \Cref{fig:search-diagram} shows how each search is performed, and how these searches differ from standard equality saturation.

The pseudo-code for the sketch-guided equality saturation algorithm is shown in \cref{fig:sketching-alg}.
The entry point is the \inlineRise{SGES} function (line \ref{line:sges}) that takes  a \inlineRise{term} and a sequence of sketches, cost models and rewrite rules (\inlineRise{params}). It repeatedly \inlineRise{search}es (line \ref{line:sges-search}) for each sketch using the associated cost model and rewrite rules, and outputs a term if found, otherwise nothing.

At the beginning of each \inlineRise{search}, we may normalize the input term (line \ref{line:normalize}) to apply destructive rewrites that are always desired before starting a purely additive equality saturation.
For our matrix multiplication running example we use $\beta\eta$ normal form.
The \inlineRise{extract} function (line \ref{line:beam-extraction}) is used to extract a term from the e-graph that satisfies the specified sketch while minimizing the specified cost model, and we describe it in the next subsection.
In this paper, we terminate equality saturation as soon as a program satisfying the current sketch is found, whether or not the cost could be further improved by a longer search.
This is because we give more value to satisfying the sketch than to minimizing the cost.
Other applications of sketch-guided equality saturation could use different stopping criteria.
The \inlineRise{found} function (line \ref{line:found}) is used to stop growing the e-graph by checking whether \inlineRise{extract} would succeed.

\subsection{Sketch-Satisfying Extraction}
\label{beam-extraction}

To \inlineRise{extract} the best program that satisfies a \sLang{} sketch $s$ from an e-graph $g$ we define a helper function $E(c, s, g)$, where $c$ is a cost function that must be monotonic and local.
While \inlineRise{extract} returns a program from an e-class $e$, the helper $E$ returns a map from e-classes to optional tuples of costs and terms.
After invoking $E$ we simply look up the e-class $e$ in the map and extract the term from the optional tuple.
For efficiency, we memoize previously computed results of $E$.
The \inlineRise{extract} function is recursively defined over the four \sLang{} cases as follows.

\smallskip

\textbf{Case 1:} $\bm{s =\ ?}$.
This case is equivalent to extracting the programs minimizing $c$ from the e-graph.
We implement this extraction as an e-class analysis (\cref{eqsat}) with data type $D = \text{Option}[(k, t)]$ and functions $make$ that constructs analysis data and $merge$ that combines analysis data from e-nodes in the same e-class:
\small
\begin{align*}
make(F(d_1, .., d_n)) = &\begin{cases}
  Some\left(\begin{array}{l}
    c(F(k_1, .., k_n)),\\
    F(t_1, .., t_n)
  \end{array}\right) & (k_i, t_i) \in d_i \\
  None & \text{otherwise}
\end{cases}\\
merge(d_1, d_2) = &\begin{cases}
  \text{if } k_1 \leq k_2 \text{ then } d_1 \text{ else } d_2 & (k_i, \_) \in d_i \\
  d_1 & (k_1, \_) \in d_1 \\
  d_2 & (k_2, \_) \in d_2 \\
  None & \text{otherwise}
\end{cases}
\end{align*}
\normalsize

\textbf{Case 2:} $\bm{s = F(s_1, .., s_n)}$.
We consider each e-class $e$ containing $F(e_1, .., e_n)$ e-nodes and the terms that should be extracted for each child e-class $e_i$.
We write $E(c,s, g)[e]$ for indexing into the map returned by $E$:
\small
\begin{align*}
E(c, F(s_1, .., s_n), g)[e] = \begin{cases}
  Some(c(F(k_1, .., k_n)), F(t_1, .., t_n)) & (k_i, t_i) \in E(c, s_i, g)[e_i] \\
  None & \text{otherwise}
\end{cases}
\end{align*}
\normalsize

\textbf{Case 3:} $\bm{s = contains(s_2)}$.
We use another e-class analysis and initialize the analysis data to $E(c, s_2, g)$ corresponding to the base case where $R(s_2) \subset R(contains(s_2))$.
To $make$ the analysis data we consider all terms that would contain terms from $s_2$ and keep the best by folding them using $merge$:
\small
\begin{align*}
make(&F(d_1, .., d_n)) = foldl~merge~None \\
  \{ &Some(c(F(k_1, .., k_j, .., k_n)), F(t_1, .., t_j, .., t_n)) \mid
     \quad i \neq j, (k_i, t_i) \in E(c, ?, g)[e_i], (k_j, t_j) \in d_j \}
\end{align*}
\normalsize

To $merge$ the analysis data, we do the same as for $s =\ ?$.

\medskip

\textbf{Case 4:} $\bm{s = s_1 \lor s_2}$. We $merge$ the results from $s_1$ and $s_2$:
\small
$$ E(c, s_1 \lor s_2, g)(e) = merge(E(c, s_1, g)(e), E(c, s_2, g)(e)) $$
\normalsize



%% file: section/eqsat-bindings.tex
\section{Efficient Equality Saturation for the Lambda Calculus} 
\label{eqsat-bindings}

\begin{table}
  \begin{tabular}{|c|c|c|}
  \hline
  \textbf{$\bm{\lambda}$ calculus} & $\bm{F}$ & $\bm{t_1, .., t_n}$ \\
  \hline
  \begin{rise}[numbers=none, frame=none]
  $\lambda$x. e
  \end{rise}%
  & lam x & e \\
  \hline
  \begin{rise}[numbers=none, frame=none]
  e$_1$ e$_2$
  \end{rise}%
  & app & e$_1$, e$_2$ \\
  \hline
  \begin{rise}[numbers=none, frame=none]
  x
  \end{rise}%
  & var x & \\
  \hline
  \end{tabular}
  \caption{Encoding of $\lambda$ calculus terms as $F(t_1, .., t_n)$ terms for the purposes of equality saturation.}
  \label{fig:rise-node-repr}
  \end{table}

\begin{figure}
\begin{align*}
    \tag{$\beta$-reduction}\label{beta-reduction}
    ({\color{cyan}\lambda x.}~b)~e \rewritesTo{}& {\color{purple}b[e/x]} \\
    \tag{$\eta$-reduction}\label{eta-reduction}
    {\color{cyan}\lambda x.}~f~x \rewritesTo{}& f &{\color{orange}\text{if}~x~\text{not free in}~f} \\
    \tag{map-fusion}\label{map-fusion}
    map~f~{\color{gray}(}map~g~{\color{gray}arg)} \rewritesTo{}& map~{\color{gray}({\color{cyan}\lambda x.}}~f~{\color{gray}(}g~{\color{gray}x))~arg} \\
    \tag{map-fission}\label{map-fission}
    map~{\color{gray}({\color{cyan}\lambda x.}}~f~gx) \rewritesTo{}& {\color{gray}{\color{cyan}\lambda y.}}~map~f~{\color{gray}(}map~{\color{gray}({\color{cyan}\lambda x.}}~gx{\color{gray})~y)}
    &{\color{orange}\text{if}~x~\text{not free in}~f}
\end{align*}
\caption{\Rise{} rewrite rules using {\color{purple}substitution}, {\color{cyan}name bindings} (lambda abstractions), and {\color{orange} freshness predicates}.}
\label{fig:example-rules}
\end{figure}

We now investigate the second direction shown in \cref{fig:overview} for scaling equality saturation to complex optimizations of functional programs, by exploring the engineering design choices required to efficiently encode a polymorphically typed lambda calculus within equality saturation.
A set of design choices are realized for the \Rise{} language in the new \kles{} implementation that is heavily inspired by the egg library \cite{willsey2021-egg}.

Lambda calculus terms can be encoded as terms of shape $F(t_1, .., t_n)$, as shown in \cref{fig:rise-node-repr}.
Variable names are not modeled directly as terms, but as operator metadata: 'lam x', 'lam y', 'var x' and 'var y' are all distinct operators.

Applying equality saturation to lambda calculus terms requires the efficient support of standard operations and rewrites.
\Cref{fig:example-rules} shows the standard rules of \ref{beta-reduction} and \ref{eta-reduction} that require dealing with substitution, name bindings and freshness predicates.
The other two rules encode \ref{map-fusion} and \ref{map-fission}, and are interesting because they introduce new name bindings on their right-hand-side.

\subsection{Substitution}
\label{substitution}

\ref{beta-reduction} requires substituting $b[e/x]$, but standard term substitution cannot be used during equality saturation as $b$ and $e$ are not terms but e-classes.
A simple way to address this is to use \emph{explicit substitution} as in egg's lambda calculus example \cite{willsey2021-egg}.
A syntactic constructor is added to represent substitution, with rewrite rules to encode its small-step behavior:
\begin{align*}
(a~b)[e/v] &\rewritesTo{} (a[e/v]~b[e/v]) \\
v[e/v] &\rewritesTo{} e
\end{align*}

Unfortunately explicit substitution adds all intermediate substitution steps to the e-graph, quickly exploding its size.
\Cref{lambda-eval} shows that this is a major problem in practice, making relatively simple rewrites unfeasible.
To avoid adding intermediate substitution steps to the e-graph, we propose \emph{extraction-based substitution} that works as follows.
\begin{enumerate}
    \item extract a term for each e-class involved in the substitution (i.e $b$ and $e$);
    \item perform standard term substitution;
    \item add the resulting term to the e-graph.
\end{enumerate}

\begin{figure}
    \includegraphics[width=0.9\linewidth]{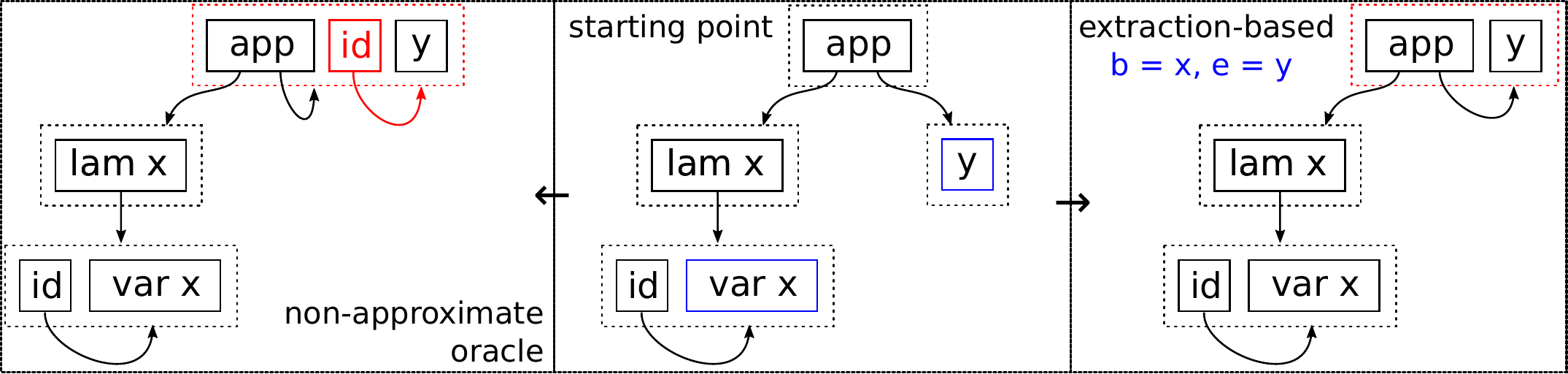}
    \caption{Example of $\beta$-reduction with extraction-based substitution (right).
    The initial e-graph (middle) represents $(\lambda x.~id~x)~y$.
    After $\beta$-reduction, the e-graph does not represent $id~y$ because $x$ has been extracted for $b$ in the rewrite rule; ignoring $id~x$.}
    \label{fig:extract-subs}
\end{figure}

\Cref{lambda-eval} demonstrates that extraction-based substitution is far more efficient than explicit substitution.
Extraction-based substitution is, however, an approximation as it computes the substitution for a subset of the terms represented by $b$ and $e$, and ignores the rest.
\Cref{fig:extract-subs} shows an example where the initial e-graph is in the middle, and the e-graph after extraction-based substitution with $b=x$ and $e=y$ on the right.
This particular choice results in an e-graph lacking the $id~x$ program that is included in the e-graph without approximation (left in~\cref{fig:extract-subs}).

In practice, we have not observed the approximation to be an issue when optimizing \Rise{} programs (\cref{evaluation}), and believe that two main reasons account for this.
Firstly, the substitution is computed on each equality saturation iteration, where different terms may be extracted, increasing coverage of the set of terms represented by $b$ and $e$.
Secondly, many of the ignored equivalences are recovered either by e-graph congruence, or by applying further rewrite rules.
Future work may investigate alternative substitution implementations to balance efficiency with non-approximation.

\subsection{Name Bindings}
\label{name-bindings}

In equality saturation inappropriate handling of name bindings easily leads to serious efficiency issues.
Consider rewrite rules like \ref{map-fusion} that create a new lambda abstraction on their right-hand side.
What name should be introduced when they are applied?
In standard term rewriting, generating a fresh name using a global counter (aka. gensym) is a common solution.
But if a new name is generated each time the rewrite rule is applied, the e-graph is quickly burdened with many $\alpha$-equivalent terms\footnote{Two $\lambda$ terms are $\alpha$-equivalent if one can be made equivalent to the other simply by renaming variables.}.

Fewer $\alpha$-equivalent terms are introduced if fresh names are generated as a function of the matched e-class identifiers.
However as the e-graph grows and e-classes are merged, e-class identifiers change, and $\alpha$-equivalent terms are still generated and duplicated in the e-graph.

De Bruijn indices~\cite{DEBRUIJN1972381} are a standard technique for representing lambda calculus terms without naming the bound variables, and avoid the need for $\alpha$ conversions.
If De Bruijn indices enable two $\alpha$-equivalent terms to be structurally equivalent, the standard e-graph congruence invariant prevents their duplication, by ensuring that equivalent e-nodes are not allocated to different e-classes.
Hence we translate our terms and rewrite rules to use De Bruijn indices instead of names, and achieve significant efficiency gains (\cref{lambda-eval}).

\paragraph{True equality modulo $\alpha$-renaming}
While De Bruijn indices give a significant performance improvement, they do not provide equality modulo $\alpha$-renaming for sub-terms.
Consider  $f~(\lambda x.~f) = \%0~(\lambda.~\%1)$, where $\%i$ are De Bruijn indices.
Although $\%0$ and $\%1$ are structurally different, they both correspond to the same variable $f$.
In practice, we have not observed this to be a significant issue when optimizing \Rise{} programs, but it does require care when comparing sub-terms that have a different number of surrounding lambdas.
Future work may investigate alternatives to De Bruijn indices, for example through hashing modulo $\alpha$-renaming \cite{maziarz2021-hash-mod-alpha-eq}, or through nominal rewriting techniques \cite{fernandez2007-nominal-rewriting}.

\paragraph{Translating name-based rules into index-based rules}
Using De Bruijn indices means that rewrite rules must manipulate terms with De Bruijn indices.
Thankfully, more user-friendly name-based rewrite rules can be automatically translated to the index-based rules used internally \cite{bonelli2000-bruijn-rewriting}.
An example demonstrating this is given in \cref{rewrite-example}.

\paragraph{Explicit or extraction-based substitution}
Both explicit substitution and extraction-based substitution are compatible with De Bruijn indices, and for explicit substitution we use the $\lambda s$ calculus \cite{kamareddine1995-lambda-v}.

\paragraph{Shifting De Bruijn indices}
De Bruijn indices must be shifted when a term is used with different surrounding lambdas (example in \cref{rewrite-example}).
As for substitution, shifting can be implemented with explicit rewrite rules, or with \emph{extraction-based index shifting}:
\begin{itemize}
    \item extract a term from the e-class whose indices need shifting;
    \item perform index shifting on the term;
    \item add the resulting term to the e-graph.
\end{itemize}
In \kles{} we use extraction-based index shifting whenever extraction-based substitution is used.

\paragraph{Avoiding Name Bindings using Combinators}
It is also possible to avoid name bindings entirely \cite{smith2021-access-patterns}.
For example, it is possible to introduce a function composition combinator `$\circ$' as in \cref{eqsat}, greatly simplifying the \ref{map-fusion} and \ref{map-fission} rules:
\begin{align}
    \tag{$\circ$-intro}
    f~(g~x) &\rewritesTo{} (f \circ g)~x \\
    \tag{map-fusion$_2$}
    map~f \circ map~g &\rewritesTo{} map~(f \circ g) \\
    \tag{map-fission$_2$}
    map~(f \circ g) &\rewritesTo{} map~f \circ map~g
\end{align}

However, this approach has its own downsides.
Associativity rules are required, which we know increases the growth rate of the e-graph \cite{willsey2021-egg}.
Only using a left-/right-most associativity rule avoids generating too many equivalent ways to parenthesize terms.
But other rewrite rules now have to take this associativity convention into account, making their definition more difficult and their matching more expensive.
In general, matching modulo associativity or commutativity are algorithmically hard problems \cite{benanav1987-complexity-matching}.

The function composition $\circ$ combinator on its own is also not sufficient to remove the need for name bindings.
At one extreme, combinatory logic could be used as any lambda calculus term can be represented, replacing function abstraction by a limited set of combinators.
However, translating a lambda calculus term into combinatory logic results in a term of size $O(n^3)$ in the worst case, where $n$ is the size of the initial term \cite{lachowski2018-complexity}.
Translating existing rewrite systems to combinatory logic would be challenging in itself.

\subsection{Freshness Predicates}
\label{predicates}

Handling predicates is not trivial in equality saturation.
The \ref{eta-reduction} has the side condition that "$\text{if } x \text{ not free in } f$", but in an e-graph $f$ is an e-class and not a term.

The predicate could be precisely handled by filtering $f$ into $f' = \{ t \mid t \in f\text{, }x~\text{not free in}~t \}$, and using $f'$ on the right-hand-side of the rule.
However, this requires splitting an e-class in two: one that satisfies the predicate, and one that does not.
We do not attempt such a split as it would reduce sharing, increase e-graph size and be incompatible with the congruence invariant unless the very notion of equality is changed.

The design of \kles{} makes the engineering trade-off to only apply the \ref{eta-reduction} rewrite rule if $\forall t \in f.~x~\text{not free in}~t$, following egg's lambda calculus example \cite{willsey2021-egg}.
Advantages are that this predicate is efficient to compute using an e-class analysis, and that there is no need to split the e-class.
The disadvantage is that it is an approximation that ignores some valid terms.

\begin{figure}
    \includegraphics[width=0.55\linewidth]{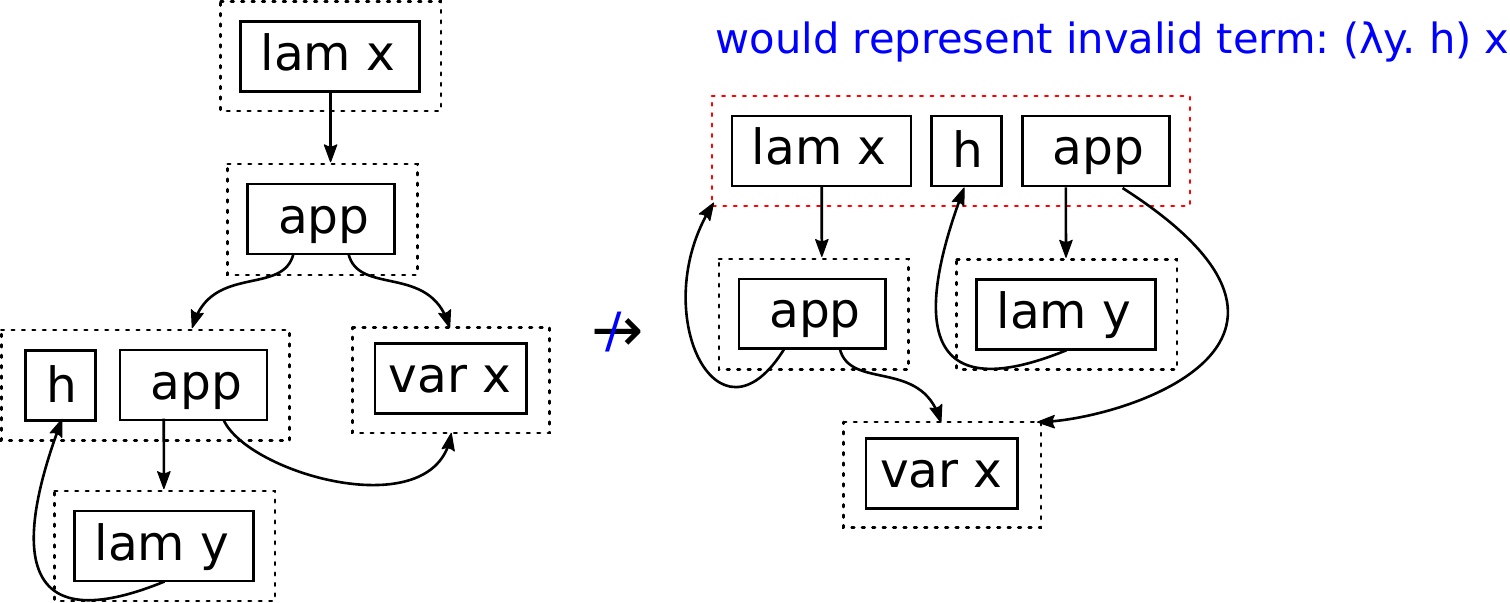}
    \caption{Example where $\eta$-reduction is not applied.
    The initial e-graph represents $\lambda x.~h~x$, but $\eta$-reduction is not triggered because $h = (\lambda y.~h)~x$ where $x$ is free.
    Using $\exists$ instead of $\forall$ in our predicate, we would obtain the e-graph on the right which represents $h$, but also invalid terms such as $(\lambda y.~h)~x$ where $x$ is no longer bound.}
    \label{fig:missed-eta-example}
\end{figure}

\Cref{fig:missed-eta-example} shows an example where \ref{eta-reduction} is not applied.
In practice, we have not observed the approximation to be an issue, e.g.\ for the results presented in \cref{evaluation}.

\subsection{Adding Polymorphic Types}
\label{types}

Typed lambda calculi are pervasive, e.g.~as the foundation for almost all functional languages, and a key consideration is how to add types to the e-graph.
More specifically, we look at how polymorphic types interact with equivalence classes.
If types can be computed in a bottom-up fashion, an e-class analysis can be used, similar to how the size and shape of tensors is computed in \cite{wang2020-spores}.
However, if polymorphic types are monomorphized, their instantiation is context-dependent and cannot be computed in a bottom-up fashion.
For example, consider the terms $(\lambda x.~x)~(0: int)$ and $(\lambda x.~x)~(0.0: float)$.
In \Rise{}, the identify function is monomorphized and has two different type instantiations that should live in different e-classes: $\lambda x.~x : \text{int} \rightarrow \text{int}$ and $\lambda x.~x : \text{float} \rightarrow \text{float}$.
Hence, in \kles{} instantiated types are embedded in the e-graph instead of computed by an analysis.
Each e-class is associated with a type that all of its e-nodes must satisfy.

In an e-graph there is a tension between sharing and the availability of contextual information in a given e-class.
Type instantiation prevents sharing, as the same polymorphic expression produces a different e-class at each type. However, instantiation provides additional contextual information, as each e-class is associated with a precise monotype.

\paragraph{Hash-consing types}
Since types are duplicated many times in the e-graph, and since structural type equality is often required, we hash-cons types for efficiency \cite{filliatre2006-hashconsing}.
Alternatively, types could be stored in the e-graph itself to provide equational reasoning at the type level, but this is not necessary for this paper.

\subsection{Compiling user-defined rewrite rules}
\label{rewrite-example}

To avoid explicit typing or explicit use of De Bruijn indices in user-defined rewrite rules, user-friendly name-based and partially typed rewrite rules are compiled internally as required.

Types are inferred with \Rise{} type inference: both sides of rewrite rules can be seen as terms where free variables correspond to pattern variables.
After inferring the types on the left-hand-side, we check that the right-hand-side is well-typed for any well-typed left-hand-side match.
When applied, typed rewrite rules match (deconstruct) types with their left-hand-side, and construct types on their right-hand-side.
Type annotations can be used to constrain the inferred types.

Bound variables are replaced with their corresponding De Bruijn index, and indices are shifted as required for terms with differing numbers of surrounding lambdas.
We illustrate with examples.

\paragraph{Example 1:  $\eta$-reduction}

\begin{align*}
\lambda x.~f~x \quad\rewritesTo{}\quad& f &\text{if}~x~\text{not free in}~f
\end{align*}

\kles{} translates this rule into:
\begin{align*}
(\lambda. f~(\%0 :\ t_0)) :\ t_0 \to\ t_1 \quad\rewritesTo{}\quad& (\varphi^{-1}_1 f) :\ t_0 \to\ t_1 &\text{if}~\%0~\text{not free in}~f
\end{align*}

The transformed rule uses a De Bruijn index $\%0$ for the bound variable, and pattern variables otherwise: $f$, $t_0$ and $t_1$. It provides index shifting through   $\varphi^{-1}_1 f$ that shifts all indices $\ge 1$ by $-1$ as a surrounding $\lambda$ has been removed.
Some types are matched on the left-hand-side ($t_0$ and $t_1$), and used to construct types on the right-hand-side ($t_0 \to t_1$).
Some types, like the type of $f$, are not matched on the left-hand-side as \kles{} avoids matching on redundant type information to some extent, assuming that the terms matched are well-typed.

\paragraph{Example 2: $\eta$-abstraction}

\begin{equation*}
f :\ t_0 \to t_1 \quad\rewritesTo{}\quad \lambda x.~f~x
\end{equation*}
$\eta$-abstraction illustrates how type annotations may be used, and are sometimes required.
\kles{} translates this rule into:
\begin{equation*}
f :\ t_0 \to t_1 \quad\rewritesTo{}\quad (\lambda.~(((\varphi^1_0 f) : t_0 \to t_1)~(\%0 : t_0)) :\ t_1) :\ t_0 \to t_1
\end{equation*}
A type error occurs if the type of $f$ is not annotated as in $f :\ t_0 \to t_1$ on the left-hand-side, since the right-hand-side requires it to be a function type. 

%% file: section/evaluation.tex
\section{Evaluation}
\label{evaluation}

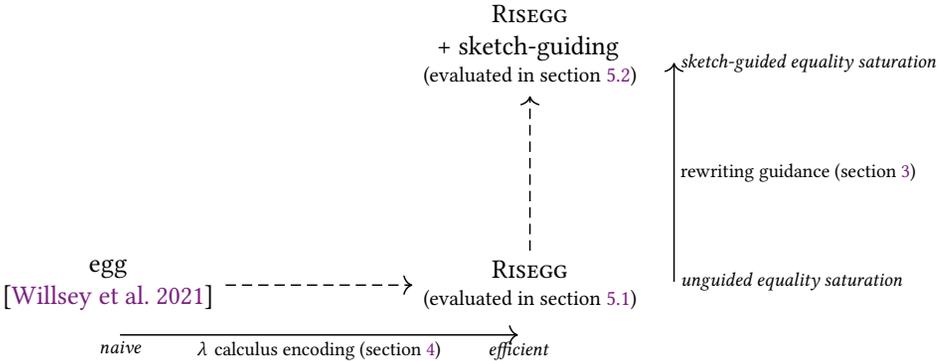
\begin{figure}[b]
    \centering
\[\begin{tikzcd}[sep=tiny]
	&&&&&&&&&&& \text{\parbox{8em}{\centering{}\kles{}\\ + sketch-guiding\\\footnotesize(evaluated in \cref{mm-eval})}} & {} \\
	\\
	\\
	{} \\
	& {} &&&& {} & {} \\
	\\
	& {} \\
	\\
	{} & {} & \text{\parbox{8em}{\centering{}egg\\\cite{willsey2021-egg}}} &&&&&&&&& \text{\parbox{8em}{\centering{}\kles{}\\\footnotesize(evaluated in \cref{lambda-eval})}} & {} \\
	{} & {} & {} &&&& {} &&&&& {} & {}
	\arrow["\textit{unguided equality saturation}"' at start, "\text{rewriting guidance (\cref{sketching})}"', "\textit{sketch-guided equality saturation}"' at end, from=9-13, to=1-13]
	\arrow["\textit{naive}"' at start, "\lambda~\text{calculus encoding (\cref{eqsat-bindings})}"', "\textit{efficient}"' at end, from=10-3, to=10-12]
	\arrow[dashed, from=9-12, to=1-12]
	\arrow[dashed, from=9-3, to=9-12]
\end{tikzcd}\]
\caption{We first evaluate our efficient $\lambda$ calculus encoding before evaluating sketch-guided equality saturation using this encoding.}
    \label{fig:eval-overview}
\end{figure}

We now evaluate our two proposed techniques for scaling equality saturation to complex optimizations of functional programs.
\Cref{fig:eval-overview} illustrates our evaluation process.
Starting from the naive lambda calculus encoding used in egg's example \cite{willsey2021-egg}, we first investigate the effectiveness of the new encoding from \cref{eqsat-bindings}, implemented in \kles{}, for unguided equality saturation (\cref{lambda-eval}).
Thereafter, we adopt the new lambda calculus encoding in \kles{}, and compare unguided and sketch-guided equality saturation to reproduce seven complex matrix multiplication optimizations previously applied with \Elevate{} rewriting strategies (\cref{mm-eval}).

\subsection{Impact of Lambda Calculus Encoding}
\label{lambda-eval}

The efficiency of equality saturation for the lambda calculus is evaluated by attempting to discover three rewrite goals using four combinations of the substitution and name binding techniques outlined in \cref{eqsat-bindings}.
The naive lambda calculus encoding uses explicit substitution and variable names.
The efficient encoding uses extraction-based substitution to avoid intermediate substitution steps, and De Bruijn indices to avoid duplicating $\alpha$-equivalent terms.
\emph{Discovering a rewrite goal} means that it is feasible to grow an e-graph starting from the initial program until the goal program is represented in the e-class of the initial program.

\paragraph{Experimental Setup}
The alternate encodings are realized for an untyped subset of the \Rise{} language in an early prototype of \kles{}.\footnote{https://github.com/Bastacyclop/egg-rise}
To measure search runtime, we use egg's built-in mechanisms, falling back to the GNU time utility in case of an out-of-memory exception.
Maximum memory residency is measured with  the GNU time utility.
The experiments are run on a laptop with an AMD Ryzen 5 PRO 2500U processor, and we limit the available RAM to 2 GB.

\paragraph{Rewrite Goals}
We use three rewrite goals with increasing complexity.

\def\reductionGoal {\goalStyle{reduction}}
The \reductionGoal{} rewrite goal in \cref{fig:reduction-rewrite} is based on a unit test from egg's lambda calculus example, and only uses \ref{eta-reduction} and \ref{beta-reduction} rules.
The lambda calculus examples from egg are relatively simple, as the rewrite rules involved do not introduce new names on their right-hand side and in most cases do not increase term size: the e-graph size does not grow explosively.

\begin{figure}[b]
\begin{rise}
($\lambda$comp.
  ($\lambda$add1.
    comp add1 (comp add1 (comp add1 (comp add1 (comp add1 (comp add1 add1)))))
  ) ($\lambda$y. y + 1)
) ($\lambda$f. $\lambda$g. $\lambda$x. f (g x))
\end{rise}
$$ \rewritesTo{}^* $$
\begin{rise}
$\lambda$x. ((((((x + 1) + 1) + 1) + 1) + 1) + 1) + 1
\end{rise}
\caption{\textbf{\reductionGoal{} rewrite goal.}
The initial program creates a \inlineRise{comp} combinator for function composition and uses it to compose the \inlineRise{add1} function with itself 7 times.
All uses of \inlineRise{comp} and \inlineRise{add1} are $\beta$-reduced in the final program, which simply applies \inlineRise{+ 1} to its input value 7 times.}
\label{fig:reduction-rewrite}
\end{figure}

\def\fission {\goalStyle{fission}}
The \fission{} rewrite goal in \cref{fig:fission-rewrite} adds the use of \ref{map-fusion} and \ref{map-fission} rewrite rules that introduce new name bindings on their right-hand-side, and interact with each other as well as \ref{beta-reduction} to create many possibilities: the e-graph size starts to explode.

\begin{figure}[b]
\begin{rise}
map ($\lambda$x. f$_5$ (f$_4$ (f$_3$ (f$_2$ (f$_1$ x)))))
\end{rise}
$$ \rewritesTo{}^* $$
\begin{rise}
$\lambda$y. map ($\lambda$x. f$_5$ (f$_4$ (f$_3$ x))) (map ($\lambda$x. (f$_2$ (f$_1$ x))) y)
\end{rise}
\caption{\textbf{\fission{} rewrite goal.}
The initial program successively applies \inlineRise!f$_1$! to \inlineRise!f$_5$! inside a \inlineRise{map} pattern.
The final program first applies \inlineRise!f$_1$! and \inlineRise!f$_2$! inside one \inlineRise{map} pattern, before applying \inlineRise!f$_3$! to \inlineRise!f$_5$! inside another \inlineRise{map} pattern.}
\label{fig:fission-rewrite}
\end{figure}

\def\binomial {\goalStyle{binomial}}
The \binomial{} rewrite goal from \cref{fig:binom-rewrite} is a real optimization that requires 6 more rewrite rules, increasing interactions between rewrite rules and aggravating the e-graph growth rate.
A binomial filter is an essential component of many image processing pipelines, where it reduces noise or detail.
It is a 2D convolution, and in \Rise{} uses a range of patterns including \inlineRise{zip}, \inlineRise{transpose} and \inlineRise{slide} that reshape arrays in various ways.
The purpose of the rewrite is to separate the 2D convolution into two 1D convolutions according to the well-known convolution kernel equation:
\begin{equation*}\label{binom-weights}
\scriptsize
    \left[\begin{matrix}
        1 & 2 & 1 \\
        2 & 4 & 2 \\
        1 & 2 & 1 \\
    \end{matrix}\right]
    =
    \left[\begin{matrix}
        1 \\
        2 \\
        1 \\
    \end{matrix}\right]
    \times
    \left[\begin{matrix}
        1 & 2 & 1 \\
    \end{matrix}\right]
\end{equation*}

This separation optimization reduces both memory accesses and arithmetic complexity.
Although more complex than the previous two rewrite goals, this rewrite goal is still relatively simple and we consider that unguided equality saturation should at least scale to this goal to be useful in practice for \Rise{}.
Prior work achieved this optimization by orchestrating 30 rewrite rules -- including 17 $\eta$/$\beta$-reductions -- with \Elevate{} rewriting strategies \cite{koehler2021-elevate-imgproc}.

\begin{figure}
  \begin{rise}
map (map $\lambda$nbh. dot (join weights2d) (join nbh))
  (map transpose (slide 3 1 (map (slide 3 1) input)))
  \end{rise}
  $$ \rewritesTo{}^* $$
  \begin{rise}
map ($\lambda$nbhL. map ($\lambda$nbhH. dot weightsH nbhH)
    (slide 3 1 (map ($\lambda$nbhV. dot weightsV nbhV) transpose nbhL))) (slide 3 1 input)
  \end{rise}
  \caption{\textbf{\binomial{} rewrite goal.}
  The initial \Rise{} program iterates over 2D neighborhoods (\inlineRise{nbh}).
  A \inlineRise{dot} product is computed between the weights (\inlineRise{weight2d} shown in \cref{binom-weights}) and each neighborhood.
  The final program iterates over two 1D neighborhoods (vertical and horizontal) instead.}
  \label{fig:binom-rewrite}
\end{figure}

\paragraph{Results}

\newcommand{\yes}{\ding{51}}
\newcommand{\no}{\ding{55}}
\begin{table}[b]
\small
\begin{tabular}{|c|c|c|c|r|r|r|r|r|} \hline
\multicolumn{2}{|c|}{$\lambda$ calculus encoding} & \multirow{2}{*}{\textbf{goal}} & \multirow{2}{*}{\textbf{found?}} & \multirow{2}{*}{\textbf{runtime}} & \multirow{2}{*}{\textbf{RAM}} & \multirow{2}{*}{\textbf{rules}} & \multicolumn{2}{c|}{e-graph size} \\ \cline{1-2}\cline{8-9}
\textbf{\footnotesize extraction?} & \textbf{\footnotesize De Bruijn?} &&&&&& \textbf{\footnotesize e-nodes} & \textbf{\footnotesize e-classes} \\ \hline \hline
\no & \no & \reductionGoal{} & \yes & 0.02s & 3 MB & 0.5K & 0.5K & 0.1K \\ \hline
\no & \no & \fission{} & \color{red}\no & 16s & $>$2000 MB &  &  &  \\ \hline
\no & \no & \binomial{} & \color{red}\no & 15s & $>$2000 MB &  &  &  \\ \hline \hline

\no & \yes & \reductionGoal{} & \yes & 0.2s & 35 MB & 28K & 53K & 25K \\ \hline
\no & \yes & \fission{} & \yes & 0.3s & 36 MB & 39K & 21K & 10K \\ \hline
\no & \yes & \binomial{} & \color{red}\no & 30s & $>$2000 MB &  &  &  \\ \hline \hline

\yes & \no & \reductionGoal{} & \yes & 0.004s & 3 MB & 0.1K & 0.3K & 0.2K \\ \hline
\yes & \no & \fission{} & \yes & 0.006s & 3 MB & 0.2K & 1K & 0.7K \\ \hline
\yes & \no & \binomial{} & \color{red}\no & 20s & $>$2000 MB &  &  &  \\ \hline \hline

\rowcolor{green!45}
\yes & \yes & \reductionGoal{} & \yes & 0.002s & 3 MB & 0.1K & 0.2K & 0.1K \\ \hline
\rowcolor{green!45}
\yes & \yes & \fission{} & \yes & 0.006s & 3 MB & 0.6K & 0.6K & 0.3K \\ \hline
\rowcolor{green!45}
\yes & \yes & \binomial{} & \yes & 0.1s & 8 MB & 5K & 3K & 1K \\ \hline
\end{tabular}
\caption{Evaluating the efficiency of lambda calculus encoding techniques on three rewrite goals.
Combining extraction-based substitution and De Bruijn indices minimizes runtime and memory consumption (green background), and is the only encoding that finds the \binomial{} rewrite goal.}
\label{fig:lambda-res}
\end{table}

\Cref{fig:lambda-res} compares the performance of equality saturation using different combinations of substitution (explicit/extraction-based) and name binding (named/De Bruijn) techniques.
It reports whether the goal is found, the search runtime, memory consumption, number of applied rewrite rules, and e-graph size.
The simple \reductionGoal{} goal is found by all combinations, although explicit substitution with De Bruijn indices is less efficient with 25K rewrite rules, 25K e-classes, and occupying 35 MB.
The \fission{} goal is not found if explicit substitution is used with named variables, exhausting the 2 GB memory and showing that this encoding is particularly inefficient.

The \binomial{} goal is only found by combining extraction-based substitution and De Bruijn indices.
With this encoding, all three rewrite goals are found by applying less than 5K rewrite rules, producing e-graphs with fewer than 3K e-nodes and 1K e-classes, and occupying less than 8 MB of memory.
For all three rewrite goals, this combination provides the fastest searches and most compact e-graphs, often by orders of magnitude.
We conclude that \emph{combining extraction-based substitution and De Bruijn indices gives an efficient encoding of lambda calculus for equality saturation}.

\subsection{Impact of Sketch Guidance}
\label{mm-eval}

This section compares unguided equality saturation to the new sketch-guided equality saturation to achieve complex optimization goals.
In the evaluation, \emph{both} equality saturation techniques use the efficient lambda calculus encoding from \cref{lambda-eval}.
Matrix multiplication is selected as the case study as it allows us to compare against published \Elevate{} strategies that specify optimizations equivalent to TVM schedules~\cite{hagedorn2020-elevate}.
TVM is a state-of-the-art deep learning compiler \cite{chen2018-tvm}, and \citet{hagedorn2020-elevate} demonstrate that expressing optimizations performed by TVM as compositions of rewrites is possible and achieves the same high performance as TVM.
We evaluate the seven matrix multiplication optimizations described in the TVM manual and presented in~\cite{hagedorn2020-elevate}.
The optimizations are typical compiler optimizations, including loop blocking, loop permutation, vectorization, and multi-threading.

In this evaluation, we compare how much runtime and memory are required for unguided equality saturation and our sketch-guided equality saturation.
For both guided and unguided equality saturation the optimization goal is specified as a sketch, that acts as the stopping criteria.
This is less restrictive than the searches for a goal program in the previous subsection as the sketch goal may be satisfied by many programs.

We validate that the result of each complex optimizations is high performance code as follows.
When a program satisfying the optimization goal sketch is found, we check that the generated C code is equivalent, modulo variable names, to the manually optimized versions that already demonstrated performance competitive with TVM.


\paragraph{Experimental Setup}
The full version of \kles{}\footnote{\url{https://github.com/rise-lang/shine/tree/sges/src/main/scala/rise/eqsat}} is implemented in Scala, and we use standard Java utilities for measurements: \texttt{System.nanoTime()} to measure search runtime, and the \texttt{Runtime} api to approximate maximum heap memory residency with regular  sampling.

The experiments are performed on two platforms.
For \Elevate{} strategies and our sketch-guided equality saturation, we use a less powerful AMD Ryzen 5 PRO 2500U with 4 GB of RAM available to the JVM.
For unguided equality saturation, we use a more powerful Intel Xeon E5-2640 v2 with 60 GB of RAM available to the JVM.

\def\baseline {\goalStyle{baseline}}
\def\blocking {\goalStyle{blocking}}
\def\vectorization {\goalStyle{vectorization}}
\def\loopperm {\goalStyle{loop-perm}}
\def\arraypacking {\goalStyle{array-packing}}
\def\cacheblocks {\goalStyle{cache-blocks}}
\def\parallel {\goalStyle{parallel}}

\begin{table}
\small
\begin{tabular}{|l|c|r|r|r|r|r|}
\hline
\textbf{goal} & \textbf{found?} & \textbf{runtime} & \textbf{RAM} & \textbf{rules} & \textbf{e-nodes} & \textbf{e-classes} \\
\hline
\baseline{} & \yes & 0.5s & 0.02 GB & 2 & 51 & 49 \\
\hline
\blocking{} & \yes & >1h & 35 GB & 5M & 4M & 2M \\
\hline
\vectorization{} & \color{red} \no & >1h & >60 GB &  &  &  \\
\hline
\loopperm{} & \color{red} \no & >1h & >60 GB &  &  &  \\
\hline
\arraypacking{} & \color{red} \no & 35mn & >60 GB &  &  &  \\
\hline
\cacheblocks{} & \color{red} \no & 35mn & >60 GB &  &  &  \\
\hline
\parallel{} & \color{red} \no & 35mn & >60 GB &  &  &  \\
\hline
\end{tabular}
\caption{Runtime and memory consumption for \textbf{unguided equality saturation} with efficient lambda calculus encoding.
Only the \baseline{} and \blocking{} optimization goals are found, with other optimizations exceeding 60 GB.}
\label{fig:s2}
\end{table}

\paragraph{Unguided Equality Saturation}
\Cref{fig:s2} shows the runtime and memory consumption required to find the optimization goals with unguided equality saturation.
The search terminates when the sketch describing the optimization goal is found in the e-graph.
Most optimization goals are not found before exhausting the 60 GB of available memory.
Only the \baseline{} and \blocking{} goals are found, and the search for \blocking{} requires more than 1h and about 35 GB of RAM.
Millions of rewrite rules are applied, and the e-graph contains millions of e-nodes and e-classes.
More complex optimizations involve more rewrite rules, creating a richer space of equivalent programs but exhausting memory faster.
As examples, \vectorization{} and \loopperm{} use vectorization rules, while \arraypacking{}, \cacheblocks{}, and \parallel{} use rules for optimizing memory storage.

\begin{table}
\small
\begin{tabular}{|l|c|c|r|r|r|r|r|}
\hline
\textbf{goal} & \textbf{sketch guides} & \textbf{found?} & \textbf{runtime} & \textbf{RAM} & \textbf{rules} & \textbf{e-nodes} & \textbf{e-classes} \\
\hline
\baseline{} & 0 & \yes & 0.5s & 0.02 GB & 2 & 51 & 49 \\
\hline
\blocking{} & 1 & \yes & 7s & 0.3 GB & 11K & 11K & 7K \\
\hline
\vectorization{} & 2 & \yes & 7s & 0.4 GB & 11K & 11K & 7K \\
\hline
\loopperm{} & 2 & \yes & 4s & 0.3 GB & 6K & 10K & 7K \\
\hline
\arraypacking{} & 3 & \yes & 5s & 0.4 GB & 9K & 10K & 7K \\
\hline
\cacheblocks{} & 3 & \yes & 5s & 0.5 GB & 9K & 10K & 7K \\
\hline
\parallel{} & 3 & \yes & 5s & 0.4 GB & 9K & 10K & 7K \\
\hline
\end{tabular}
\caption{Runtime and memory consumption for \textbf{sketch-guided equality saturation} with efficient lambda calculus encoding.
All optimizations are found in seconds using less than 0.5 GB of memory, and requiring at most 3 sketch guides.}
\label{fig:s3}
\end{table}

\paragraph{Sketch-Guided Equality Saturation}
\Cref{fig:s3} shows the runtime and memory consumption for sketch-guided equality saturation, where sketches guide the optimization process and break a single equality saturation search into multiple.
All optimizations are found in less than 10s, using less than 0.5 GB of RAM.
Interestingly, the number of rewrite rules applied by sketch-guided equality saturation is in the same order of magnitude as for the manual \Elevate{} strategies reported in~\cite{hagedorn2020-elevate}.
On one hand, equality saturation applies more rules than necessary because of its explorative nature.
On the other hand, \Elevate{} strategies apply more rules than necessary because they re-apply the same rule to the same sub-expression and do not necessarily orchestrate the shortest possible rewrite path.
The e-graphs contain no more than $10^{4}$ e-nodes and e-classes, two orders of magnitude less than the $10^{6}$ required for \blocking{} without sketch-guidance.

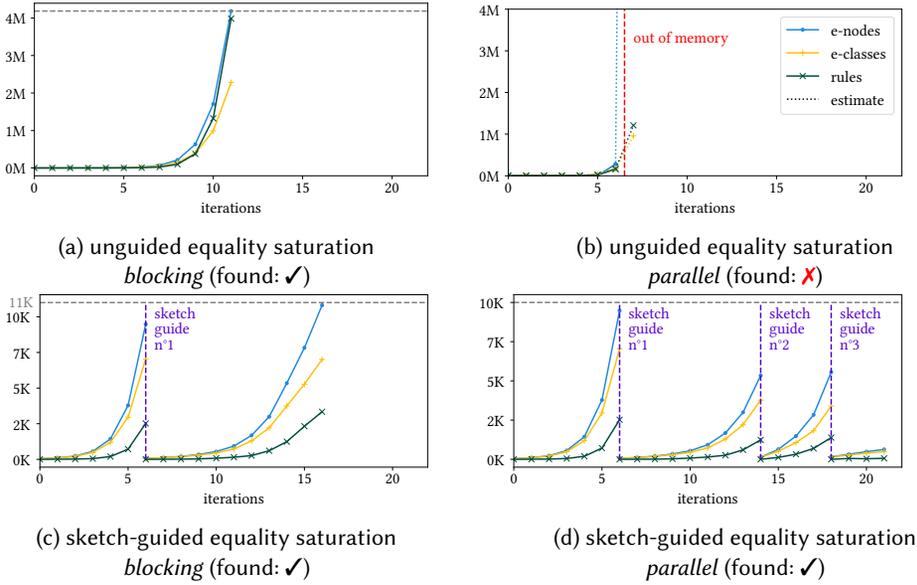
\begin{figure}
\captionsetup{justification=centering}
    \centering
    \begin{subfigure}[b]{0.49\linewidth}
    \centering
    \resizebox{!}{8em}{\input{media/unguided-blocking.pgf}}
    \caption{unguided equality saturation\\ \blocking{} (found: \yes)}
    \end{subfigure}
    \begin{subfigure}[b]{0.49\linewidth}
    \resizebox{!}{8em}{\input{media/unguided-parallel.pgf}}
    \caption{unguided equality saturation\\ \parallel{} (found: {\color{red}\no})}
    \end{subfigure}\\
    \begin{subfigure}[b]{0.49\linewidth}
    \centering
    \resizebox{!}{8em}{\input{media/guided-blocking.pgf}}
    \caption{sketch-guided equality saturation\\ \blocking{} (found: \yes)}
    \end{subfigure}
    \begin{subfigure}[b]{0.49\linewidth}
    \resizebox{!}{8em}{\input{media/guided-parallel.pgf}}
    \caption{sketch-guided equality saturation\\ \parallel{} (found: \yes)}
    \end{subfigure}
\captionsetup{justification=justified, singlelinecheck=off}
    \caption{The evolution of the e-graph, and the number of rewrite rules applied, during searches for two optimization goals.
    Sketch guides are depicted with purple vertical lines.
    Note that the scale of the y-axes for unguided graphs (a) and (b) is millions, while for guided graphs (c) and (d) it is thousands.}
    \label{fig:evolution-over-iterations}
\end{figure}

\paragraph{E-Graph Evolution in Guided and Unguided Search}
\Cref{fig:evolution-over-iterations} plots the growth of the e-graphs during unguided and sketch-guided equality saturation searches for the \blocking{} and \parallel{} optimization goals from Tables~\ref{fig:s2} and~\ref{fig:s3}.
The e-graphs produced by unguided equality saturation grow exponentially with each search iteration. The e-graph contains millions of e-nodes and e-classes, and millions of rules have been applied, within a small number of iterations (less than 10). Such rapid growth limits the scalability of unguided search, for example in the 7th iteration of the \parallel{} search the e-graph exhausts 60GB memory.

While the e-graphs produced by sketch-guided equality saturation typically also grow exponentially with each iteration, sketches are satisfied within a small number of iterations thanks to an appropriate selection of sketch guides.
In consequence the number of rewrites and crucially the maximum size of the e-graphs is \emph{three orders of magnitudes smaller} than for unguided search: no more than 11K in our example searches.
Once a program satisfying a sketch guide is found, a new search is started for the next sketch using that program, growing a fresh e-graph that remains small.
Hence sketch-guided equality saturation can scale to more complex optimizations of functional programs, such as \parallel{}. 
Conceptually factoring optimizations into a long sequence of sketch-guided searches means that there is no limit on the complexity of the optimizations, i.e. the number of rewrites, that may be searched for.

The search for the final \parallel{} sketch goal shows linear rather than exponential growth. This is likely because the set of rewrite rules used by the search have little interaction. We will discuss the impact of choosing suitable sets of rewrite rules shortly.

\paragraph{Sketches Guiding the Search}
\Cref{fig:sketch-logical} shows how each optimization goal is described by logical optimization steps, each corresponding to a sketch describing the program after the step is applied.
It transpires that the \goalStyle{split} sketch in \cref{mm-blocking-split-sketch} is a useful first guide for all optimization goals.
While the sketch sizes range from 7 to 12, programs are of size 90 to 124, showing that a sketch elides around 90\% of the program.
Even when 4 sketches must be written, the total size of the sketches is still small: the largest total being 38.
Intricate aspects of the optimized program never need to be specified in the sketches, for example array reshaping patterns such as \inlineRise{split}, \inlineRise{join} and \inlineRise{transpose}.

\begin{table}
  \small
\begin{tabular}{|l|l|l|r|r|} \hline
\textbf{goal} & \textbf{sketch guides} & \textbf{sketch goal} & \textbf{sketch sizes} & \textbf{program size}\\ \hline
\blocking{} & \goalStyle{split} & \goalStyle{reorder$_1$} & 7 & 90 \\ \hline
\vectorization{} & \goalStyle{split} + \goalStyle{reorder$_1$} & \goalStyle{lower$_1$} & 7 & 124 \\ \hline
\loopperm{} & \goalStyle{split} + \goalStyle{reorder$_2$} & \goalStyle{lower$_2$} & 7 & 104 \\ \hline
\arraypacking{} & \goalStyle{split} + \goalStyle{reorder$_2$} + \goalStyle{store} & \goalStyle{lower$_3$} & 7-12 & 121 \\ \hline
\cacheblocks{} & \goalStyle{split} + \goalStyle{reorder$_2$} + \goalStyle{store} & \goalStyle{lower$_4$} & 7-12 & 121 \\ \hline
\parallel{} & \goalStyle{split} + \goalStyle{reorder$_2$} + \goalStyle{store} & \goalStyle{lower$_5$} & 7-12 & 121 \\ \hline
\end{tabular}
\caption{Decomposition of each optimization goal into logical steps. A sketch is defined for each logical step. In this table, sketch size counts operators such as \inlineRiseSketch{containsMap}, program size counts operators such as \inlineRise{map}, lambdas, variables and constants: $\lambda$ applications are not counted.}
\label{fig:sketch-logical}
\end{table}

\paragraph{Choice of Rules and Cost Model}
Besides the sketches, programmers also specify the rules used in each search and a cost model.
For the \texttt{split} sketch, 8 rules are required explaining how to split \inlineRise{map} and \inlineRise{reduce}.
The \texttt{reorder} sketches require 9 rules that swap various nestings of \inlineRise{map} and \inlineRise{reduce}.
The \texttt{store} sketch requires 4 rules and the \texttt{lower} sketches 10 rules including \ref{map-fusion}, 6 rules for vectorization, 1 rule for loop unrolling and 1 rule for loop parallelization.
If we naively use all rules for the blocking search, the search runtime increases by about 25$\times$, still finding the optimizations in minutes but showing the importance of selecting a small set of rules.

We use a simple cost model that minimizes weighted term size.
Rules and cost models may be reused and packaged into libraries for recurring logical steps.




%% file: media/unguided-blocking.pgf
\begingroup%
\makeatletter%
\begin{pgfpicture}%
\pgfpathrectangle{\pgfpointorigin}{\pgfqpoint{6.000000in}{3.000000in}}%
\pgfusepath{use as bounding box, clip}%
\begin{pgfscope}%
\pgfsetbuttcap%
\pgfsetmiterjoin%
\definecolor{currentfill}{rgb}{1.000000,1.000000,1.000000}%
\pgfsetfillcolor{currentfill}%
\pgfsetlinewidth{0.000000pt}%
\definecolor{currentstroke}{rgb}{1.000000,1.000000,1.000000}%
\pgfsetstrokecolor{currentstroke}%
\pgfsetdash{}{0pt}%
\pgfpathmoveto{\pgfqpoint{0.000000in}{0.000000in}}%
\pgfpathlineto{\pgfqpoint{6.000000in}{0.000000in}}%
\pgfpathlineto{\pgfqpoint{6.000000in}{3.000000in}}%
\pgfpathlineto{\pgfqpoint{0.000000in}{3.000000in}}%
\pgfpathlineto{\pgfqpoint{0.000000in}{0.000000in}}%
\pgfpathclose%
\pgfusepath{fill}%
\end{pgfscope}%
\begin{pgfscope}%
\pgfsetbuttcap%
\pgfsetmiterjoin%
\definecolor{currentfill}{rgb}{1.000000,1.000000,1.000000}%
\pgfsetfillcolor{currentfill}%
\pgfsetlinewidth{0.000000pt}%
\definecolor{currentstroke}{rgb}{0.000000,0.000000,0.000000}%
\pgfsetstrokecolor{currentstroke}%
\pgfsetstrokeopacity{0.000000}%
\pgfsetdash{}{0pt}%
\pgfpathmoveto{\pgfqpoint{0.411301in}{0.579475in}}%
\pgfpathlineto{\pgfqpoint{6.000000in}{0.579475in}}%
\pgfpathlineto{\pgfqpoint{6.000000in}{3.000000in}}%
\pgfpathlineto{\pgfqpoint{0.411301in}{3.000000in}}%
\pgfpathlineto{\pgfqpoint{0.411301in}{0.579475in}}%
\pgfpathclose%
\pgfusepath{fill}%
\end{pgfscope}%
\begin{pgfscope}%
\pgfsetbuttcap%
\pgfsetroundjoin%
\definecolor{currentfill}{rgb}{0.000000,0.000000,0.000000}%
\pgfsetfillcolor{currentfill}%
\pgfsetlinewidth{0.803000pt}%
\definecolor{currentstroke}{rgb}{0.000000,0.000000,0.000000}%
\pgfsetstrokecolor{currentstroke}%
\pgfsetdash{}{0pt}%
\pgfsys@defobject{currentmarker}{\pgfqpoint{0.000000in}{-0.048611in}}{\pgfqpoint{0.000000in}{0.000000in}}{%
\pgfpathmoveto{\pgfqpoint{0.000000in}{0.000000in}}%
\pgfpathlineto{\pgfqpoint{0.000000in}{-0.048611in}}%
\pgfusepath{stroke,fill}%
}%
\begin{pgfscope}%
\pgfsys@transformshift{0.411301in}{0.579475in}%
\pgfsys@useobject{currentmarker}{}%
\end{pgfscope}%
\end{pgfscope}%
\begin{pgfscope}%
\definecolor{textcolor}{rgb}{0.000000,0.000000,0.000000}%
\pgfsetstrokecolor{textcolor}%
\pgfsetfillcolor{textcolor}%
\pgftext[x=0.411301in,y=0.482253in,,top]{\color{textcolor}\rmfamily\fontsize{16.000000}{19.200000}\selectfont \(\displaystyle {0}\)}%
\end{pgfscope}%
\begin{pgfscope}%
\pgfsetbuttcap%
\pgfsetroundjoin%
\definecolor{currentfill}{rgb}{0.000000,0.000000,0.000000}%
\pgfsetfillcolor{currentfill}%
\pgfsetlinewidth{0.803000pt}%
\definecolor{currentstroke}{rgb}{0.000000,0.000000,0.000000}%
\pgfsetstrokecolor{currentstroke}%
\pgfsetdash{}{0pt}%
\pgfsys@defobject{currentmarker}{\pgfqpoint{0.000000in}{-0.048611in}}{\pgfqpoint{0.000000in}{0.000000in}}{%
\pgfpathmoveto{\pgfqpoint{0.000000in}{0.000000in}}%
\pgfpathlineto{\pgfqpoint{0.000000in}{-0.048611in}}%
\pgfusepath{stroke,fill}%
}%
\begin{pgfscope}%
\pgfsys@transformshift{1.681460in}{0.579475in}%
\pgfsys@useobject{currentmarker}{}%
\end{pgfscope}%
\end{pgfscope}%
\begin{pgfscope}%
\definecolor{textcolor}{rgb}{0.000000,0.000000,0.000000}%
\pgfsetstrokecolor{textcolor}%
\pgfsetfillcolor{textcolor}%
\pgftext[x=1.681460in,y=0.482253in,,top]{\color{textcolor}\rmfamily\fontsize{16.000000}{19.200000}\selectfont \(\displaystyle {5}\)}%
\end{pgfscope}%
\begin{pgfscope}%
\pgfsetbuttcap%
\pgfsetroundjoin%
\definecolor{currentfill}{rgb}{0.000000,0.000000,0.000000}%
\pgfsetfillcolor{currentfill}%
\pgfsetlinewidth{0.803000pt}%
\definecolor{currentstroke}{rgb}{0.000000,0.000000,0.000000}%
\pgfsetstrokecolor{currentstroke}%
\pgfsetdash{}{0pt}%
\pgfsys@defobject{currentmarker}{\pgfqpoint{0.000000in}{-0.048611in}}{\pgfqpoint{0.000000in}{0.000000in}}{%
\pgfpathmoveto{\pgfqpoint{0.000000in}{0.000000in}}%
\pgfpathlineto{\pgfqpoint{0.000000in}{-0.048611in}}%
\pgfusepath{stroke,fill}%
}%
\begin{pgfscope}%
\pgfsys@transformshift{2.951619in}{0.579475in}%
\pgfsys@useobject{currentmarker}{}%
\end{pgfscope}%
\end{pgfscope}%
\begin{pgfscope}%
\definecolor{textcolor}{rgb}{0.000000,0.000000,0.000000}%
\pgfsetstrokecolor{textcolor}%
\pgfsetfillcolor{textcolor}%
\pgftext[x=2.951619in,y=0.482253in,,top]{\color{textcolor}\rmfamily\fontsize{16.000000}{19.200000}\selectfont \(\displaystyle {10}\)}%
\end{pgfscope}%
\begin{pgfscope}%
\pgfsetbuttcap%
\pgfsetroundjoin%
\definecolor{currentfill}{rgb}{0.000000,0.000000,0.000000}%
\pgfsetfillcolor{currentfill}%
\pgfsetlinewidth{0.803000pt}%
\definecolor{currentstroke}{rgb}{0.000000,0.000000,0.000000}%
\pgfsetstrokecolor{currentstroke}%
\pgfsetdash{}{0pt}%
\pgfsys@defobject{currentmarker}{\pgfqpoint{0.000000in}{-0.048611in}}{\pgfqpoint{0.000000in}{0.000000in}}{%
\pgfpathmoveto{\pgfqpoint{0.000000in}{0.000000in}}%
\pgfpathlineto{\pgfqpoint{0.000000in}{-0.048611in}}%
\pgfusepath{stroke,fill}%
}%
\begin{pgfscope}%
\pgfsys@transformshift{4.221778in}{0.579475in}%
\pgfsys@useobject{currentmarker}{}%
\end{pgfscope}%
\end{pgfscope}%
\begin{pgfscope}%
\definecolor{textcolor}{rgb}{0.000000,0.000000,0.000000}%
\pgfsetstrokecolor{textcolor}%
\pgfsetfillcolor{textcolor}%
\pgftext[x=4.221778in,y=0.482253in,,top]{\color{textcolor}\rmfamily\fontsize{16.000000}{19.200000}\selectfont \(\displaystyle {15}\)}%
\end{pgfscope}%
\begin{pgfscope}%
\pgfsetbuttcap%
\pgfsetroundjoin%
\definecolor{currentfill}{rgb}{0.000000,0.000000,0.000000}%
\pgfsetfillcolor{currentfill}%
\pgfsetlinewidth{0.803000pt}%
\definecolor{currentstroke}{rgb}{0.000000,0.000000,0.000000}%
\pgfsetstrokecolor{currentstroke}%
\pgfsetdash{}{0pt}%
\pgfsys@defobject{currentmarker}{\pgfqpoint{0.000000in}{-0.048611in}}{\pgfqpoint{0.000000in}{0.000000in}}{%
\pgfpathmoveto{\pgfqpoint{0.000000in}{0.000000in}}%
\pgfpathlineto{\pgfqpoint{0.000000in}{-0.048611in}}%
\pgfusepath{stroke,fill}%
}%
\begin{pgfscope}%
\pgfsys@transformshift{5.491936in}{0.579475in}%
\pgfsys@useobject{currentmarker}{}%
\end{pgfscope}%
\end{pgfscope}%
\begin{pgfscope}%
\definecolor{textcolor}{rgb}{0.000000,0.000000,0.000000}%
\pgfsetstrokecolor{textcolor}%
\pgfsetfillcolor{textcolor}%
\pgftext[x=5.491936in,y=0.482253in,,top]{\color{textcolor}\rmfamily\fontsize{16.000000}{19.200000}\selectfont \(\displaystyle {20}\)}%
\end{pgfscope}%
\begin{pgfscope}%
\definecolor{textcolor}{rgb}{0.000000,0.000000,0.000000}%
\pgfsetstrokecolor{textcolor}%
\pgfsetfillcolor{textcolor}%
\pgftext[x=3.205651in,y=0.213349in,,top]{\color{textcolor}\rmfamily\fontsize{16.000000}{19.200000}\selectfont iterations}%
\end{pgfscope}%
\begin{pgfscope}%
\pgfsetbuttcap%
\pgfsetroundjoin%
\definecolor{currentfill}{rgb}{0.000000,0.000000,0.000000}%
\pgfsetfillcolor{currentfill}%
\pgfsetlinewidth{0.803000pt}%
\definecolor{currentstroke}{rgb}{0.000000,0.000000,0.000000}%
\pgfsetstrokecolor{currentstroke}%
\pgfsetdash{}{0pt}%
\pgfsys@defobject{currentmarker}{\pgfqpoint{-0.048611in}{0.000000in}}{\pgfqpoint{-0.000000in}{0.000000in}}{%
\pgfpathmoveto{\pgfqpoint{-0.000000in}{0.000000in}}%
\pgfpathlineto{\pgfqpoint{-0.048611in}{0.000000in}}%
\pgfusepath{stroke,fill}%
}%
\begin{pgfscope}%
\pgfsys@transformshift{0.411301in}{0.689499in}%
\pgfsys@useobject{currentmarker}{}%
\end{pgfscope}%
\end{pgfscope}%
\begin{pgfscope}%
\definecolor{textcolor}{rgb}{0.000000,0.000000,0.000000}%
\pgfsetstrokecolor{textcolor}%
\pgfsetfillcolor{textcolor}%
\pgftext[x=0.000000in, y=0.606166in, left, base]{\color{textcolor}\rmfamily\fontsize{16.000000}{19.200000}\selectfont 0M}%
\end{pgfscope}%
\begin{pgfscope}%
\pgfsetbuttcap%
\pgfsetroundjoin%
\definecolor{currentfill}{rgb}{0.000000,0.000000,0.000000}%
\pgfsetfillcolor{currentfill}%
\pgfsetlinewidth{0.803000pt}%
\definecolor{currentstroke}{rgb}{0.000000,0.000000,0.000000}%
\pgfsetstrokecolor{currentstroke}%
\pgfsetdash{}{0pt}%
\pgfsys@defobject{currentmarker}{\pgfqpoint{-0.048611in}{0.000000in}}{\pgfqpoint{-0.000000in}{0.000000in}}{%
\pgfpathmoveto{\pgfqpoint{-0.000000in}{0.000000in}}%
\pgfpathlineto{\pgfqpoint{-0.048611in}{0.000000in}}%
\pgfusepath{stroke,fill}%
}%
\begin{pgfscope}%
\pgfsys@transformshift{0.411301in}{1.215179in}%
\pgfsys@useobject{currentmarker}{}%
\end{pgfscope}%
\end{pgfscope}%
\begin{pgfscope}%
\definecolor{textcolor}{rgb}{0.000000,0.000000,0.000000}%
\pgfsetstrokecolor{textcolor}%
\pgfsetfillcolor{textcolor}%
\pgftext[x=0.000000in, y=1.131845in, left, base]{\color{textcolor}\rmfamily\fontsize{16.000000}{19.200000}\selectfont 1M}%
\end{pgfscope}%
\begin{pgfscope}%
\pgfsetbuttcap%
\pgfsetroundjoin%
\definecolor{currentfill}{rgb}{0.000000,0.000000,0.000000}%
\pgfsetfillcolor{currentfill}%
\pgfsetlinewidth{0.803000pt}%
\definecolor{currentstroke}{rgb}{0.000000,0.000000,0.000000}%
\pgfsetstrokecolor{currentstroke}%
\pgfsetdash{}{0pt}%
\pgfsys@defobject{currentmarker}{\pgfqpoint{-0.048611in}{0.000000in}}{\pgfqpoint{-0.000000in}{0.000000in}}{%
\pgfpathmoveto{\pgfqpoint{-0.000000in}{0.000000in}}%
\pgfpathlineto{\pgfqpoint{-0.048611in}{0.000000in}}%
\pgfusepath{stroke,fill}%
}%
\begin{pgfscope}%
\pgfsys@transformshift{0.411301in}{1.740858in}%
\pgfsys@useobject{currentmarker}{}%
\end{pgfscope}%
\end{pgfscope}%
\begin{pgfscope}%
\definecolor{textcolor}{rgb}{0.000000,0.000000,0.000000}%
\pgfsetstrokecolor{textcolor}%
\pgfsetfillcolor{textcolor}%
\pgftext[x=0.000000in, y=1.657525in, left, base]{\color{textcolor}\rmfamily\fontsize{16.000000}{19.200000}\selectfont 2M}%
\end{pgfscope}%
\begin{pgfscope}%
\pgfsetbuttcap%
\pgfsetroundjoin%
\definecolor{currentfill}{rgb}{0.000000,0.000000,0.000000}%
\pgfsetfillcolor{currentfill}%
\pgfsetlinewidth{0.803000pt}%
\definecolor{currentstroke}{rgb}{0.000000,0.000000,0.000000}%
\pgfsetstrokecolor{currentstroke}%
\pgfsetdash{}{0pt}%
\pgfsys@defobject{currentmarker}{\pgfqpoint{-0.048611in}{0.000000in}}{\pgfqpoint{-0.000000in}{0.000000in}}{%
\pgfpathmoveto{\pgfqpoint{-0.000000in}{0.000000in}}%
\pgfpathlineto{\pgfqpoint{-0.048611in}{0.000000in}}%
\pgfusepath{stroke,fill}%
}%
\begin{pgfscope}%
\pgfsys@transformshift{0.411301in}{2.266538in}%
\pgfsys@useobject{currentmarker}{}%
\end{pgfscope}%
\end{pgfscope}%
\begin{pgfscope}%
\definecolor{textcolor}{rgb}{0.000000,0.000000,0.000000}%
\pgfsetstrokecolor{textcolor}%
\pgfsetfillcolor{textcolor}%
\pgftext[x=0.000000in, y=2.183205in, left, base]{\color{textcolor}\rmfamily\fontsize{16.000000}{19.200000}\selectfont 3M}%
\end{pgfscope}%
\begin{pgfscope}%
\pgfsetbuttcap%
\pgfsetroundjoin%
\definecolor{currentfill}{rgb}{0.000000,0.000000,0.000000}%
\pgfsetfillcolor{currentfill}%
\pgfsetlinewidth{0.803000pt}%
\definecolor{currentstroke}{rgb}{0.000000,0.000000,0.000000}%
\pgfsetstrokecolor{currentstroke}%
\pgfsetdash{}{0pt}%
\pgfsys@defobject{currentmarker}{\pgfqpoint{-0.048611in}{0.000000in}}{\pgfqpoint{-0.000000in}{0.000000in}}{%
\pgfpathmoveto{\pgfqpoint{-0.000000in}{0.000000in}}%
\pgfpathlineto{\pgfqpoint{-0.048611in}{0.000000in}}%
\pgfusepath{stroke,fill}%
}%
\begin{pgfscope}%
\pgfsys@transformshift{0.411301in}{2.792218in}%
\pgfsys@useobject{currentmarker}{}%
\end{pgfscope}%
\end{pgfscope}%
\begin{pgfscope}%
\definecolor{textcolor}{rgb}{0.000000,0.000000,0.000000}%
\pgfsetstrokecolor{textcolor}%
\pgfsetfillcolor{textcolor}%
\pgftext[x=0.000000in, y=2.708885in, left, base]{\color{textcolor}\rmfamily\fontsize{16.000000}{19.200000}\selectfont 4M}%
\end{pgfscope}%
\begin{pgfscope}%
\pgfpathrectangle{\pgfqpoint{0.411301in}{0.579475in}}{\pgfqpoint{5.588699in}{2.420525in}}%
\pgfusepath{clip}%
\pgfsetrectcap%
\pgfsetroundjoin%
\pgfsetlinewidth{1.505625pt}%
\definecolor{currentstroke}{rgb}{0.117647,0.533333,0.898039}%
\pgfsetstrokecolor{currentstroke}%
\pgfsetdash{}{0pt}%
\pgfpathmoveto{\pgfqpoint{0.411301in}{0.689520in}}%
\pgfpathlineto{\pgfqpoint{0.665333in}{0.689551in}}%
\pgfpathlineto{\pgfqpoint{0.919365in}{0.689634in}}%
\pgfpathlineto{\pgfqpoint{1.173396in}{0.689890in}}%
\pgfpathlineto{\pgfqpoint{1.427428in}{0.690652in}}%
\pgfpathlineto{\pgfqpoint{1.681460in}{0.693062in}}%
\pgfpathlineto{\pgfqpoint{1.935492in}{0.700647in}}%
\pgfpathlineto{\pgfqpoint{2.189523in}{0.725229in}}%
\pgfpathlineto{\pgfqpoint{2.443555in}{0.800648in}}%
\pgfpathlineto{\pgfqpoint{2.697587in}{1.022099in}}%
\pgfpathlineto{\pgfqpoint{2.951619in}{1.585872in}}%
\pgfpathlineto{\pgfqpoint{3.205651in}{2.889976in}}%
\pgfusepath{stroke}%
\end{pgfscope}%
\begin{pgfscope}%
\pgfpathrectangle{\pgfqpoint{0.411301in}{0.579475in}}{\pgfqpoint{5.588699in}{2.420525in}}%
\pgfusepath{clip}%
\pgfsetbuttcap%
\pgfsetroundjoin%
\definecolor{currentfill}{rgb}{0.117647,0.533333,0.898039}%
\pgfsetfillcolor{currentfill}%
\pgfsetlinewidth{1.003750pt}%
\definecolor{currentstroke}{rgb}{0.117647,0.533333,0.898039}%
\pgfsetstrokecolor{currentstroke}%
\pgfsetdash{}{0pt}%
\pgfsys@defobject{currentmarker}{\pgfqpoint{-0.020833in}{-0.020833in}}{\pgfqpoint{0.020833in}{0.020833in}}{%
\pgfpathmoveto{\pgfqpoint{0.000000in}{-0.020833in}}%
\pgfpathcurveto{\pgfqpoint{0.005525in}{-0.020833in}}{\pgfqpoint{0.010825in}{-0.018638in}}{\pgfqpoint{0.014731in}{-0.014731in}}%
\pgfpathcurveto{\pgfqpoint{0.018638in}{-0.010825in}}{\pgfqpoint{0.020833in}{-0.005525in}}{\pgfqpoint{0.020833in}{0.000000in}}%
\pgfpathcurveto{\pgfqpoint{0.020833in}{0.005525in}}{\pgfqpoint{0.018638in}{0.010825in}}{\pgfqpoint{0.014731in}{0.014731in}}%
\pgfpathcurveto{\pgfqpoint{0.010825in}{0.018638in}}{\pgfqpoint{0.005525in}{0.020833in}}{\pgfqpoint{0.000000in}{0.020833in}}%
\pgfpathcurveto{\pgfqpoint{-0.005525in}{0.020833in}}{\pgfqpoint{-0.010825in}{0.018638in}}{\pgfqpoint{-0.014731in}{0.014731in}}%
\pgfpathcurveto{\pgfqpoint{-0.018638in}{0.010825in}}{\pgfqpoint{-0.020833in}{0.005525in}}{\pgfqpoint{-0.020833in}{0.000000in}}%
\pgfpathcurveto{\pgfqpoint{-0.020833in}{-0.005525in}}{\pgfqpoint{-0.018638in}{-0.010825in}}{\pgfqpoint{-0.014731in}{-0.014731in}}%
\pgfpathcurveto{\pgfqpoint{-0.010825in}{-0.018638in}}{\pgfqpoint{-0.005525in}{-0.020833in}}{\pgfqpoint{0.000000in}{-0.020833in}}%
\pgfpathlineto{\pgfqpoint{0.000000in}{-0.020833in}}%
\pgfpathclose%
\pgfusepath{stroke,fill}%
}%
\begin{pgfscope}%
\pgfsys@transformshift{0.411301in}{0.689520in}%
\pgfsys@useobject{currentmarker}{}%
\end{pgfscope}%
\begin{pgfscope}%
\pgfsys@transformshift{0.665333in}{0.689551in}%
\pgfsys@useobject{currentmarker}{}%
\end{pgfscope}%
\begin{pgfscope}%
\pgfsys@transformshift{0.919365in}{0.689634in}%
\pgfsys@useobject{currentmarker}{}%
\end{pgfscope}%
\begin{pgfscope}%
\pgfsys@transformshift{1.173396in}{0.689890in}%
\pgfsys@useobject{currentmarker}{}%
\end{pgfscope}%
\begin{pgfscope}%
\pgfsys@transformshift{1.427428in}{0.690652in}%
\pgfsys@useobject{currentmarker}{}%
\end{pgfscope}%
\begin{pgfscope}%
\pgfsys@transformshift{1.681460in}{0.693062in}%
\pgfsys@useobject{currentmarker}{}%
\end{pgfscope}%
\begin{pgfscope}%
\pgfsys@transformshift{1.935492in}{0.700647in}%
\pgfsys@useobject{currentmarker}{}%
\end{pgfscope}%
\begin{pgfscope}%
\pgfsys@transformshift{2.189523in}{0.725229in}%
\pgfsys@useobject{currentmarker}{}%
\end{pgfscope}%
\begin{pgfscope}%
\pgfsys@transformshift{2.443555in}{0.800648in}%
\pgfsys@useobject{currentmarker}{}%
\end{pgfscope}%
\begin{pgfscope}%
\pgfsys@transformshift{2.697587in}{1.022099in}%
\pgfsys@useobject{currentmarker}{}%
\end{pgfscope}%
\begin{pgfscope}%
\pgfsys@transformshift{2.951619in}{1.585872in}%
\pgfsys@useobject{currentmarker}{}%
\end{pgfscope}%
\begin{pgfscope}%
\pgfsys@transformshift{3.205651in}{2.889976in}%
\pgfsys@useobject{currentmarker}{}%
\end{pgfscope}%
\end{pgfscope}%
\begin{pgfscope}%
\pgfpathrectangle{\pgfqpoint{0.411301in}{0.579475in}}{\pgfqpoint{5.588699in}{2.420525in}}%
\pgfusepath{clip}%
\pgfsetrectcap%
\pgfsetroundjoin%
\pgfsetlinewidth{1.505625pt}%
\definecolor{currentstroke}{rgb}{1.000000,0.756863,0.027451}%
\pgfsetstrokecolor{currentstroke}%
\pgfsetdash{}{0pt}%
\pgfpathmoveto{\pgfqpoint{0.411301in}{0.689520in}}%
\pgfpathlineto{\pgfqpoint{0.665333in}{0.689548in}}%
\pgfpathlineto{\pgfqpoint{0.919365in}{0.689621in}}%
\pgfpathlineto{\pgfqpoint{1.173396in}{0.689840in}}%
\pgfpathlineto{\pgfqpoint{1.427428in}{0.690458in}}%
\pgfpathlineto{\pgfqpoint{1.681460in}{0.692308in}}%
\pgfpathlineto{\pgfqpoint{1.935492in}{0.697799in}}%
\pgfpathlineto{\pgfqpoint{2.189523in}{0.714592in}}%
\pgfpathlineto{\pgfqpoint{2.443555in}{0.762724in}}%
\pgfpathlineto{\pgfqpoint{2.697587in}{0.894920in}}%
\pgfpathlineto{\pgfqpoint{2.951619in}{1.208597in}}%
\pgfpathlineto{\pgfqpoint{3.205651in}{1.887890in}}%
\pgfusepath{stroke}%
\end{pgfscope}%
\begin{pgfscope}%
\pgfpathrectangle{\pgfqpoint{0.411301in}{0.579475in}}{\pgfqpoint{5.588699in}{2.420525in}}%
\pgfusepath{clip}%
\pgfsetbuttcap%
\pgfsetroundjoin%
\definecolor{currentfill}{rgb}{1.000000,0.756863,0.027451}%
\pgfsetfillcolor{currentfill}%
\pgfsetlinewidth{1.003750pt}%
\definecolor{currentstroke}{rgb}{1.000000,0.756863,0.027451}%
\pgfsetstrokecolor{currentstroke}%
\pgfsetdash{}{0pt}%
\pgfsys@defobject{currentmarker}{\pgfqpoint{-0.041667in}{-0.041667in}}{\pgfqpoint{0.041667in}{0.041667in}}{%
\pgfpathmoveto{\pgfqpoint{-0.041667in}{0.000000in}}%
\pgfpathlineto{\pgfqpoint{0.041667in}{0.000000in}}%
\pgfpathmoveto{\pgfqpoint{0.000000in}{-0.041667in}}%
\pgfpathlineto{\pgfqpoint{0.000000in}{0.041667in}}%
\pgfusepath{stroke,fill}%
}%
\begin{pgfscope}%
\pgfsys@transformshift{0.411301in}{0.689520in}%
\pgfsys@useobject{currentmarker}{}%
\end{pgfscope}%
\begin{pgfscope}%
\pgfsys@transformshift{0.665333in}{0.689548in}%
\pgfsys@useobject{currentmarker}{}%
\end{pgfscope}%
\begin{pgfscope}%
\pgfsys@transformshift{0.919365in}{0.689621in}%
\pgfsys@useobject{currentmarker}{}%
\end{pgfscope}%
\begin{pgfscope}%
\pgfsys@transformshift{1.173396in}{0.689840in}%
\pgfsys@useobject{currentmarker}{}%
\end{pgfscope}%
\begin{pgfscope}%
\pgfsys@transformshift{1.427428in}{0.690458in}%
\pgfsys@useobject{currentmarker}{}%
\end{pgfscope}%
\begin{pgfscope}%
\pgfsys@transformshift{1.681460in}{0.692308in}%
\pgfsys@useobject{currentmarker}{}%
\end{pgfscope}%
\begin{pgfscope}%
\pgfsys@transformshift{1.935492in}{0.697799in}%
\pgfsys@useobject{currentmarker}{}%
\end{pgfscope}%
\begin{pgfscope}%
\pgfsys@transformshift{2.189523in}{0.714592in}%
\pgfsys@useobject{currentmarker}{}%
\end{pgfscope}%
\begin{pgfscope}%
\pgfsys@transformshift{2.443555in}{0.762724in}%
\pgfsys@useobject{currentmarker}{}%
\end{pgfscope}%
\begin{pgfscope}%
\pgfsys@transformshift{2.697587in}{0.894920in}%
\pgfsys@useobject{currentmarker}{}%
\end{pgfscope}%
\begin{pgfscope}%
\pgfsys@transformshift{2.951619in}{1.208597in}%
\pgfsys@useobject{currentmarker}{}%
\end{pgfscope}%
\begin{pgfscope}%
\pgfsys@transformshift{3.205651in}{1.887890in}%
\pgfsys@useobject{currentmarker}{}%
\end{pgfscope}%
\end{pgfscope}%
\begin{pgfscope}%
\pgfpathrectangle{\pgfqpoint{0.411301in}{0.579475in}}{\pgfqpoint{5.588699in}{2.420525in}}%
\pgfusepath{clip}%
\pgfsetrectcap%
\pgfsetroundjoin%
\pgfsetlinewidth{1.505625pt}%
\definecolor{currentstroke}{rgb}{0.000000,0.301961,0.250980}%
\pgfsetstrokecolor{currentstroke}%
\pgfsetdash{}{0pt}%
\pgfpathmoveto{\pgfqpoint{0.411301in}{0.689499in}}%
\pgfpathlineto{\pgfqpoint{0.665333in}{0.689502in}}%
\pgfpathlineto{\pgfqpoint{0.919365in}{0.689509in}}%
\pgfpathlineto{\pgfqpoint{1.173396in}{0.689536in}}%
\pgfpathlineto{\pgfqpoint{1.427428in}{0.689655in}}%
\pgfpathlineto{\pgfqpoint{1.681460in}{0.690172in}}%
\pgfpathlineto{\pgfqpoint{1.935492in}{0.692451in}}%
\pgfpathlineto{\pgfqpoint{2.189523in}{0.701710in}}%
\pgfpathlineto{\pgfqpoint{2.443555in}{0.740343in}}%
\pgfpathlineto{\pgfqpoint{2.697587in}{0.887019in}}%
\pgfpathlineto{\pgfqpoint{2.951619in}{1.385065in}}%
\pgfpathlineto{\pgfqpoint{3.205651in}{2.784411in}}%
\pgfusepath{stroke}%
\end{pgfscope}%
\begin{pgfscope}%
\pgfpathrectangle{\pgfqpoint{0.411301in}{0.579475in}}{\pgfqpoint{5.588699in}{2.420525in}}%
\pgfusepath{clip}%
\pgfsetbuttcap%
\pgfsetroundjoin%
\definecolor{currentfill}{rgb}{0.000000,0.301961,0.250980}%
\pgfsetfillcolor{currentfill}%
\pgfsetlinewidth{1.003750pt}%
\definecolor{currentstroke}{rgb}{0.000000,0.301961,0.250980}%
\pgfsetstrokecolor{currentstroke}%
\pgfsetdash{}{0pt}%
\pgfsys@defobject{currentmarker}{\pgfqpoint{-0.041667in}{-0.041667in}}{\pgfqpoint{0.041667in}{0.041667in}}{%
\pgfpathmoveto{\pgfqpoint{-0.041667in}{-0.041667in}}%
\pgfpathlineto{\pgfqpoint{0.041667in}{0.041667in}}%
\pgfpathmoveto{\pgfqpoint{-0.041667in}{0.041667in}}%
\pgfpathlineto{\pgfqpoint{0.041667in}{-0.041667in}}%
\pgfusepath{stroke,fill}%
}%
\begin{pgfscope}%
\pgfsys@transformshift{0.411301in}{0.689499in}%
\pgfsys@useobject{currentmarker}{}%
\end{pgfscope}%
\begin{pgfscope}%
\pgfsys@transformshift{0.665333in}{0.689502in}%
\pgfsys@useobject{currentmarker}{}%
\end{pgfscope}%
\begin{pgfscope}%
\pgfsys@transformshift{0.919365in}{0.689509in}%
\pgfsys@useobject{currentmarker}{}%
\end{pgfscope}%
\begin{pgfscope}%
\pgfsys@transformshift{1.173396in}{0.689536in}%
\pgfsys@useobject{currentmarker}{}%
\end{pgfscope}%
\begin{pgfscope}%
\pgfsys@transformshift{1.427428in}{0.689655in}%
\pgfsys@useobject{currentmarker}{}%
\end{pgfscope}%
\begin{pgfscope}%
\pgfsys@transformshift{1.681460in}{0.690172in}%
\pgfsys@useobject{currentmarker}{}%
\end{pgfscope}%
\begin{pgfscope}%
\pgfsys@transformshift{1.935492in}{0.692451in}%
\pgfsys@useobject{currentmarker}{}%
\end{pgfscope}%
\begin{pgfscope}%
\pgfsys@transformshift{2.189523in}{0.701710in}%
\pgfsys@useobject{currentmarker}{}%
\end{pgfscope}%
\begin{pgfscope}%
\pgfsys@transformshift{2.443555in}{0.740343in}%
\pgfsys@useobject{currentmarker}{}%
\end{pgfscope}%
\begin{pgfscope}%
\pgfsys@transformshift{2.697587in}{0.887019in}%
\pgfsys@useobject{currentmarker}{}%
\end{pgfscope}%
\begin{pgfscope}%
\pgfsys@transformshift{2.951619in}{1.385065in}%
\pgfsys@useobject{currentmarker}{}%
\end{pgfscope}%
\begin{pgfscope}%
\pgfsys@transformshift{3.205651in}{2.784411in}%
\pgfsys@useobject{currentmarker}{}%
\end{pgfscope}%
\end{pgfscope}%
\begin{pgfscope}%
\pgfpathrectangle{\pgfqpoint{0.411301in}{0.579475in}}{\pgfqpoint{5.588699in}{2.420525in}}%
\pgfusepath{clip}%
\pgfsetbuttcap%
\pgfsetroundjoin%
\pgfsetlinewidth{1.505625pt}%
\definecolor{currentstroke}{rgb}{0.501961,0.501961,0.501961}%
\pgfsetstrokecolor{currentstroke}%
\pgfsetdash{{5.550000pt}{2.400000pt}}{0.000000pt}%
\pgfpathmoveto{\pgfqpoint{0.411301in}{2.889995in}}%
\pgfpathlineto{\pgfqpoint{6.000000in}{2.889995in}}%
\pgfusepath{stroke}%
\end{pgfscope}%
\begin{pgfscope}%
\pgfsetrectcap%
\pgfsetmiterjoin%
\pgfsetlinewidth{0.803000pt}%
\definecolor{currentstroke}{rgb}{0.000000,0.000000,0.000000}%
\pgfsetstrokecolor{currentstroke}%
\pgfsetdash{}{0pt}%
\pgfpathmoveto{\pgfqpoint{0.411301in}{0.579475in}}%
\pgfpathlineto{\pgfqpoint{0.411301in}{3.000000in}}%
\pgfusepath{stroke}%
\end{pgfscope}%
\begin{pgfscope}%
\pgfsetrectcap%
\pgfsetmiterjoin%
\pgfsetlinewidth{0.803000pt}%
\definecolor{currentstroke}{rgb}{0.000000,0.000000,0.000000}%
\pgfsetstrokecolor{currentstroke}%
\pgfsetdash{}{0pt}%
\pgfpathmoveto{\pgfqpoint{6.000000in}{0.579475in}}%
\pgfpathlineto{\pgfqpoint{6.000000in}{3.000000in}}%
\pgfusepath{stroke}%
\end{pgfscope}%
\begin{pgfscope}%
\pgfsetrectcap%
\pgfsetmiterjoin%
\pgfsetlinewidth{0.803000pt}%
\definecolor{currentstroke}{rgb}{0.000000,0.000000,0.000000}%
\pgfsetstrokecolor{currentstroke}%
\pgfsetdash{}{0pt}%
\pgfpathmoveto{\pgfqpoint{0.411301in}{0.579475in}}%
\pgfpathlineto{\pgfqpoint{6.000000in}{0.579475in}}%
\pgfusepath{stroke}%
\end{pgfscope}%
\begin{pgfscope}%
\pgfsetrectcap%
\pgfsetmiterjoin%
\pgfsetlinewidth{0.803000pt}%
\definecolor{currentstroke}{rgb}{0.000000,0.000000,0.000000}%
\pgfsetstrokecolor{currentstroke}%
\pgfsetdash{}{0pt}%
\pgfpathmoveto{\pgfqpoint{0.411301in}{3.000000in}}%
\pgfpathlineto{\pgfqpoint{6.000000in}{3.000000in}}%
\pgfusepath{stroke}%
\end{pgfscope}%
\end{pgfpicture}%
\makeatother%
\endgroup%

%% file: media/unguided-parallel.pgf
\begingroup%
\makeatletter%
\begin{pgfpicture}%
\pgfpathrectangle{\pgfpointorigin}{\pgfqpoint{6.000000in}{3.000000in}}%
\pgfusepath{use as bounding box, clip}%
\begin{pgfscope}%
\pgfsetbuttcap%
\pgfsetmiterjoin%
\definecolor{currentfill}{rgb}{1.000000,1.000000,1.000000}%
\pgfsetfillcolor{currentfill}%
\pgfsetlinewidth{0.000000pt}%
\definecolor{currentstroke}{rgb}{1.000000,1.000000,1.000000}%
\pgfsetstrokecolor{currentstroke}%
\pgfsetdash{}{0pt}%
\pgfpathmoveto{\pgfqpoint{0.000000in}{0.000000in}}%
\pgfpathlineto{\pgfqpoint{6.000000in}{0.000000in}}%
\pgfpathlineto{\pgfqpoint{6.000000in}{3.000000in}}%
\pgfpathlineto{\pgfqpoint{0.000000in}{3.000000in}}%
\pgfpathlineto{\pgfqpoint{0.000000in}{0.000000in}}%
\pgfpathclose%
\pgfusepath{fill}%
\end{pgfscope}%
\begin{pgfscope}%
\pgfsetbuttcap%
\pgfsetmiterjoin%
\definecolor{currentfill}{rgb}{1.000000,1.000000,1.000000}%
\pgfsetfillcolor{currentfill}%
\pgfsetlinewidth{0.000000pt}%
\definecolor{currentstroke}{rgb}{0.000000,0.000000,0.000000}%
\pgfsetstrokecolor{currentstroke}%
\pgfsetstrokeopacity{0.000000}%
\pgfsetdash{}{0pt}%
\pgfpathmoveto{\pgfqpoint{0.411301in}{0.579475in}}%
\pgfpathlineto{\pgfqpoint{6.000000in}{0.579475in}}%
\pgfpathlineto{\pgfqpoint{6.000000in}{2.916667in}}%
\pgfpathlineto{\pgfqpoint{0.411301in}{2.916667in}}%
\pgfpathlineto{\pgfqpoint{0.411301in}{0.579475in}}%
\pgfpathclose%
\pgfusepath{fill}%
\end{pgfscope}%
\begin{pgfscope}%
\pgfsetbuttcap%
\pgfsetroundjoin%
\definecolor{currentfill}{rgb}{0.000000,0.000000,0.000000}%
\pgfsetfillcolor{currentfill}%
\pgfsetlinewidth{0.803000pt}%
\definecolor{currentstroke}{rgb}{0.000000,0.000000,0.000000}%
\pgfsetstrokecolor{currentstroke}%
\pgfsetdash{}{0pt}%
\pgfsys@defobject{currentmarker}{\pgfqpoint{0.000000in}{-0.048611in}}{\pgfqpoint{0.000000in}{0.000000in}}{%
\pgfpathmoveto{\pgfqpoint{0.000000in}{0.000000in}}%
\pgfpathlineto{\pgfqpoint{0.000000in}{-0.048611in}}%
\pgfusepath{stroke,fill}%
}%
\begin{pgfscope}%
\pgfsys@transformshift{0.411301in}{0.579475in}%
\pgfsys@useobject{currentmarker}{}%
\end{pgfscope}%
\end{pgfscope}%
\begin{pgfscope}%
\definecolor{textcolor}{rgb}{0.000000,0.000000,0.000000}%
\pgfsetstrokecolor{textcolor}%
\pgfsetfillcolor{textcolor}%
\pgftext[x=0.411301in,y=0.482253in,,top]{\color{textcolor}\rmfamily\fontsize{16.000000}{19.200000}\selectfont \(\displaystyle {0}\)}%
\end{pgfscope}%
\begin{pgfscope}%
\pgfsetbuttcap%
\pgfsetroundjoin%
\definecolor{currentfill}{rgb}{0.000000,0.000000,0.000000}%
\pgfsetfillcolor{currentfill}%
\pgfsetlinewidth{0.803000pt}%
\definecolor{currentstroke}{rgb}{0.000000,0.000000,0.000000}%
\pgfsetstrokecolor{currentstroke}%
\pgfsetdash{}{0pt}%
\pgfsys@defobject{currentmarker}{\pgfqpoint{0.000000in}{-0.048611in}}{\pgfqpoint{0.000000in}{0.000000in}}{%
\pgfpathmoveto{\pgfqpoint{0.000000in}{0.000000in}}%
\pgfpathlineto{\pgfqpoint{0.000000in}{-0.048611in}}%
\pgfusepath{stroke,fill}%
}%
\begin{pgfscope}%
\pgfsys@transformshift{1.681460in}{0.579475in}%
\pgfsys@useobject{currentmarker}{}%
\end{pgfscope}%
\end{pgfscope}%
\begin{pgfscope}%
\definecolor{textcolor}{rgb}{0.000000,0.000000,0.000000}%
\pgfsetstrokecolor{textcolor}%
\pgfsetfillcolor{textcolor}%
\pgftext[x=1.681460in,y=0.482253in,,top]{\color{textcolor}\rmfamily\fontsize{16.000000}{19.200000}\selectfont \(\displaystyle {5}\)}%
\end{pgfscope}%
\begin{pgfscope}%
\pgfsetbuttcap%
\pgfsetroundjoin%
\definecolor{currentfill}{rgb}{0.000000,0.000000,0.000000}%
\pgfsetfillcolor{currentfill}%
\pgfsetlinewidth{0.803000pt}%
\definecolor{currentstroke}{rgb}{0.000000,0.000000,0.000000}%
\pgfsetstrokecolor{currentstroke}%
\pgfsetdash{}{0pt}%
\pgfsys@defobject{currentmarker}{\pgfqpoint{0.000000in}{-0.048611in}}{\pgfqpoint{0.000000in}{0.000000in}}{%
\pgfpathmoveto{\pgfqpoint{0.000000in}{0.000000in}}%
\pgfpathlineto{\pgfqpoint{0.000000in}{-0.048611in}}%
\pgfusepath{stroke,fill}%
}%
\begin{pgfscope}%
\pgfsys@transformshift{2.951619in}{0.579475in}%
\pgfsys@useobject{currentmarker}{}%
\end{pgfscope}%
\end{pgfscope}%
\begin{pgfscope}%
\definecolor{textcolor}{rgb}{0.000000,0.000000,0.000000}%
\pgfsetstrokecolor{textcolor}%
\pgfsetfillcolor{textcolor}%
\pgftext[x=2.951619in,y=0.482253in,,top]{\color{textcolor}\rmfamily\fontsize{16.000000}{19.200000}\selectfont \(\displaystyle {10}\)}%
\end{pgfscope}%
\begin{pgfscope}%
\pgfsetbuttcap%
\pgfsetroundjoin%
\definecolor{currentfill}{rgb}{0.000000,0.000000,0.000000}%
\pgfsetfillcolor{currentfill}%
\pgfsetlinewidth{0.803000pt}%
\definecolor{currentstroke}{rgb}{0.000000,0.000000,0.000000}%
\pgfsetstrokecolor{currentstroke}%
\pgfsetdash{}{0pt}%
\pgfsys@defobject{currentmarker}{\pgfqpoint{0.000000in}{-0.048611in}}{\pgfqpoint{0.000000in}{0.000000in}}{%
\pgfpathmoveto{\pgfqpoint{0.000000in}{0.000000in}}%
\pgfpathlineto{\pgfqpoint{0.000000in}{-0.048611in}}%
\pgfusepath{stroke,fill}%
}%
\begin{pgfscope}%
\pgfsys@transformshift{4.221778in}{0.579475in}%
\pgfsys@useobject{currentmarker}{}%
\end{pgfscope}%
\end{pgfscope}%
\begin{pgfscope}%
\definecolor{textcolor}{rgb}{0.000000,0.000000,0.000000}%
\pgfsetstrokecolor{textcolor}%
\pgfsetfillcolor{textcolor}%
\pgftext[x=4.221778in,y=0.482253in,,top]{\color{textcolor}\rmfamily\fontsize{16.000000}{19.200000}\selectfont \(\displaystyle {15}\)}%
\end{pgfscope}%
\begin{pgfscope}%
\pgfsetbuttcap%
\pgfsetroundjoin%
\definecolor{currentfill}{rgb}{0.000000,0.000000,0.000000}%
\pgfsetfillcolor{currentfill}%
\pgfsetlinewidth{0.803000pt}%
\definecolor{currentstroke}{rgb}{0.000000,0.000000,0.000000}%
\pgfsetstrokecolor{currentstroke}%
\pgfsetdash{}{0pt}%
\pgfsys@defobject{currentmarker}{\pgfqpoint{0.000000in}{-0.048611in}}{\pgfqpoint{0.000000in}{0.000000in}}{%
\pgfpathmoveto{\pgfqpoint{0.000000in}{0.000000in}}%
\pgfpathlineto{\pgfqpoint{0.000000in}{-0.048611in}}%
\pgfusepath{stroke,fill}%
}%
\begin{pgfscope}%
\pgfsys@transformshift{5.491936in}{0.579475in}%
\pgfsys@useobject{currentmarker}{}%
\end{pgfscope}%
\end{pgfscope}%
\begin{pgfscope}%
\definecolor{textcolor}{rgb}{0.000000,0.000000,0.000000}%
\pgfsetstrokecolor{textcolor}%
\pgfsetfillcolor{textcolor}%
\pgftext[x=5.491936in,y=0.482253in,,top]{\color{textcolor}\rmfamily\fontsize{16.000000}{19.200000}\selectfont \(\displaystyle {20}\)}%
\end{pgfscope}%
\begin{pgfscope}%
\definecolor{textcolor}{rgb}{0.000000,0.000000,0.000000}%
\pgfsetstrokecolor{textcolor}%
\pgfsetfillcolor{textcolor}%
\pgftext[x=3.205651in,y=0.213349in,,top]{\color{textcolor}\rmfamily\fontsize{16.000000}{19.200000}\selectfont iterations}%
\end{pgfscope}%
\begin{pgfscope}%
\pgfsetbuttcap%
\pgfsetroundjoin%
\definecolor{currentfill}{rgb}{0.000000,0.000000,0.000000}%
\pgfsetfillcolor{currentfill}%
\pgfsetlinewidth{0.803000pt}%
\definecolor{currentstroke}{rgb}{0.000000,0.000000,0.000000}%
\pgfsetstrokecolor{currentstroke}%
\pgfsetdash{}{0pt}%
\pgfsys@defobject{currentmarker}{\pgfqpoint{-0.048611in}{0.000000in}}{\pgfqpoint{-0.000000in}{0.000000in}}{%
\pgfpathmoveto{\pgfqpoint{-0.000000in}{0.000000in}}%
\pgfpathlineto{\pgfqpoint{-0.048611in}{0.000000in}}%
\pgfusepath{stroke,fill}%
}%
\begin{pgfscope}%
\pgfsys@transformshift{0.411301in}{0.579475in}%
\pgfsys@useobject{currentmarker}{}%
\end{pgfscope}%
\end{pgfscope}%
\begin{pgfscope}%
\definecolor{textcolor}{rgb}{0.000000,0.000000,0.000000}%
\pgfsetstrokecolor{textcolor}%
\pgfsetfillcolor{textcolor}%
\pgftext[x=0.000000in, y=0.496142in, left, base]{\color{textcolor}\rmfamily\fontsize{16.000000}{19.200000}\selectfont 0M}%
\end{pgfscope}%
\begin{pgfscope}%
\pgfsetbuttcap%
\pgfsetroundjoin%
\definecolor{currentfill}{rgb}{0.000000,0.000000,0.000000}%
\pgfsetfillcolor{currentfill}%
\pgfsetlinewidth{0.803000pt}%
\definecolor{currentstroke}{rgb}{0.000000,0.000000,0.000000}%
\pgfsetstrokecolor{currentstroke}%
\pgfsetdash{}{0pt}%
\pgfsys@defobject{currentmarker}{\pgfqpoint{-0.048611in}{0.000000in}}{\pgfqpoint{-0.000000in}{0.000000in}}{%
\pgfpathmoveto{\pgfqpoint{-0.000000in}{0.000000in}}%
\pgfpathlineto{\pgfqpoint{-0.048611in}{0.000000in}}%
\pgfusepath{stroke,fill}%
}%
\begin{pgfscope}%
\pgfsys@transformshift{0.411301in}{1.163773in}%
\pgfsys@useobject{currentmarker}{}%
\end{pgfscope}%
\end{pgfscope}%
\begin{pgfscope}%
\definecolor{textcolor}{rgb}{0.000000,0.000000,0.000000}%
\pgfsetstrokecolor{textcolor}%
\pgfsetfillcolor{textcolor}%
\pgftext[x=0.000000in, y=1.080440in, left, base]{\color{textcolor}\rmfamily\fontsize{16.000000}{19.200000}\selectfont 1M}%
\end{pgfscope}%
\begin{pgfscope}%
\pgfsetbuttcap%
\pgfsetroundjoin%
\definecolor{currentfill}{rgb}{0.000000,0.000000,0.000000}%
\pgfsetfillcolor{currentfill}%
\pgfsetlinewidth{0.803000pt}%
\definecolor{currentstroke}{rgb}{0.000000,0.000000,0.000000}%
\pgfsetstrokecolor{currentstroke}%
\pgfsetdash{}{0pt}%
\pgfsys@defobject{currentmarker}{\pgfqpoint{-0.048611in}{0.000000in}}{\pgfqpoint{-0.000000in}{0.000000in}}{%
\pgfpathmoveto{\pgfqpoint{-0.000000in}{0.000000in}}%
\pgfpathlineto{\pgfqpoint{-0.048611in}{0.000000in}}%
\pgfusepath{stroke,fill}%
}%
\begin{pgfscope}%
\pgfsys@transformshift{0.411301in}{1.748071in}%
\pgfsys@useobject{currentmarker}{}%
\end{pgfscope}%
\end{pgfscope}%
\begin{pgfscope}%
\definecolor{textcolor}{rgb}{0.000000,0.000000,0.000000}%
\pgfsetstrokecolor{textcolor}%
\pgfsetfillcolor{textcolor}%
\pgftext[x=0.000000in, y=1.664738in, left, base]{\color{textcolor}\rmfamily\fontsize{16.000000}{19.200000}\selectfont 2M}%
\end{pgfscope}%
\begin{pgfscope}%
\pgfsetbuttcap%
\pgfsetroundjoin%
\definecolor{currentfill}{rgb}{0.000000,0.000000,0.000000}%
\pgfsetfillcolor{currentfill}%
\pgfsetlinewidth{0.803000pt}%
\definecolor{currentstroke}{rgb}{0.000000,0.000000,0.000000}%
\pgfsetstrokecolor{currentstroke}%
\pgfsetdash{}{0pt}%
\pgfsys@defobject{currentmarker}{\pgfqpoint{-0.048611in}{0.000000in}}{\pgfqpoint{-0.000000in}{0.000000in}}{%
\pgfpathmoveto{\pgfqpoint{-0.000000in}{0.000000in}}%
\pgfpathlineto{\pgfqpoint{-0.048611in}{0.000000in}}%
\pgfusepath{stroke,fill}%
}%
\begin{pgfscope}%
\pgfsys@transformshift{0.411301in}{2.332369in}%
\pgfsys@useobject{currentmarker}{}%
\end{pgfscope}%
\end{pgfscope}%
\begin{pgfscope}%
\definecolor{textcolor}{rgb}{0.000000,0.000000,0.000000}%
\pgfsetstrokecolor{textcolor}%
\pgfsetfillcolor{textcolor}%
\pgftext[x=0.000000in, y=2.249035in, left, base]{\color{textcolor}\rmfamily\fontsize{16.000000}{19.200000}\selectfont 3M}%
\end{pgfscope}%
\begin{pgfscope}%
\pgfsetbuttcap%
\pgfsetroundjoin%
\definecolor{currentfill}{rgb}{0.000000,0.000000,0.000000}%
\pgfsetfillcolor{currentfill}%
\pgfsetlinewidth{0.803000pt}%
\definecolor{currentstroke}{rgb}{0.000000,0.000000,0.000000}%
\pgfsetstrokecolor{currentstroke}%
\pgfsetdash{}{0pt}%
\pgfsys@defobject{currentmarker}{\pgfqpoint{-0.048611in}{0.000000in}}{\pgfqpoint{-0.000000in}{0.000000in}}{%
\pgfpathmoveto{\pgfqpoint{-0.000000in}{0.000000in}}%
\pgfpathlineto{\pgfqpoint{-0.048611in}{0.000000in}}%
\pgfusepath{stroke,fill}%
}%
\begin{pgfscope}%
\pgfsys@transformshift{0.411301in}{2.916667in}%
\pgfsys@useobject{currentmarker}{}%
\end{pgfscope}%
\end{pgfscope}%
\begin{pgfscope}%
\definecolor{textcolor}{rgb}{0.000000,0.000000,0.000000}%
\pgfsetstrokecolor{textcolor}%
\pgfsetfillcolor{textcolor}%
\pgftext[x=0.000000in, y=2.833333in, left, base]{\color{textcolor}\rmfamily\fontsize{16.000000}{19.200000}\selectfont 4M}%
\end{pgfscope}%
\begin{pgfscope}%
\pgfpathrectangle{\pgfqpoint{0.411301in}{0.579475in}}{\pgfqpoint{5.588699in}{2.337192in}}%
\pgfusepath{clip}%
\pgfsetrectcap%
\pgfsetroundjoin%
\pgfsetlinewidth{1.505625pt}%
\definecolor{currentstroke}{rgb}{0.117647,0.533333,0.898039}%
\pgfsetstrokecolor{currentstroke}%
\pgfsetdash{}{0pt}%
\pgfpathmoveto{\pgfqpoint{0.411301in}{0.579498in}}%
\pgfpathlineto{\pgfqpoint{0.665333in}{0.579629in}}%
\pgfpathlineto{\pgfqpoint{0.919365in}{0.580081in}}%
\pgfpathlineto{\pgfqpoint{1.173396in}{0.581549in}}%
\pgfpathlineto{\pgfqpoint{1.427428in}{0.586665in}}%
\pgfpathlineto{\pgfqpoint{1.681460in}{0.582398in}}%
\pgfpathlineto{\pgfqpoint{1.935492in}{0.742503in}}%
\pgfusepath{stroke}%
\end{pgfscope}%
\begin{pgfscope}%
\pgfpathrectangle{\pgfqpoint{0.411301in}{0.579475in}}{\pgfqpoint{5.588699in}{2.337192in}}%
\pgfusepath{clip}%
\pgfsetbuttcap%
\pgfsetroundjoin%
\definecolor{currentfill}{rgb}{0.117647,0.533333,0.898039}%
\pgfsetfillcolor{currentfill}%
\pgfsetlinewidth{1.003750pt}%
\definecolor{currentstroke}{rgb}{0.117647,0.533333,0.898039}%
\pgfsetstrokecolor{currentstroke}%
\pgfsetdash{}{0pt}%
\pgfsys@defobject{currentmarker}{\pgfqpoint{-0.020833in}{-0.020833in}}{\pgfqpoint{0.020833in}{0.020833in}}{%
\pgfpathmoveto{\pgfqpoint{0.000000in}{-0.020833in}}%
\pgfpathcurveto{\pgfqpoint{0.005525in}{-0.020833in}}{\pgfqpoint{0.010825in}{-0.018638in}}{\pgfqpoint{0.014731in}{-0.014731in}}%
\pgfpathcurveto{\pgfqpoint{0.018638in}{-0.010825in}}{\pgfqpoint{0.020833in}{-0.005525in}}{\pgfqpoint{0.020833in}{0.000000in}}%
\pgfpathcurveto{\pgfqpoint{0.020833in}{0.005525in}}{\pgfqpoint{0.018638in}{0.010825in}}{\pgfqpoint{0.014731in}{0.014731in}}%
\pgfpathcurveto{\pgfqpoint{0.010825in}{0.018638in}}{\pgfqpoint{0.005525in}{0.020833in}}{\pgfqpoint{0.000000in}{0.020833in}}%
\pgfpathcurveto{\pgfqpoint{-0.005525in}{0.020833in}}{\pgfqpoint{-0.010825in}{0.018638in}}{\pgfqpoint{-0.014731in}{0.014731in}}%
\pgfpathcurveto{\pgfqpoint{-0.018638in}{0.010825in}}{\pgfqpoint{-0.020833in}{0.005525in}}{\pgfqpoint{-0.020833in}{0.000000in}}%
\pgfpathcurveto{\pgfqpoint{-0.020833in}{-0.005525in}}{\pgfqpoint{-0.018638in}{-0.010825in}}{\pgfqpoint{-0.014731in}{-0.014731in}}%
\pgfpathcurveto{\pgfqpoint{-0.010825in}{-0.018638in}}{\pgfqpoint{-0.005525in}{-0.020833in}}{\pgfqpoint{0.000000in}{-0.020833in}}%
\pgfpathlineto{\pgfqpoint{0.000000in}{-0.020833in}}%
\pgfpathclose%
\pgfusepath{stroke,fill}%
}%
\begin{pgfscope}%
\pgfsys@transformshift{0.411301in}{0.579498in}%
\pgfsys@useobject{currentmarker}{}%
\end{pgfscope}%
\begin{pgfscope}%
\pgfsys@transformshift{0.665333in}{0.579629in}%
\pgfsys@useobject{currentmarker}{}%
\end{pgfscope}%
\begin{pgfscope}%
\pgfsys@transformshift{0.919365in}{0.580081in}%
\pgfsys@useobject{currentmarker}{}%
\end{pgfscope}%
\begin{pgfscope}%
\pgfsys@transformshift{1.173396in}{0.581549in}%
\pgfsys@useobject{currentmarker}{}%
\end{pgfscope}%
\begin{pgfscope}%
\pgfsys@transformshift{1.427428in}{0.586665in}%
\pgfsys@useobject{currentmarker}{}%
\end{pgfscope}%
\begin{pgfscope}%
\pgfsys@transformshift{1.681460in}{0.582398in}%
\pgfsys@useobject{currentmarker}{}%
\end{pgfscope}%
\begin{pgfscope}%
\pgfsys@transformshift{1.935492in}{0.742503in}%
\pgfsys@useobject{currentmarker}{}%
\end{pgfscope}%
\end{pgfscope}%
\begin{pgfscope}%
\pgfpathrectangle{\pgfqpoint{0.411301in}{0.579475in}}{\pgfqpoint{5.588699in}{2.337192in}}%
\pgfusepath{clip}%
\pgfsetrectcap%
\pgfsetroundjoin%
\pgfsetlinewidth{1.505625pt}%
\definecolor{currentstroke}{rgb}{1.000000,0.756863,0.027451}%
\pgfsetstrokecolor{currentstroke}%
\pgfsetdash{}{0pt}%
\pgfpathmoveto{\pgfqpoint{0.411301in}{0.579498in}}%
\pgfpathlineto{\pgfqpoint{0.665333in}{0.579599in}}%
\pgfpathlineto{\pgfqpoint{0.919365in}{0.579910in}}%
\pgfpathlineto{\pgfqpoint{1.173396in}{0.580860in}}%
\pgfpathlineto{\pgfqpoint{1.427428in}{0.584097in}}%
\pgfpathlineto{\pgfqpoint{1.681460in}{0.597608in}}%
\pgfpathlineto{\pgfqpoint{1.935492in}{0.680392in}}%
\pgfusepath{stroke}%
\end{pgfscope}%
\begin{pgfscope}%
\pgfpathrectangle{\pgfqpoint{0.411301in}{0.579475in}}{\pgfqpoint{5.588699in}{2.337192in}}%
\pgfusepath{clip}%
\pgfsetbuttcap%
\pgfsetroundjoin%
\definecolor{currentfill}{rgb}{1.000000,0.756863,0.027451}%
\pgfsetfillcolor{currentfill}%
\pgfsetlinewidth{1.003750pt}%
\definecolor{currentstroke}{rgb}{1.000000,0.756863,0.027451}%
\pgfsetstrokecolor{currentstroke}%
\pgfsetdash{}{0pt}%
\pgfsys@defobject{currentmarker}{\pgfqpoint{-0.041667in}{-0.041667in}}{\pgfqpoint{0.041667in}{0.041667in}}{%
\pgfpathmoveto{\pgfqpoint{-0.041667in}{0.000000in}}%
\pgfpathlineto{\pgfqpoint{0.041667in}{0.000000in}}%
\pgfpathmoveto{\pgfqpoint{0.000000in}{-0.041667in}}%
\pgfpathlineto{\pgfqpoint{0.000000in}{0.041667in}}%
\pgfusepath{stroke,fill}%
}%
\begin{pgfscope}%
\pgfsys@transformshift{0.411301in}{0.579498in}%
\pgfsys@useobject{currentmarker}{}%
\end{pgfscope}%
\begin{pgfscope}%
\pgfsys@transformshift{0.665333in}{0.579599in}%
\pgfsys@useobject{currentmarker}{}%
\end{pgfscope}%
\begin{pgfscope}%
\pgfsys@transformshift{0.919365in}{0.579910in}%
\pgfsys@useobject{currentmarker}{}%
\end{pgfscope}%
\begin{pgfscope}%
\pgfsys@transformshift{1.173396in}{0.580860in}%
\pgfsys@useobject{currentmarker}{}%
\end{pgfscope}%
\begin{pgfscope}%
\pgfsys@transformshift{1.427428in}{0.584097in}%
\pgfsys@useobject{currentmarker}{}%
\end{pgfscope}%
\begin{pgfscope}%
\pgfsys@transformshift{1.681460in}{0.597608in}%
\pgfsys@useobject{currentmarker}{}%
\end{pgfscope}%
\begin{pgfscope}%
\pgfsys@transformshift{1.935492in}{0.680392in}%
\pgfsys@useobject{currentmarker}{}%
\end{pgfscope}%
\end{pgfscope}%
\begin{pgfscope}%
\pgfpathrectangle{\pgfqpoint{0.411301in}{0.579475in}}{\pgfqpoint{5.588699in}{2.337192in}}%
\pgfusepath{clip}%
\pgfsetrectcap%
\pgfsetroundjoin%
\pgfsetlinewidth{1.505625pt}%
\definecolor{currentstroke}{rgb}{0.000000,0.301961,0.250980}%
\pgfsetstrokecolor{currentstroke}%
\pgfsetdash{}{0pt}%
\pgfpathmoveto{\pgfqpoint{0.411301in}{0.579475in}}%
\pgfpathlineto{\pgfqpoint{0.665333in}{0.579504in}}%
\pgfpathlineto{\pgfqpoint{0.919365in}{0.579632in}}%
\pgfpathlineto{\pgfqpoint{1.173396in}{0.580200in}}%
\pgfpathlineto{\pgfqpoint{1.427428in}{0.582224in}}%
\pgfpathlineto{\pgfqpoint{1.681460in}{0.592361in}}%
\pgfpathlineto{\pgfqpoint{1.935492in}{0.674818in}}%
\pgfusepath{stroke}%
\end{pgfscope}%
\begin{pgfscope}%
\pgfpathrectangle{\pgfqpoint{0.411301in}{0.579475in}}{\pgfqpoint{5.588699in}{2.337192in}}%
\pgfusepath{clip}%
\pgfsetbuttcap%
\pgfsetroundjoin%
\definecolor{currentfill}{rgb}{0.000000,0.301961,0.250980}%
\pgfsetfillcolor{currentfill}%
\pgfsetlinewidth{1.003750pt}%
\definecolor{currentstroke}{rgb}{0.000000,0.301961,0.250980}%
\pgfsetstrokecolor{currentstroke}%
\pgfsetdash{}{0pt}%
\pgfsys@defobject{currentmarker}{\pgfqpoint{-0.041667in}{-0.041667in}}{\pgfqpoint{0.041667in}{0.041667in}}{%
\pgfpathmoveto{\pgfqpoint{-0.041667in}{-0.041667in}}%
\pgfpathlineto{\pgfqpoint{0.041667in}{0.041667in}}%
\pgfpathmoveto{\pgfqpoint{-0.041667in}{0.041667in}}%
\pgfpathlineto{\pgfqpoint{0.041667in}{-0.041667in}}%
\pgfusepath{stroke,fill}%
}%
\begin{pgfscope}%
\pgfsys@transformshift{0.411301in}{0.579475in}%
\pgfsys@useobject{currentmarker}{}%
\end{pgfscope}%
\begin{pgfscope}%
\pgfsys@transformshift{0.665333in}{0.579504in}%
\pgfsys@useobject{currentmarker}{}%
\end{pgfscope}%
\begin{pgfscope}%
\pgfsys@transformshift{0.919365in}{0.579632in}%
\pgfsys@useobject{currentmarker}{}%
\end{pgfscope}%
\begin{pgfscope}%
\pgfsys@transformshift{1.173396in}{0.580200in}%
\pgfsys@useobject{currentmarker}{}%
\end{pgfscope}%
\begin{pgfscope}%
\pgfsys@transformshift{1.427428in}{0.582224in}%
\pgfsys@useobject{currentmarker}{}%
\end{pgfscope}%
\begin{pgfscope}%
\pgfsys@transformshift{1.681460in}{0.592361in}%
\pgfsys@useobject{currentmarker}{}%
\end{pgfscope}%
\begin{pgfscope}%
\pgfsys@transformshift{1.935492in}{0.674818in}%
\pgfsys@useobject{currentmarker}{}%
\end{pgfscope}%
\end{pgfscope}%
\begin{pgfscope}%
\pgfpathrectangle{\pgfqpoint{0.411301in}{0.579475in}}{\pgfqpoint{5.588699in}{2.337192in}}%
\pgfusepath{clip}%
\pgfsetbuttcap%
\pgfsetroundjoin%
\pgfsetlinewidth{1.505625pt}%
\definecolor{currentstroke}{rgb}{0.117647,0.533333,0.898039}%
\pgfsetstrokecolor{currentstroke}%
\pgfsetdash{{1.500000pt}{2.475000pt}}{0.000000pt}%
\pgfpathmoveto{\pgfqpoint{0.411301in}{0.581470in}}%
\pgfpathlineto{\pgfqpoint{0.665333in}{0.581470in}}%
\pgfpathlineto{\pgfqpoint{0.919365in}{0.581470in}}%
\pgfpathlineto{\pgfqpoint{1.173396in}{0.581471in}}%
\pgfpathlineto{\pgfqpoint{1.427428in}{0.581477in}}%
\pgfpathlineto{\pgfqpoint{1.681460in}{0.582462in}}%
\pgfpathlineto{\pgfqpoint{1.935492in}{0.742502in}}%
\pgfpathlineto{\pgfqpoint{1.956837in}{2.926667in}}%
\pgfusepath{stroke}%
\end{pgfscope}%
\begin{pgfscope}%
\pgfpathrectangle{\pgfqpoint{0.411301in}{0.579475in}}{\pgfqpoint{5.588699in}{2.337192in}}%
\pgfusepath{clip}%
\pgfsetbuttcap%
\pgfsetroundjoin%
\definecolor{currentfill}{rgb}{0.117647,0.533333,0.898039}%
\pgfsetfillcolor{currentfill}%
\pgfsetlinewidth{1.003750pt}%
\definecolor{currentstroke}{rgb}{0.117647,0.533333,0.898039}%
\pgfsetstrokecolor{currentstroke}%
\pgfsetdash{}{0pt}%
\pgfsys@defobject{currentmarker}{\pgfqpoint{-0.020833in}{-0.020833in}}{\pgfqpoint{0.020833in}{0.020833in}}{%
\pgfpathmoveto{\pgfqpoint{0.000000in}{-0.020833in}}%
\pgfpathcurveto{\pgfqpoint{0.005525in}{-0.020833in}}{\pgfqpoint{0.010825in}{-0.018638in}}{\pgfqpoint{0.014731in}{-0.014731in}}%
\pgfpathcurveto{\pgfqpoint{0.018638in}{-0.010825in}}{\pgfqpoint{0.020833in}{-0.005525in}}{\pgfqpoint{0.020833in}{0.000000in}}%
\pgfpathcurveto{\pgfqpoint{0.020833in}{0.005525in}}{\pgfqpoint{0.018638in}{0.010825in}}{\pgfqpoint{0.014731in}{0.014731in}}%
\pgfpathcurveto{\pgfqpoint{0.010825in}{0.018638in}}{\pgfqpoint{0.005525in}{0.020833in}}{\pgfqpoint{0.000000in}{0.020833in}}%
\pgfpathcurveto{\pgfqpoint{-0.005525in}{0.020833in}}{\pgfqpoint{-0.010825in}{0.018638in}}{\pgfqpoint{-0.014731in}{0.014731in}}%
\pgfpathcurveto{\pgfqpoint{-0.018638in}{0.010825in}}{\pgfqpoint{-0.020833in}{0.005525in}}{\pgfqpoint{-0.020833in}{0.000000in}}%
\pgfpathcurveto{\pgfqpoint{-0.020833in}{-0.005525in}}{\pgfqpoint{-0.018638in}{-0.010825in}}{\pgfqpoint{-0.014731in}{-0.014731in}}%
\pgfpathcurveto{\pgfqpoint{-0.010825in}{-0.018638in}}{\pgfqpoint{-0.005525in}{-0.020833in}}{\pgfqpoint{0.000000in}{-0.020833in}}%
\pgfpathlineto{\pgfqpoint{0.000000in}{-0.020833in}}%
\pgfpathclose%
\pgfusepath{stroke,fill}%
}%
\begin{pgfscope}%
\pgfsys@transformshift{0.411301in}{0.581470in}%
\pgfsys@useobject{currentmarker}{}%
\end{pgfscope}%
\begin{pgfscope}%
\pgfsys@transformshift{0.665333in}{0.581470in}%
\pgfsys@useobject{currentmarker}{}%
\end{pgfscope}%
\begin{pgfscope}%
\pgfsys@transformshift{0.919365in}{0.581470in}%
\pgfsys@useobject{currentmarker}{}%
\end{pgfscope}%
\begin{pgfscope}%
\pgfsys@transformshift{1.173396in}{0.581471in}%
\pgfsys@useobject{currentmarker}{}%
\end{pgfscope}%
\begin{pgfscope}%
\pgfsys@transformshift{1.427428in}{0.581477in}%
\pgfsys@useobject{currentmarker}{}%
\end{pgfscope}%
\begin{pgfscope}%
\pgfsys@transformshift{1.681460in}{0.582462in}%
\pgfsys@useobject{currentmarker}{}%
\end{pgfscope}%
\begin{pgfscope}%
\pgfsys@transformshift{1.935492in}{0.742502in}%
\pgfsys@useobject{currentmarker}{}%
\end{pgfscope}%
\begin{pgfscope}%
\pgfsys@transformshift{2.189523in}{26.736265in}%
\pgfsys@useobject{currentmarker}{}%
\end{pgfscope}%
\end{pgfscope}%
\begin{pgfscope}%
\pgfpathrectangle{\pgfqpoint{0.411301in}{0.579475in}}{\pgfqpoint{5.588699in}{2.337192in}}%
\pgfusepath{clip}%
\pgfsetbuttcap%
\pgfsetroundjoin%
\pgfsetlinewidth{1.505625pt}%
\definecolor{currentstroke}{rgb}{1.000000,0.756863,0.027451}%
\pgfsetstrokecolor{currentstroke}%
\pgfsetdash{{1.500000pt}{2.475000pt}}{0.000000pt}%
\pgfpathmoveto{\pgfqpoint{0.411301in}{0.579943in}}%
\pgfpathlineto{\pgfqpoint{0.665333in}{0.579958in}}%
\pgfpathlineto{\pgfqpoint{0.919365in}{0.580044in}}%
\pgfpathlineto{\pgfqpoint{1.173396in}{0.580520in}}%
\pgfpathlineto{\pgfqpoint{1.427428in}{0.583175in}}%
\pgfpathlineto{\pgfqpoint{1.681460in}{0.597966in}}%
\pgfpathlineto{\pgfqpoint{1.935492in}{0.680359in}}%
\pgfpathlineto{\pgfqpoint{2.189523in}{1.139343in}}%
\pgfusepath{stroke}%
\end{pgfscope}%
\begin{pgfscope}%
\pgfpathrectangle{\pgfqpoint{0.411301in}{0.579475in}}{\pgfqpoint{5.588699in}{2.337192in}}%
\pgfusepath{clip}%
\pgfsetbuttcap%
\pgfsetroundjoin%
\definecolor{currentfill}{rgb}{1.000000,0.756863,0.027451}%
\pgfsetfillcolor{currentfill}%
\pgfsetlinewidth{1.003750pt}%
\definecolor{currentstroke}{rgb}{1.000000,0.756863,0.027451}%
\pgfsetstrokecolor{currentstroke}%
\pgfsetdash{}{0pt}%
\pgfsys@defobject{currentmarker}{\pgfqpoint{-0.041667in}{-0.041667in}}{\pgfqpoint{0.041667in}{0.041667in}}{%
\pgfpathmoveto{\pgfqpoint{-0.041667in}{0.000000in}}%
\pgfpathlineto{\pgfqpoint{0.041667in}{0.000000in}}%
\pgfpathmoveto{\pgfqpoint{0.000000in}{-0.041667in}}%
\pgfpathlineto{\pgfqpoint{0.000000in}{0.041667in}}%
\pgfusepath{stroke,fill}%
}%
\begin{pgfscope}%
\pgfsys@transformshift{0.411301in}{0.579943in}%
\pgfsys@useobject{currentmarker}{}%
\end{pgfscope}%
\begin{pgfscope}%
\pgfsys@transformshift{0.665333in}{0.579958in}%
\pgfsys@useobject{currentmarker}{}%
\end{pgfscope}%
\begin{pgfscope}%
\pgfsys@transformshift{0.919365in}{0.580044in}%
\pgfsys@useobject{currentmarker}{}%
\end{pgfscope}%
\begin{pgfscope}%
\pgfsys@transformshift{1.173396in}{0.580520in}%
\pgfsys@useobject{currentmarker}{}%
\end{pgfscope}%
\begin{pgfscope}%
\pgfsys@transformshift{1.427428in}{0.583175in}%
\pgfsys@useobject{currentmarker}{}%
\end{pgfscope}%
\begin{pgfscope}%
\pgfsys@transformshift{1.681460in}{0.597966in}%
\pgfsys@useobject{currentmarker}{}%
\end{pgfscope}%
\begin{pgfscope}%
\pgfsys@transformshift{1.935492in}{0.680359in}%
\pgfsys@useobject{currentmarker}{}%
\end{pgfscope}%
\begin{pgfscope}%
\pgfsys@transformshift{2.189523in}{1.139343in}%
\pgfsys@useobject{currentmarker}{}%
\end{pgfscope}%
\end{pgfscope}%
\begin{pgfscope}%
\pgfpathrectangle{\pgfqpoint{0.411301in}{0.579475in}}{\pgfqpoint{5.588699in}{2.337192in}}%
\pgfusepath{clip}%
\pgfsetbuttcap%
\pgfsetroundjoin%
\pgfsetlinewidth{1.505625pt}%
\definecolor{currentstroke}{rgb}{0.000000,0.301961,0.250980}%
\pgfsetstrokecolor{currentstroke}%
\pgfsetdash{{1.500000pt}{2.475000pt}}{0.000000pt}%
\pgfpathmoveto{\pgfqpoint{0.411301in}{0.579771in}}%
\pgfpathlineto{\pgfqpoint{0.665333in}{0.579775in}}%
\pgfpathlineto{\pgfqpoint{0.919365in}{0.579802in}}%
\pgfpathlineto{\pgfqpoint{1.173396in}{0.580003in}}%
\pgfpathlineto{\pgfqpoint{1.427428in}{0.581493in}}%
\pgfpathlineto{\pgfqpoint{1.681460in}{0.592566in}}%
\pgfpathlineto{\pgfqpoint{1.935492in}{0.674804in}}%
\pgfpathlineto{\pgfqpoint{2.189523in}{1.285571in}}%
\pgfusepath{stroke}%
\end{pgfscope}%
\begin{pgfscope}%
\pgfpathrectangle{\pgfqpoint{0.411301in}{0.579475in}}{\pgfqpoint{5.588699in}{2.337192in}}%
\pgfusepath{clip}%
\pgfsetbuttcap%
\pgfsetroundjoin%
\definecolor{currentfill}{rgb}{0.000000,0.301961,0.250980}%
\pgfsetfillcolor{currentfill}%
\pgfsetlinewidth{1.003750pt}%
\definecolor{currentstroke}{rgb}{0.000000,0.301961,0.250980}%
\pgfsetstrokecolor{currentstroke}%
\pgfsetdash{}{0pt}%
\pgfsys@defobject{currentmarker}{\pgfqpoint{-0.041667in}{-0.041667in}}{\pgfqpoint{0.041667in}{0.041667in}}{%
\pgfpathmoveto{\pgfqpoint{-0.041667in}{-0.041667in}}%
\pgfpathlineto{\pgfqpoint{0.041667in}{0.041667in}}%
\pgfpathmoveto{\pgfqpoint{-0.041667in}{0.041667in}}%
\pgfpathlineto{\pgfqpoint{0.041667in}{-0.041667in}}%
\pgfusepath{stroke,fill}%
}%
\begin{pgfscope}%
\pgfsys@transformshift{0.411301in}{0.579771in}%
\pgfsys@useobject{currentmarker}{}%
\end{pgfscope}%
\begin{pgfscope}%
\pgfsys@transformshift{0.665333in}{0.579775in}%
\pgfsys@useobject{currentmarker}{}%
\end{pgfscope}%
\begin{pgfscope}%
\pgfsys@transformshift{0.919365in}{0.579802in}%
\pgfsys@useobject{currentmarker}{}%
\end{pgfscope}%
\begin{pgfscope}%
\pgfsys@transformshift{1.173396in}{0.580003in}%
\pgfsys@useobject{currentmarker}{}%
\end{pgfscope}%
\begin{pgfscope}%
\pgfsys@transformshift{1.427428in}{0.581493in}%
\pgfsys@useobject{currentmarker}{}%
\end{pgfscope}%
\begin{pgfscope}%
\pgfsys@transformshift{1.681460in}{0.592566in}%
\pgfsys@useobject{currentmarker}{}%
\end{pgfscope}%
\begin{pgfscope}%
\pgfsys@transformshift{1.935492in}{0.674804in}%
\pgfsys@useobject{currentmarker}{}%
\end{pgfscope}%
\begin{pgfscope}%
\pgfsys@transformshift{2.189523in}{1.285571in}%
\pgfsys@useobject{currentmarker}{}%
\end{pgfscope}%
\end{pgfscope}%
\begin{pgfscope}%
\pgfpathrectangle{\pgfqpoint{0.411301in}{0.579475in}}{\pgfqpoint{5.588699in}{2.337192in}}%
\pgfusepath{clip}%
\pgfsetbuttcap%
\pgfsetroundjoin%
\pgfsetlinewidth{1.505625pt}%
\definecolor{currentstroke}{rgb}{1.000000,0.000000,0.000000}%
\pgfsetstrokecolor{currentstroke}%
\pgfsetdash{{5.550000pt}{2.400000pt}}{0.000000pt}%
\pgfpathmoveto{\pgfqpoint{2.062508in}{0.579475in}}%
\pgfpathlineto{\pgfqpoint{2.062508in}{2.916667in}}%
\pgfusepath{stroke}%
\end{pgfscope}%
\begin{pgfscope}%
\pgfsetrectcap%
\pgfsetmiterjoin%
\pgfsetlinewidth{0.803000pt}%
\definecolor{currentstroke}{rgb}{0.000000,0.000000,0.000000}%
\pgfsetstrokecolor{currentstroke}%
\pgfsetdash{}{0pt}%
\pgfpathmoveto{\pgfqpoint{0.411301in}{0.579475in}}%
\pgfpathlineto{\pgfqpoint{0.411301in}{2.916667in}}%
\pgfusepath{stroke}%
\end{pgfscope}%
\begin{pgfscope}%
\pgfsetrectcap%
\pgfsetmiterjoin%
\pgfsetlinewidth{0.803000pt}%
\definecolor{currentstroke}{rgb}{0.000000,0.000000,0.000000}%
\pgfsetstrokecolor{currentstroke}%
\pgfsetdash{}{0pt}%
\pgfpathmoveto{\pgfqpoint{6.000000in}{0.579475in}}%
\pgfpathlineto{\pgfqpoint{6.000000in}{2.916667in}}%
\pgfusepath{stroke}%
\end{pgfscope}%
\begin{pgfscope}%
\pgfsetrectcap%
\pgfsetmiterjoin%
\pgfsetlinewidth{0.803000pt}%
\definecolor{currentstroke}{rgb}{0.000000,0.000000,0.000000}%
\pgfsetstrokecolor{currentstroke}%
\pgfsetdash{}{0pt}%
\pgfpathmoveto{\pgfqpoint{0.411301in}{0.579475in}}%
\pgfpathlineto{\pgfqpoint{6.000000in}{0.579475in}}%
\pgfusepath{stroke}%
\end{pgfscope}%
\begin{pgfscope}%
\pgfsetrectcap%
\pgfsetmiterjoin%
\pgfsetlinewidth{0.803000pt}%
\definecolor{currentstroke}{rgb}{0.000000,0.000000,0.000000}%
\pgfsetstrokecolor{currentstroke}%
\pgfsetdash{}{0pt}%
\pgfpathmoveto{\pgfqpoint{0.411301in}{2.916667in}}%
\pgfpathlineto{\pgfqpoint{6.000000in}{2.916667in}}%
\pgfusepath{stroke}%
\end{pgfscope}%
\begin{pgfscope}%
\definecolor{textcolor}{rgb}{1.000000,0.000000,0.000000}%
\pgfsetstrokecolor{textcolor}%
\pgfsetfillcolor{textcolor}%
\pgftext[x=2.189523in,y=2.449228in,left,base]{\color{textcolor}\rmfamily\fontsize{16.000000}{19.200000}\selectfont out of memory}%
\end{pgfscope}%
\begin{pgfscope}%
\pgfsetbuttcap%
\pgfsetmiterjoin%
\definecolor{currentfill}{rgb}{1.000000,1.000000,1.000000}%
\pgfsetfillcolor{currentfill}%
\pgfsetfillopacity{0.800000}%
\pgfsetlinewidth{1.003750pt}%
\definecolor{currentstroke}{rgb}{0.800000,0.800000,0.800000}%
\pgfsetstrokecolor{currentstroke}%
\pgfsetstrokeopacity{0.800000}%
\pgfsetdash{}{0pt}%
\pgfpathmoveto{\pgfqpoint{4.326764in}{1.441050in}}%
\pgfpathlineto{\pgfqpoint{5.844444in}{1.441050in}}%
\pgfpathquadraticcurveto{\pgfqpoint{5.888889in}{1.441050in}}{\pgfqpoint{5.888889in}{1.485494in}}%
\pgfpathlineto{\pgfqpoint{5.888889in}{2.761111in}}%
\pgfpathquadraticcurveto{\pgfqpoint{5.888889in}{2.805556in}}{\pgfqpoint{5.844444in}{2.805556in}}%
\pgfpathlineto{\pgfqpoint{4.326764in}{2.805556in}}%
\pgfpathquadraticcurveto{\pgfqpoint{4.282320in}{2.805556in}}{\pgfqpoint{4.282320in}{2.761111in}}%
\pgfpathlineto{\pgfqpoint{4.282320in}{1.485494in}}%
\pgfpathquadraticcurveto{\pgfqpoint{4.282320in}{1.441050in}}{\pgfqpoint{4.326764in}{1.441050in}}%
\pgfpathlineto{\pgfqpoint{4.326764in}{1.441050in}}%
\pgfpathclose%
\pgfusepath{stroke,fill}%
\end{pgfscope}%
\begin{pgfscope}%
\pgfsetrectcap%
\pgfsetroundjoin%
\pgfsetlinewidth{1.505625pt}%
\definecolor{currentstroke}{rgb}{0.117647,0.533333,0.898039}%
\pgfsetstrokecolor{currentstroke}%
\pgfsetdash{}{0pt}%
\pgfpathmoveto{\pgfqpoint{4.371209in}{2.627778in}}%
\pgfpathlineto{\pgfqpoint{4.593431in}{2.627778in}}%
\pgfpathlineto{\pgfqpoint{4.815653in}{2.627778in}}%
\pgfusepath{stroke}%
\end{pgfscope}%
\begin{pgfscope}%
\pgfsetbuttcap%
\pgfsetroundjoin%
\definecolor{currentfill}{rgb}{0.117647,0.533333,0.898039}%
\pgfsetfillcolor{currentfill}%
\pgfsetlinewidth{1.003750pt}%
\definecolor{currentstroke}{rgb}{0.117647,0.533333,0.898039}%
\pgfsetstrokecolor{currentstroke}%
\pgfsetdash{}{0pt}%
\pgfsys@defobject{currentmarker}{\pgfqpoint{-0.020833in}{-0.020833in}}{\pgfqpoint{0.020833in}{0.020833in}}{%
\pgfpathmoveto{\pgfqpoint{0.000000in}{-0.020833in}}%
\pgfpathcurveto{\pgfqpoint{0.005525in}{-0.020833in}}{\pgfqpoint{0.010825in}{-0.018638in}}{\pgfqpoint{0.014731in}{-0.014731in}}%
\pgfpathcurveto{\pgfqpoint{0.018638in}{-0.010825in}}{\pgfqpoint{0.020833in}{-0.005525in}}{\pgfqpoint{0.020833in}{0.000000in}}%
\pgfpathcurveto{\pgfqpoint{0.020833in}{0.005525in}}{\pgfqpoint{0.018638in}{0.010825in}}{\pgfqpoint{0.014731in}{0.014731in}}%
\pgfpathcurveto{\pgfqpoint{0.010825in}{0.018638in}}{\pgfqpoint{0.005525in}{0.020833in}}{\pgfqpoint{0.000000in}{0.020833in}}%
\pgfpathcurveto{\pgfqpoint{-0.005525in}{0.020833in}}{\pgfqpoint{-0.010825in}{0.018638in}}{\pgfqpoint{-0.014731in}{0.014731in}}%
\pgfpathcurveto{\pgfqpoint{-0.018638in}{0.010825in}}{\pgfqpoint{-0.020833in}{0.005525in}}{\pgfqpoint{-0.020833in}{0.000000in}}%
\pgfpathcurveto{\pgfqpoint{-0.020833in}{-0.005525in}}{\pgfqpoint{-0.018638in}{-0.010825in}}{\pgfqpoint{-0.014731in}{-0.014731in}}%
\pgfpathcurveto{\pgfqpoint{-0.010825in}{-0.018638in}}{\pgfqpoint{-0.005525in}{-0.020833in}}{\pgfqpoint{0.000000in}{-0.020833in}}%
\pgfpathlineto{\pgfqpoint{0.000000in}{-0.020833in}}%
\pgfpathclose%
\pgfusepath{stroke,fill}%
}%
\begin{pgfscope}%
\pgfsys@transformshift{4.593431in}{2.627778in}%
\pgfsys@useobject{currentmarker}{}%
\end{pgfscope}%
\end{pgfscope}%
\begin{pgfscope}%
\definecolor{textcolor}{rgb}{0.000000,0.000000,0.000000}%
\pgfsetstrokecolor{textcolor}%
\pgfsetfillcolor{textcolor}%
\pgftext[x=4.993431in,y=2.550000in,left,base]{\color{textcolor}\rmfamily\fontsize{16.000000}{19.200000}\selectfont e-nodes}%
\end{pgfscope}%
\begin{pgfscope}%
\pgfsetrectcap%
\pgfsetroundjoin%
\pgfsetlinewidth{1.505625pt}%
\definecolor{currentstroke}{rgb}{1.000000,0.756863,0.027451}%
\pgfsetstrokecolor{currentstroke}%
\pgfsetdash{}{0pt}%
\pgfpathmoveto{\pgfqpoint{4.371209in}{2.303318in}}%
\pgfpathlineto{\pgfqpoint{4.593431in}{2.303318in}}%
\pgfpathlineto{\pgfqpoint{4.815653in}{2.303318in}}%
\pgfusepath{stroke}%
\end{pgfscope}%
\begin{pgfscope}%
\pgfsetbuttcap%
\pgfsetroundjoin%
\definecolor{currentfill}{rgb}{1.000000,0.756863,0.027451}%
\pgfsetfillcolor{currentfill}%
\pgfsetlinewidth{1.003750pt}%
\definecolor{currentstroke}{rgb}{1.000000,0.756863,0.027451}%
\pgfsetstrokecolor{currentstroke}%
\pgfsetdash{}{0pt}%
\pgfsys@defobject{currentmarker}{\pgfqpoint{-0.041667in}{-0.041667in}}{\pgfqpoint{0.041667in}{0.041667in}}{%
\pgfpathmoveto{\pgfqpoint{-0.041667in}{0.000000in}}%
\pgfpathlineto{\pgfqpoint{0.041667in}{0.000000in}}%
\pgfpathmoveto{\pgfqpoint{0.000000in}{-0.041667in}}%
\pgfpathlineto{\pgfqpoint{0.000000in}{0.041667in}}%
\pgfusepath{stroke,fill}%
}%
\begin{pgfscope}%
\pgfsys@transformshift{4.593431in}{2.303318in}%
\pgfsys@useobject{currentmarker}{}%
\end{pgfscope}%
\end{pgfscope}%
\begin{pgfscope}%
\definecolor{textcolor}{rgb}{0.000000,0.000000,0.000000}%
\pgfsetstrokecolor{textcolor}%
\pgfsetfillcolor{textcolor}%
\pgftext[x=4.993431in,y=2.225540in,left,base]{\color{textcolor}\rmfamily\fontsize{16.000000}{19.200000}\selectfont e-classes}%
\end{pgfscope}%
\begin{pgfscope}%
\pgfsetrectcap%
\pgfsetroundjoin%
\pgfsetlinewidth{1.505625pt}%
\definecolor{currentstroke}{rgb}{0.000000,0.301961,0.250980}%
\pgfsetstrokecolor{currentstroke}%
\pgfsetdash{}{0pt}%
\pgfpathmoveto{\pgfqpoint{4.371209in}{1.978858in}}%
\pgfpathlineto{\pgfqpoint{4.593431in}{1.978858in}}%
\pgfpathlineto{\pgfqpoint{4.815653in}{1.978858in}}%
\pgfusepath{stroke}%
\end{pgfscope}%
\begin{pgfscope}%
\pgfsetbuttcap%
\pgfsetroundjoin%
\definecolor{currentfill}{rgb}{0.000000,0.301961,0.250980}%
\pgfsetfillcolor{currentfill}%
\pgfsetlinewidth{1.003750pt}%
\definecolor{currentstroke}{rgb}{0.000000,0.301961,0.250980}%
\pgfsetstrokecolor{currentstroke}%
\pgfsetdash{}{0pt}%
\pgfsys@defobject{currentmarker}{\pgfqpoint{-0.041667in}{-0.041667in}}{\pgfqpoint{0.041667in}{0.041667in}}{%
\pgfpathmoveto{\pgfqpoint{-0.041667in}{-0.041667in}}%
\pgfpathlineto{\pgfqpoint{0.041667in}{0.041667in}}%
\pgfpathmoveto{\pgfqpoint{-0.041667in}{0.041667in}}%
\pgfpathlineto{\pgfqpoint{0.041667in}{-0.041667in}}%
\pgfusepath{stroke,fill}%
}%
\begin{pgfscope}%
\pgfsys@transformshift{4.593431in}{1.978858in}%
\pgfsys@useobject{currentmarker}{}%
\end{pgfscope}%
\end{pgfscope}%
\begin{pgfscope}%
\definecolor{textcolor}{rgb}{0.000000,0.000000,0.000000}%
\pgfsetstrokecolor{textcolor}%
\pgfsetfillcolor{textcolor}%
\pgftext[x=4.993431in,y=1.901081in,left,base]{\color{textcolor}\rmfamily\fontsize{16.000000}{19.200000}\selectfont rules}%
\end{pgfscope}%
\begin{pgfscope}%
\pgfsetbuttcap%
\pgfsetroundjoin%
\pgfsetlinewidth{1.505625pt}%
\definecolor{currentstroke}{rgb}{0.000000,0.000000,0.000000}%
\pgfsetstrokecolor{currentstroke}%
\pgfsetdash{{1.500000pt}{2.475000pt}}{0.000000pt}%
\pgfpathmoveto{\pgfqpoint{4.371209in}{1.654399in}}%
\pgfpathlineto{\pgfqpoint{4.593431in}{1.654399in}}%
\pgfpathlineto{\pgfqpoint{4.815653in}{1.654399in}}%
\pgfusepath{stroke}%
\end{pgfscope}%
\begin{pgfscope}%
\definecolor{textcolor}{rgb}{0.000000,0.000000,0.000000}%
\pgfsetstrokecolor{textcolor}%
\pgfsetfillcolor{textcolor}%
\pgftext[x=4.993431in,y=1.576621in,left,base]{\color{textcolor}\rmfamily\fontsize{16.000000}{19.200000}\selectfont estimate}%
\end{pgfscope}%
\end{pgfpicture}%
\makeatother%
\endgroup%

%% file: media/guided-blocking.pgf
\begingroup%
\makeatletter%
\begin{pgfpicture}%
\pgfpathrectangle{\pgfpointorigin}{\pgfqpoint{6.000000in}{3.000000in}}%
\pgfusepath{use as bounding box, clip}%
\begin{pgfscope}%
\pgfsetbuttcap%
\pgfsetmiterjoin%
\definecolor{currentfill}{rgb}{1.000000,1.000000,1.000000}%
\pgfsetfillcolor{currentfill}%
\pgfsetlinewidth{0.000000pt}%
\definecolor{currentstroke}{rgb}{1.000000,1.000000,1.000000}%
\pgfsetstrokecolor{currentstroke}%
\pgfsetdash{}{0pt}%
\pgfpathmoveto{\pgfqpoint{0.000000in}{0.000000in}}%
\pgfpathlineto{\pgfqpoint{6.000000in}{0.000000in}}%
\pgfpathlineto{\pgfqpoint{6.000000in}{3.000000in}}%
\pgfpathlineto{\pgfqpoint{0.000000in}{3.000000in}}%
\pgfpathlineto{\pgfqpoint{0.000000in}{0.000000in}}%
\pgfpathclose%
\pgfusepath{fill}%
\end{pgfscope}%
\begin{pgfscope}%
\pgfsetbuttcap%
\pgfsetmiterjoin%
\definecolor{currentfill}{rgb}{1.000000,1.000000,1.000000}%
\pgfsetfillcolor{currentfill}%
\pgfsetlinewidth{0.000000pt}%
\definecolor{currentstroke}{rgb}{0.000000,0.000000,0.000000}%
\pgfsetstrokecolor{currentstroke}%
\pgfsetstrokeopacity{0.000000}%
\pgfsetdash{}{0pt}%
\pgfpathmoveto{\pgfqpoint{0.490023in}{0.579475in}}%
\pgfpathlineto{\pgfqpoint{6.000000in}{0.579475in}}%
\pgfpathlineto{\pgfqpoint{6.000000in}{2.998326in}}%
\pgfpathlineto{\pgfqpoint{0.490023in}{2.998326in}}%
\pgfpathlineto{\pgfqpoint{0.490023in}{0.579475in}}%
\pgfpathclose%
\pgfusepath{fill}%
\end{pgfscope}%
\begin{pgfscope}%
\pgfsetbuttcap%
\pgfsetroundjoin%
\definecolor{currentfill}{rgb}{0.000000,0.000000,0.000000}%
\pgfsetfillcolor{currentfill}%
\pgfsetlinewidth{0.803000pt}%
\definecolor{currentstroke}{rgb}{0.000000,0.000000,0.000000}%
\pgfsetstrokecolor{currentstroke}%
\pgfsetdash{}{0pt}%
\pgfsys@defobject{currentmarker}{\pgfqpoint{0.000000in}{-0.048611in}}{\pgfqpoint{0.000000in}{0.000000in}}{%
\pgfpathmoveto{\pgfqpoint{0.000000in}{0.000000in}}%
\pgfpathlineto{\pgfqpoint{0.000000in}{-0.048611in}}%
\pgfusepath{stroke,fill}%
}%
\begin{pgfscope}%
\pgfsys@transformshift{0.490023in}{0.579475in}%
\pgfsys@useobject{currentmarker}{}%
\end{pgfscope}%
\end{pgfscope}%
\begin{pgfscope}%
\definecolor{textcolor}{rgb}{0.000000,0.000000,0.000000}%
\pgfsetstrokecolor{textcolor}%
\pgfsetfillcolor{textcolor}%
\pgftext[x=0.490023in,y=0.482253in,,top]{\color{textcolor}\rmfamily\fontsize{16.000000}{19.200000}\selectfont \(\displaystyle {0}\)}%
\end{pgfscope}%
\begin{pgfscope}%
\pgfsetbuttcap%
\pgfsetroundjoin%
\definecolor{currentfill}{rgb}{0.000000,0.000000,0.000000}%
\pgfsetfillcolor{currentfill}%
\pgfsetlinewidth{0.803000pt}%
\definecolor{currentstroke}{rgb}{0.000000,0.000000,0.000000}%
\pgfsetstrokecolor{currentstroke}%
\pgfsetdash{}{0pt}%
\pgfsys@defobject{currentmarker}{\pgfqpoint{0.000000in}{-0.048611in}}{\pgfqpoint{0.000000in}{0.000000in}}{%
\pgfpathmoveto{\pgfqpoint{0.000000in}{0.000000in}}%
\pgfpathlineto{\pgfqpoint{0.000000in}{-0.048611in}}%
\pgfusepath{stroke,fill}%
}%
\begin{pgfscope}%
\pgfsys@transformshift{1.742291in}{0.579475in}%
\pgfsys@useobject{currentmarker}{}%
\end{pgfscope}%
\end{pgfscope}%
\begin{pgfscope}%
\definecolor{textcolor}{rgb}{0.000000,0.000000,0.000000}%
\pgfsetstrokecolor{textcolor}%
\pgfsetfillcolor{textcolor}%
\pgftext[x=1.742291in,y=0.482253in,,top]{\color{textcolor}\rmfamily\fontsize{16.000000}{19.200000}\selectfont \(\displaystyle {5}\)}%
\end{pgfscope}%
\begin{pgfscope}%
\pgfsetbuttcap%
\pgfsetroundjoin%
\definecolor{currentfill}{rgb}{0.000000,0.000000,0.000000}%
\pgfsetfillcolor{currentfill}%
\pgfsetlinewidth{0.803000pt}%
\definecolor{currentstroke}{rgb}{0.000000,0.000000,0.000000}%
\pgfsetstrokecolor{currentstroke}%
\pgfsetdash{}{0pt}%
\pgfsys@defobject{currentmarker}{\pgfqpoint{0.000000in}{-0.048611in}}{\pgfqpoint{0.000000in}{0.000000in}}{%
\pgfpathmoveto{\pgfqpoint{0.000000in}{0.000000in}}%
\pgfpathlineto{\pgfqpoint{0.000000in}{-0.048611in}}%
\pgfusepath{stroke,fill}%
}%
\begin{pgfscope}%
\pgfsys@transformshift{2.994558in}{0.579475in}%
\pgfsys@useobject{currentmarker}{}%
\end{pgfscope}%
\end{pgfscope}%
\begin{pgfscope}%
\definecolor{textcolor}{rgb}{0.000000,0.000000,0.000000}%
\pgfsetstrokecolor{textcolor}%
\pgfsetfillcolor{textcolor}%
\pgftext[x=2.994558in,y=0.482253in,,top]{\color{textcolor}\rmfamily\fontsize{16.000000}{19.200000}\selectfont \(\displaystyle {10}\)}%
\end{pgfscope}%
\begin{pgfscope}%
\pgfsetbuttcap%
\pgfsetroundjoin%
\definecolor{currentfill}{rgb}{0.000000,0.000000,0.000000}%
\pgfsetfillcolor{currentfill}%
\pgfsetlinewidth{0.803000pt}%
\definecolor{currentstroke}{rgb}{0.000000,0.000000,0.000000}%
\pgfsetstrokecolor{currentstroke}%
\pgfsetdash{}{0pt}%
\pgfsys@defobject{currentmarker}{\pgfqpoint{0.000000in}{-0.048611in}}{\pgfqpoint{0.000000in}{0.000000in}}{%
\pgfpathmoveto{\pgfqpoint{0.000000in}{0.000000in}}%
\pgfpathlineto{\pgfqpoint{0.000000in}{-0.048611in}}%
\pgfusepath{stroke,fill}%
}%
\begin{pgfscope}%
\pgfsys@transformshift{4.246826in}{0.579475in}%
\pgfsys@useobject{currentmarker}{}%
\end{pgfscope}%
\end{pgfscope}%
\begin{pgfscope}%
\definecolor{textcolor}{rgb}{0.000000,0.000000,0.000000}%
\pgfsetstrokecolor{textcolor}%
\pgfsetfillcolor{textcolor}%
\pgftext[x=4.246826in,y=0.482253in,,top]{\color{textcolor}\rmfamily\fontsize{16.000000}{19.200000}\selectfont \(\displaystyle {15}\)}%
\end{pgfscope}%
\begin{pgfscope}%
\pgfsetbuttcap%
\pgfsetroundjoin%
\definecolor{currentfill}{rgb}{0.000000,0.000000,0.000000}%
\pgfsetfillcolor{currentfill}%
\pgfsetlinewidth{0.803000pt}%
\definecolor{currentstroke}{rgb}{0.000000,0.000000,0.000000}%
\pgfsetstrokecolor{currentstroke}%
\pgfsetdash{}{0pt}%
\pgfsys@defobject{currentmarker}{\pgfqpoint{0.000000in}{-0.048611in}}{\pgfqpoint{0.000000in}{0.000000in}}{%
\pgfpathmoveto{\pgfqpoint{0.000000in}{0.000000in}}%
\pgfpathlineto{\pgfqpoint{0.000000in}{-0.048611in}}%
\pgfusepath{stroke,fill}%
}%
\begin{pgfscope}%
\pgfsys@transformshift{5.499093in}{0.579475in}%
\pgfsys@useobject{currentmarker}{}%
\end{pgfscope}%
\end{pgfscope}%
\begin{pgfscope}%
\definecolor{textcolor}{rgb}{0.000000,0.000000,0.000000}%
\pgfsetstrokecolor{textcolor}%
\pgfsetfillcolor{textcolor}%
\pgftext[x=5.499093in,y=0.482253in,,top]{\color{textcolor}\rmfamily\fontsize{16.000000}{19.200000}\selectfont \(\displaystyle {20}\)}%
\end{pgfscope}%
\begin{pgfscope}%
\definecolor{textcolor}{rgb}{0.000000,0.000000,0.000000}%
\pgfsetstrokecolor{textcolor}%
\pgfsetfillcolor{textcolor}%
\pgftext[x=3.245012in,y=0.213349in,,top]{\color{textcolor}\rmfamily\fontsize{16.000000}{19.200000}\selectfont iterations}%
\end{pgfscope}%
\begin{pgfscope}%
\pgfsetbuttcap%
\pgfsetroundjoin%
\definecolor{currentfill}{rgb}{0.000000,0.000000,0.000000}%
\pgfsetfillcolor{currentfill}%
\pgfsetlinewidth{0.803000pt}%
\definecolor{currentstroke}{rgb}{0.000000,0.000000,0.000000}%
\pgfsetstrokecolor{currentstroke}%
\pgfsetdash{}{0pt}%
\pgfsys@defobject{currentmarker}{\pgfqpoint{-0.048611in}{0.000000in}}{\pgfqpoint{-0.000000in}{0.000000in}}{%
\pgfpathmoveto{\pgfqpoint{-0.000000in}{0.000000in}}%
\pgfpathlineto{\pgfqpoint{-0.048611in}{0.000000in}}%
\pgfusepath{stroke,fill}%
}%
\begin{pgfscope}%
\pgfsys@transformshift{0.490023in}{0.689423in}%
\pgfsys@useobject{currentmarker}{}%
\end{pgfscope}%
\end{pgfscope}%
\begin{pgfscope}%
\definecolor{textcolor}{rgb}{0.000000,0.000000,0.000000}%
\pgfsetstrokecolor{textcolor}%
\pgfsetfillcolor{textcolor}%
\pgftext[x=0.110068in, y=0.606089in, left, base]{\color{textcolor}\rmfamily\fontsize{16.000000}{19.200000}\selectfont 0K}%
\end{pgfscope}%
\begin{pgfscope}%
\pgfsetbuttcap%
\pgfsetroundjoin%
\definecolor{currentfill}{rgb}{0.000000,0.000000,0.000000}%
\pgfsetfillcolor{currentfill}%
\pgfsetlinewidth{0.803000pt}%
\definecolor{currentstroke}{rgb}{0.000000,0.000000,0.000000}%
\pgfsetstrokecolor{currentstroke}%
\pgfsetdash{}{0pt}%
\pgfsys@defobject{currentmarker}{\pgfqpoint{-0.048611in}{0.000000in}}{\pgfqpoint{-0.000000in}{0.000000in}}{%
\pgfpathmoveto{\pgfqpoint{-0.000000in}{0.000000in}}%
\pgfpathlineto{\pgfqpoint{-0.048611in}{0.000000in}}%
\pgfusepath{stroke,fill}%
}%
\begin{pgfscope}%
\pgfsys@transformshift{0.490023in}{1.189185in}%
\pgfsys@useobject{currentmarker}{}%
\end{pgfscope}%
\end{pgfscope}%
\begin{pgfscope}%
\definecolor{textcolor}{rgb}{0.000000,0.000000,0.000000}%
\pgfsetstrokecolor{textcolor}%
\pgfsetfillcolor{textcolor}%
\pgftext[x=0.110068in, y=1.105852in, left, base]{\color{textcolor}\rmfamily\fontsize{16.000000}{19.200000}\selectfont 2K}%
\end{pgfscope}%
\begin{pgfscope}%
\pgfsetbuttcap%
\pgfsetroundjoin%
\definecolor{currentfill}{rgb}{0.000000,0.000000,0.000000}%
\pgfsetfillcolor{currentfill}%
\pgfsetlinewidth{0.803000pt}%
\definecolor{currentstroke}{rgb}{0.000000,0.000000,0.000000}%
\pgfsetstrokecolor{currentstroke}%
\pgfsetdash{}{0pt}%
\pgfsys@defobject{currentmarker}{\pgfqpoint{-0.048611in}{0.000000in}}{\pgfqpoint{-0.000000in}{0.000000in}}{%
\pgfpathmoveto{\pgfqpoint{-0.000000in}{0.000000in}}%
\pgfpathlineto{\pgfqpoint{-0.048611in}{0.000000in}}%
\pgfusepath{stroke,fill}%
}%
\begin{pgfscope}%
\pgfsys@transformshift{0.490023in}{1.688948in}%
\pgfsys@useobject{currentmarker}{}%
\end{pgfscope}%
\end{pgfscope}%
\begin{pgfscope}%
\definecolor{textcolor}{rgb}{0.000000,0.000000,0.000000}%
\pgfsetstrokecolor{textcolor}%
\pgfsetfillcolor{textcolor}%
\pgftext[x=0.110068in, y=1.605615in, left, base]{\color{textcolor}\rmfamily\fontsize{16.000000}{19.200000}\selectfont 5K}%
\end{pgfscope}%
\begin{pgfscope}%
\pgfsetbuttcap%
\pgfsetroundjoin%
\definecolor{currentfill}{rgb}{0.000000,0.000000,0.000000}%
\pgfsetfillcolor{currentfill}%
\pgfsetlinewidth{0.803000pt}%
\definecolor{currentstroke}{rgb}{0.000000,0.000000,0.000000}%
\pgfsetstrokecolor{currentstroke}%
\pgfsetdash{}{0pt}%
\pgfsys@defobject{currentmarker}{\pgfqpoint{-0.048611in}{0.000000in}}{\pgfqpoint{-0.000000in}{0.000000in}}{%
\pgfpathmoveto{\pgfqpoint{-0.000000in}{0.000000in}}%
\pgfpathlineto{\pgfqpoint{-0.048611in}{0.000000in}}%
\pgfusepath{stroke,fill}%
}%
\begin{pgfscope}%
\pgfsys@transformshift{0.490023in}{2.188710in}%
\pgfsys@useobject{currentmarker}{}%
\end{pgfscope}%
\end{pgfscope}%
\begin{pgfscope}%
\definecolor{textcolor}{rgb}{0.000000,0.000000,0.000000}%
\pgfsetstrokecolor{textcolor}%
\pgfsetfillcolor{textcolor}%
\pgftext[x=0.110068in, y=2.105377in, left, base]{\color{textcolor}\rmfamily\fontsize{16.000000}{19.200000}\selectfont 7K}%
\end{pgfscope}%
\begin{pgfscope}%
\pgfsetbuttcap%
\pgfsetroundjoin%
\definecolor{currentfill}{rgb}{0.000000,0.000000,0.000000}%
\pgfsetfillcolor{currentfill}%
\pgfsetlinewidth{0.803000pt}%
\definecolor{currentstroke}{rgb}{0.000000,0.000000,0.000000}%
\pgfsetstrokecolor{currentstroke}%
\pgfsetdash{}{0pt}%
\pgfsys@defobject{currentmarker}{\pgfqpoint{-0.048611in}{0.000000in}}{\pgfqpoint{-0.000000in}{0.000000in}}{%
\pgfpathmoveto{\pgfqpoint{-0.000000in}{0.000000in}}%
\pgfpathlineto{\pgfqpoint{-0.048611in}{0.000000in}}%
\pgfusepath{stroke,fill}%
}%
\begin{pgfscope}%
\pgfsys@transformshift{0.490023in}{2.688473in}%
\pgfsys@useobject{currentmarker}{}%
\end{pgfscope}%
\end{pgfscope}%
\begin{pgfscope}%
\definecolor{textcolor}{rgb}{0.000000,0.000000,0.000000}%
\pgfsetstrokecolor{textcolor}%
\pgfsetfillcolor{textcolor}%
\pgftext[x=0.000000in, y=2.605140in, left, base]{\color{textcolor}\rmfamily\fontsize{16.000000}{19.200000}\selectfont 10K}%
\end{pgfscope}%
\begin{pgfscope}%
\pgfpathrectangle{\pgfqpoint{0.490023in}{0.579475in}}{\pgfqpoint{5.509977in}{2.418851in}}%
\pgfusepath{clip}%
\pgfsetrectcap%
\pgfsetroundjoin%
\pgfsetlinewidth{1.505625pt}%
\definecolor{currentstroke}{rgb}{0.117647,0.533333,0.898039}%
\pgfsetstrokecolor{currentstroke}%
\pgfsetdash{}{0pt}%
\pgfpathmoveto{\pgfqpoint{0.490023in}{0.697419in}}%
\pgfpathlineto{\pgfqpoint{0.740477in}{0.709413in}}%
\pgfpathlineto{\pgfqpoint{0.990930in}{0.734401in}}%
\pgfpathlineto{\pgfqpoint{1.241384in}{0.801370in}}%
\pgfpathlineto{\pgfqpoint{1.491837in}{0.976286in}}%
\pgfpathlineto{\pgfqpoint{1.742291in}{1.445464in}}%
\pgfpathlineto{\pgfqpoint{1.992744in}{2.583923in}}%
\pgfpathmoveto{\pgfqpoint{1.992744in}{0.702616in}}%
\pgfpathlineto{\pgfqpoint{2.243198in}{0.711412in}}%
\pgfpathlineto{\pgfqpoint{2.493651in}{0.729804in}}%
\pgfpathlineto{\pgfqpoint{2.744105in}{0.757390in}}%
\pgfpathlineto{\pgfqpoint{2.994558in}{0.798371in}}%
\pgfpathlineto{\pgfqpoint{3.245012in}{0.874935in}}%
\pgfpathlineto{\pgfqpoint{3.495465in}{1.025063in}}%
\pgfpathlineto{\pgfqpoint{3.745919in}{1.287738in}}%
\pgfpathlineto{\pgfqpoint{3.996372in}{1.757515in}}%
\pgfpathlineto{\pgfqpoint{4.246826in}{2.254679in}}%
\pgfpathlineto{\pgfqpoint{4.497279in}{2.850996in}}%
\pgfusepath{stroke}%
\end{pgfscope}%
\begin{pgfscope}%
\pgfpathrectangle{\pgfqpoint{0.490023in}{0.579475in}}{\pgfqpoint{5.509977in}{2.418851in}}%
\pgfusepath{clip}%
\pgfsetbuttcap%
\pgfsetroundjoin%
\definecolor{currentfill}{rgb}{0.117647,0.533333,0.898039}%
\pgfsetfillcolor{currentfill}%
\pgfsetlinewidth{1.003750pt}%
\definecolor{currentstroke}{rgb}{0.117647,0.533333,0.898039}%
\pgfsetstrokecolor{currentstroke}%
\pgfsetdash{}{0pt}%
\pgfsys@defobject{currentmarker}{\pgfqpoint{-0.020833in}{-0.020833in}}{\pgfqpoint{0.020833in}{0.020833in}}{%
\pgfpathmoveto{\pgfqpoint{0.000000in}{-0.020833in}}%
\pgfpathcurveto{\pgfqpoint{0.005525in}{-0.020833in}}{\pgfqpoint{0.010825in}{-0.018638in}}{\pgfqpoint{0.014731in}{-0.014731in}}%
\pgfpathcurveto{\pgfqpoint{0.018638in}{-0.010825in}}{\pgfqpoint{0.020833in}{-0.005525in}}{\pgfqpoint{0.020833in}{0.000000in}}%
\pgfpathcurveto{\pgfqpoint{0.020833in}{0.005525in}}{\pgfqpoint{0.018638in}{0.010825in}}{\pgfqpoint{0.014731in}{0.014731in}}%
\pgfpathcurveto{\pgfqpoint{0.010825in}{0.018638in}}{\pgfqpoint{0.005525in}{0.020833in}}{\pgfqpoint{0.000000in}{0.020833in}}%
\pgfpathcurveto{\pgfqpoint{-0.005525in}{0.020833in}}{\pgfqpoint{-0.010825in}{0.018638in}}{\pgfqpoint{-0.014731in}{0.014731in}}%
\pgfpathcurveto{\pgfqpoint{-0.018638in}{0.010825in}}{\pgfqpoint{-0.020833in}{0.005525in}}{\pgfqpoint{-0.020833in}{0.000000in}}%
\pgfpathcurveto{\pgfqpoint{-0.020833in}{-0.005525in}}{\pgfqpoint{-0.018638in}{-0.010825in}}{\pgfqpoint{-0.014731in}{-0.014731in}}%
\pgfpathcurveto{\pgfqpoint{-0.010825in}{-0.018638in}}{\pgfqpoint{-0.005525in}{-0.020833in}}{\pgfqpoint{0.000000in}{-0.020833in}}%
\pgfpathlineto{\pgfqpoint{0.000000in}{-0.020833in}}%
\pgfpathclose%
\pgfusepath{stroke,fill}%
}%
\begin{pgfscope}%
\pgfsys@transformshift{0.490023in}{0.697419in}%
\pgfsys@useobject{currentmarker}{}%
\end{pgfscope}%
\begin{pgfscope}%
\pgfsys@transformshift{0.740477in}{0.709413in}%
\pgfsys@useobject{currentmarker}{}%
\end{pgfscope}%
\begin{pgfscope}%
\pgfsys@transformshift{0.990930in}{0.734401in}%
\pgfsys@useobject{currentmarker}{}%
\end{pgfscope}%
\begin{pgfscope}%
\pgfsys@transformshift{1.241384in}{0.801370in}%
\pgfsys@useobject{currentmarker}{}%
\end{pgfscope}%
\begin{pgfscope}%
\pgfsys@transformshift{1.491837in}{0.976286in}%
\pgfsys@useobject{currentmarker}{}%
\end{pgfscope}%
\begin{pgfscope}%
\pgfsys@transformshift{1.742291in}{1.445464in}%
\pgfsys@useobject{currentmarker}{}%
\end{pgfscope}%
\begin{pgfscope}%
\pgfsys@transformshift{1.992744in}{2.583923in}%
\pgfsys@useobject{currentmarker}{}%
\end{pgfscope}%
\begin{pgfscope}%
\pgfsys@transformshift{1.992744in}{0.702616in}%
\pgfsys@useobject{currentmarker}{}%
\end{pgfscope}%
\begin{pgfscope}%
\pgfsys@transformshift{2.243198in}{0.711412in}%
\pgfsys@useobject{currentmarker}{}%
\end{pgfscope}%
\begin{pgfscope}%
\pgfsys@transformshift{2.493651in}{0.729804in}%
\pgfsys@useobject{currentmarker}{}%
\end{pgfscope}%
\begin{pgfscope}%
\pgfsys@transformshift{2.744105in}{0.757390in}%
\pgfsys@useobject{currentmarker}{}%
\end{pgfscope}%
\begin{pgfscope}%
\pgfsys@transformshift{2.994558in}{0.798371in}%
\pgfsys@useobject{currentmarker}{}%
\end{pgfscope}%
\begin{pgfscope}%
\pgfsys@transformshift{3.245012in}{0.874935in}%
\pgfsys@useobject{currentmarker}{}%
\end{pgfscope}%
\begin{pgfscope}%
\pgfsys@transformshift{3.495465in}{1.025063in}%
\pgfsys@useobject{currentmarker}{}%
\end{pgfscope}%
\begin{pgfscope}%
\pgfsys@transformshift{3.745919in}{1.287738in}%
\pgfsys@useobject{currentmarker}{}%
\end{pgfscope}%
\begin{pgfscope}%
\pgfsys@transformshift{3.996372in}{1.757515in}%
\pgfsys@useobject{currentmarker}{}%
\end{pgfscope}%
\begin{pgfscope}%
\pgfsys@transformshift{4.246826in}{2.254679in}%
\pgfsys@useobject{currentmarker}{}%
\end{pgfscope}%
\begin{pgfscope}%
\pgfsys@transformshift{4.497279in}{2.850996in}%
\pgfsys@useobject{currentmarker}{}%
\end{pgfscope}%
\end{pgfscope}%
\begin{pgfscope}%
\pgfpathrectangle{\pgfqpoint{0.490023in}{0.579475in}}{\pgfqpoint{5.509977in}{2.418851in}}%
\pgfusepath{clip}%
\pgfsetrectcap%
\pgfsetroundjoin%
\pgfsetlinewidth{1.505625pt}%
\definecolor{currentstroke}{rgb}{1.000000,0.756863,0.027451}%
\pgfsetstrokecolor{currentstroke}%
\pgfsetdash{}{0pt}%
\pgfpathmoveto{\pgfqpoint{0.490023in}{0.697419in}}%
\pgfpathlineto{\pgfqpoint{0.740477in}{0.708214in}}%
\pgfpathlineto{\pgfqpoint{0.990930in}{0.729804in}}%
\pgfpathlineto{\pgfqpoint{1.241384in}{0.786577in}}%
\pgfpathlineto{\pgfqpoint{1.491837in}{0.926510in}}%
\pgfpathlineto{\pgfqpoint{1.742291in}{1.280542in}}%
\pgfpathlineto{\pgfqpoint{1.992744in}{2.099953in}}%
\pgfpathmoveto{\pgfqpoint{1.992744in}{0.702616in}}%
\pgfpathlineto{\pgfqpoint{2.243198in}{0.710613in}}%
\pgfpathlineto{\pgfqpoint{2.493651in}{0.727005in}}%
\pgfpathlineto{\pgfqpoint{2.744105in}{0.749194in}}%
\pgfpathlineto{\pgfqpoint{2.994558in}{0.778580in}}%
\pgfpathlineto{\pgfqpoint{3.245012in}{0.834354in}}%
\pgfpathlineto{\pgfqpoint{3.495465in}{0.948100in}}%
\pgfpathlineto{\pgfqpoint{3.745919in}{1.130813in}}%
\pgfpathlineto{\pgfqpoint{3.996372in}{1.438467in}}%
\pgfpathlineto{\pgfqpoint{4.246826in}{1.743922in}}%
\pgfpathlineto{\pgfqpoint{4.497279in}{2.090757in}}%
\pgfusepath{stroke}%
\end{pgfscope}%
\begin{pgfscope}%
\pgfpathrectangle{\pgfqpoint{0.490023in}{0.579475in}}{\pgfqpoint{5.509977in}{2.418851in}}%
\pgfusepath{clip}%
\pgfsetbuttcap%
\pgfsetroundjoin%
\definecolor{currentfill}{rgb}{1.000000,0.756863,0.027451}%
\pgfsetfillcolor{currentfill}%
\pgfsetlinewidth{1.003750pt}%
\definecolor{currentstroke}{rgb}{1.000000,0.756863,0.027451}%
\pgfsetstrokecolor{currentstroke}%
\pgfsetdash{}{0pt}%
\pgfsys@defobject{currentmarker}{\pgfqpoint{-0.041667in}{-0.041667in}}{\pgfqpoint{0.041667in}{0.041667in}}{%
\pgfpathmoveto{\pgfqpoint{-0.041667in}{0.000000in}}%
\pgfpathlineto{\pgfqpoint{0.041667in}{0.000000in}}%
\pgfpathmoveto{\pgfqpoint{0.000000in}{-0.041667in}}%
\pgfpathlineto{\pgfqpoint{0.000000in}{0.041667in}}%
\pgfusepath{stroke,fill}%
}%
\begin{pgfscope}%
\pgfsys@transformshift{0.490023in}{0.697419in}%
\pgfsys@useobject{currentmarker}{}%
\end{pgfscope}%
\begin{pgfscope}%
\pgfsys@transformshift{0.740477in}{0.708214in}%
\pgfsys@useobject{currentmarker}{}%
\end{pgfscope}%
\begin{pgfscope}%
\pgfsys@transformshift{0.990930in}{0.729804in}%
\pgfsys@useobject{currentmarker}{}%
\end{pgfscope}%
\begin{pgfscope}%
\pgfsys@transformshift{1.241384in}{0.786577in}%
\pgfsys@useobject{currentmarker}{}%
\end{pgfscope}%
\begin{pgfscope}%
\pgfsys@transformshift{1.491837in}{0.926510in}%
\pgfsys@useobject{currentmarker}{}%
\end{pgfscope}%
\begin{pgfscope}%
\pgfsys@transformshift{1.742291in}{1.280542in}%
\pgfsys@useobject{currentmarker}{}%
\end{pgfscope}%
\begin{pgfscope}%
\pgfsys@transformshift{1.992744in}{2.099953in}%
\pgfsys@useobject{currentmarker}{}%
\end{pgfscope}%
\begin{pgfscope}%
\pgfsys@transformshift{1.992744in}{0.702616in}%
\pgfsys@useobject{currentmarker}{}%
\end{pgfscope}%
\begin{pgfscope}%
\pgfsys@transformshift{2.243198in}{0.710613in}%
\pgfsys@useobject{currentmarker}{}%
\end{pgfscope}%
\begin{pgfscope}%
\pgfsys@transformshift{2.493651in}{0.727005in}%
\pgfsys@useobject{currentmarker}{}%
\end{pgfscope}%
\begin{pgfscope}%
\pgfsys@transformshift{2.744105in}{0.749194in}%
\pgfsys@useobject{currentmarker}{}%
\end{pgfscope}%
\begin{pgfscope}%
\pgfsys@transformshift{2.994558in}{0.778580in}%
\pgfsys@useobject{currentmarker}{}%
\end{pgfscope}%
\begin{pgfscope}%
\pgfsys@transformshift{3.245012in}{0.834354in}%
\pgfsys@useobject{currentmarker}{}%
\end{pgfscope}%
\begin{pgfscope}%
\pgfsys@transformshift{3.495465in}{0.948100in}%
\pgfsys@useobject{currentmarker}{}%
\end{pgfscope}%
\begin{pgfscope}%
\pgfsys@transformshift{3.745919in}{1.130813in}%
\pgfsys@useobject{currentmarker}{}%
\end{pgfscope}%
\begin{pgfscope}%
\pgfsys@transformshift{3.996372in}{1.438467in}%
\pgfsys@useobject{currentmarker}{}%
\end{pgfscope}%
\begin{pgfscope}%
\pgfsys@transformshift{4.246826in}{1.743922in}%
\pgfsys@useobject{currentmarker}{}%
\end{pgfscope}%
\begin{pgfscope}%
\pgfsys@transformshift{4.497279in}{2.090757in}%
\pgfsys@useobject{currentmarker}{}%
\end{pgfscope}%
\end{pgfscope}%
\begin{pgfscope}%
\pgfpathrectangle{\pgfqpoint{0.490023in}{0.579475in}}{\pgfqpoint{5.509977in}{2.418851in}}%
\pgfusepath{clip}%
\pgfsetrectcap%
\pgfsetroundjoin%
\pgfsetlinewidth{1.505625pt}%
\definecolor{currentstroke}{rgb}{0.000000,0.301961,0.250980}%
\pgfsetstrokecolor{currentstroke}%
\pgfsetdash{}{0pt}%
\pgfpathmoveto{\pgfqpoint{0.490023in}{0.689423in}}%
\pgfpathlineto{\pgfqpoint{0.740477in}{0.690622in}}%
\pgfpathlineto{\pgfqpoint{0.990930in}{0.692821in}}%
\pgfpathlineto{\pgfqpoint{1.241384in}{0.699618in}}%
\pgfpathlineto{\pgfqpoint{1.491837in}{0.728204in}}%
\pgfpathlineto{\pgfqpoint{1.742291in}{0.832755in}}%
\pgfpathlineto{\pgfqpoint{1.992744in}{1.193583in}}%
\pgfpathmoveto{\pgfqpoint{1.992744in}{0.689423in}}%
\pgfpathlineto{\pgfqpoint{2.243198in}{0.690222in}}%
\pgfpathlineto{\pgfqpoint{2.493651in}{0.691622in}}%
\pgfpathlineto{\pgfqpoint{2.744105in}{0.695820in}}%
\pgfpathlineto{\pgfqpoint{2.994558in}{0.705215in}}%
\pgfpathlineto{\pgfqpoint{3.245012in}{0.718809in}}%
\pgfpathlineto{\pgfqpoint{3.495465in}{0.741598in}}%
\pgfpathlineto{\pgfqpoint{3.745919in}{0.810165in}}%
\pgfpathlineto{\pgfqpoint{3.996372in}{0.936905in}}%
\pgfpathlineto{\pgfqpoint{4.246826in}{1.153802in}}%
\pgfpathlineto{\pgfqpoint{4.497279in}{1.356906in}}%
\pgfusepath{stroke}%
\end{pgfscope}%
\begin{pgfscope}%
\pgfpathrectangle{\pgfqpoint{0.490023in}{0.579475in}}{\pgfqpoint{5.509977in}{2.418851in}}%
\pgfusepath{clip}%
\pgfsetbuttcap%
\pgfsetroundjoin%
\definecolor{currentfill}{rgb}{0.000000,0.301961,0.250980}%
\pgfsetfillcolor{currentfill}%
\pgfsetlinewidth{1.003750pt}%
\definecolor{currentstroke}{rgb}{0.000000,0.301961,0.250980}%
\pgfsetstrokecolor{currentstroke}%
\pgfsetdash{}{0pt}%
\pgfsys@defobject{currentmarker}{\pgfqpoint{-0.041667in}{-0.041667in}}{\pgfqpoint{0.041667in}{0.041667in}}{%
\pgfpathmoveto{\pgfqpoint{-0.041667in}{-0.041667in}}%
\pgfpathlineto{\pgfqpoint{0.041667in}{0.041667in}}%
\pgfpathmoveto{\pgfqpoint{-0.041667in}{0.041667in}}%
\pgfpathlineto{\pgfqpoint{0.041667in}{-0.041667in}}%
\pgfusepath{stroke,fill}%
}%
\begin{pgfscope}%
\pgfsys@transformshift{0.490023in}{0.689423in}%
\pgfsys@useobject{currentmarker}{}%
\end{pgfscope}%
\begin{pgfscope}%
\pgfsys@transformshift{0.740477in}{0.690622in}%
\pgfsys@useobject{currentmarker}{}%
\end{pgfscope}%
\begin{pgfscope}%
\pgfsys@transformshift{0.990930in}{0.692821in}%
\pgfsys@useobject{currentmarker}{}%
\end{pgfscope}%
\begin{pgfscope}%
\pgfsys@transformshift{1.241384in}{0.699618in}%
\pgfsys@useobject{currentmarker}{}%
\end{pgfscope}%
\begin{pgfscope}%
\pgfsys@transformshift{1.491837in}{0.728204in}%
\pgfsys@useobject{currentmarker}{}%
\end{pgfscope}%
\begin{pgfscope}%
\pgfsys@transformshift{1.742291in}{0.832755in}%
\pgfsys@useobject{currentmarker}{}%
\end{pgfscope}%
\begin{pgfscope}%
\pgfsys@transformshift{1.992744in}{1.193583in}%
\pgfsys@useobject{currentmarker}{}%
\end{pgfscope}%
\begin{pgfscope}%
\pgfsys@transformshift{1.992744in}{0.689423in}%
\pgfsys@useobject{currentmarker}{}%
\end{pgfscope}%
\begin{pgfscope}%
\pgfsys@transformshift{2.243198in}{0.690222in}%
\pgfsys@useobject{currentmarker}{}%
\end{pgfscope}%
\begin{pgfscope}%
\pgfsys@transformshift{2.493651in}{0.691622in}%
\pgfsys@useobject{currentmarker}{}%
\end{pgfscope}%
\begin{pgfscope}%
\pgfsys@transformshift{2.744105in}{0.695820in}%
\pgfsys@useobject{currentmarker}{}%
\end{pgfscope}%
\begin{pgfscope}%
\pgfsys@transformshift{2.994558in}{0.705215in}%
\pgfsys@useobject{currentmarker}{}%
\end{pgfscope}%
\begin{pgfscope}%
\pgfsys@transformshift{3.245012in}{0.718809in}%
\pgfsys@useobject{currentmarker}{}%
\end{pgfscope}%
\begin{pgfscope}%
\pgfsys@transformshift{3.495465in}{0.741598in}%
\pgfsys@useobject{currentmarker}{}%
\end{pgfscope}%
\begin{pgfscope}%
\pgfsys@transformshift{3.745919in}{0.810165in}%
\pgfsys@useobject{currentmarker}{}%
\end{pgfscope}%
\begin{pgfscope}%
\pgfsys@transformshift{3.996372in}{0.936905in}%
\pgfsys@useobject{currentmarker}{}%
\end{pgfscope}%
\begin{pgfscope}%
\pgfsys@transformshift{4.246826in}{1.153802in}%
\pgfsys@useobject{currentmarker}{}%
\end{pgfscope}%
\begin{pgfscope}%
\pgfsys@transformshift{4.497279in}{1.356906in}%
\pgfsys@useobject{currentmarker}{}%
\end{pgfscope}%
\end{pgfscope}%
\begin{pgfscope}%
\pgfpathrectangle{\pgfqpoint{0.490023in}{0.579475in}}{\pgfqpoint{5.509977in}{2.418851in}}%
\pgfusepath{clip}%
\pgfsetbuttcap%
\pgfsetroundjoin%
\pgfsetlinewidth{1.505625pt}%
\definecolor{currentstroke}{rgb}{0.501961,0.501961,0.501961}%
\pgfsetstrokecolor{currentstroke}%
\pgfsetdash{{5.550000pt}{2.400000pt}}{0.000000pt}%
\pgfpathmoveto{\pgfqpoint{0.490023in}{2.888378in}}%
\pgfpathlineto{\pgfqpoint{6.000000in}{2.888378in}}%
\pgfusepath{stroke}%
\end{pgfscope}%
\begin{pgfscope}%
\pgfpathrectangle{\pgfqpoint{0.490023in}{0.579475in}}{\pgfqpoint{5.509977in}{2.418851in}}%
\pgfusepath{clip}%
\pgfsetbuttcap%
\pgfsetroundjoin%
\pgfsetlinewidth{1.505625pt}%
\definecolor{currentstroke}{rgb}{0.333333,0.000000,0.831373}%
\pgfsetstrokecolor{currentstroke}%
\pgfsetdash{{5.550000pt}{2.400000pt}}{0.000000pt}%
\pgfpathmoveto{\pgfqpoint{1.992744in}{0.689423in}}%
\pgfpathlineto{\pgfqpoint{1.992744in}{2.888378in}}%
\pgfusepath{stroke}%
\end{pgfscope}%
\begin{pgfscope}%
\pgfsetrectcap%
\pgfsetmiterjoin%
\pgfsetlinewidth{0.803000pt}%
\definecolor{currentstroke}{rgb}{0.000000,0.000000,0.000000}%
\pgfsetstrokecolor{currentstroke}%
\pgfsetdash{}{0pt}%
\pgfpathmoveto{\pgfqpoint{0.490023in}{0.579475in}}%
\pgfpathlineto{\pgfqpoint{0.490023in}{2.998326in}}%
\pgfusepath{stroke}%
\end{pgfscope}%
\begin{pgfscope}%
\pgfsetrectcap%
\pgfsetmiterjoin%
\pgfsetlinewidth{0.803000pt}%
\definecolor{currentstroke}{rgb}{0.000000,0.000000,0.000000}%
\pgfsetstrokecolor{currentstroke}%
\pgfsetdash{}{0pt}%
\pgfpathmoveto{\pgfqpoint{6.000000in}{0.579475in}}%
\pgfpathlineto{\pgfqpoint{6.000000in}{2.998326in}}%
\pgfusepath{stroke}%
\end{pgfscope}%
\begin{pgfscope}%
\pgfsetrectcap%
\pgfsetmiterjoin%
\pgfsetlinewidth{0.803000pt}%
\definecolor{currentstroke}{rgb}{0.000000,0.000000,0.000000}%
\pgfsetstrokecolor{currentstroke}%
\pgfsetdash{}{0pt}%
\pgfpathmoveto{\pgfqpoint{0.490023in}{0.579475in}}%
\pgfpathlineto{\pgfqpoint{6.000000in}{0.579475in}}%
\pgfusepath{stroke}%
\end{pgfscope}%
\begin{pgfscope}%
\pgfsetrectcap%
\pgfsetmiterjoin%
\pgfsetlinewidth{0.803000pt}%
\definecolor{currentstroke}{rgb}{0.000000,0.000000,0.000000}%
\pgfsetstrokecolor{currentstroke}%
\pgfsetdash{}{0pt}%
\pgfpathmoveto{\pgfqpoint{0.490023in}{2.998326in}}%
\pgfpathlineto{\pgfqpoint{6.000000in}{2.998326in}}%
\pgfusepath{stroke}%
\end{pgfscope}%
\begin{pgfscope}%
\definecolor{textcolor}{rgb}{0.501961,0.501961,0.501961}%
\pgfsetstrokecolor{textcolor}%
\pgfsetfillcolor{textcolor}%
\pgftext[x=0.392801in,y=2.888378in,right,]{\color{textcolor}\rmfamily\fontsize{16.000000}{19.200000}\selectfont 11K}%
\end{pgfscope}%
\begin{pgfscope}%
\definecolor{textcolor}{rgb}{0.333333,0.000000,0.831373}%
\pgfsetstrokecolor{textcolor}%
\pgfsetfillcolor{textcolor}%
\pgftext[x=2.117971in,y=2.668482in,left,base]{\color{textcolor}\rmfamily\fontsize{16.000000}{19.200000}\selectfont sketch}%
\end{pgfscope}%
\begin{pgfscope}%
\definecolor{textcolor}{rgb}{0.333333,0.000000,0.831373}%
\pgfsetstrokecolor{textcolor}%
\pgfsetfillcolor{textcolor}%
\pgftext[x=2.117971in,y=2.448587in,left,base]{\color{textcolor}\rmfamily\fontsize{16.000000}{19.200000}\selectfont guide}%
\end{pgfscope}%
\begin{pgfscope}%
\definecolor{textcolor}{rgb}{0.333333,0.000000,0.831373}%
\pgfsetstrokecolor{textcolor}%
\pgfsetfillcolor{textcolor}%
\pgftext[x=2.117971in,y=2.228691in,left,base]{\color{textcolor}\rmfamily\fontsize{16.000000}{19.200000}\selectfont n°1}%
\end{pgfscope}%
\end{pgfpicture}%
\makeatother%
\endgroup%

%% file: media/guided-parallel.pgf
\begingroup%
\makeatletter%
\begin{pgfpicture}%
\pgfpathrectangle{\pgfpointorigin}{\pgfqpoint{6.000000in}{3.000000in}}%
\pgfusepath{use as bounding box, clip}%
\begin{pgfscope}%
\pgfsetbuttcap%
\pgfsetmiterjoin%
\definecolor{currentfill}{rgb}{1.000000,1.000000,1.000000}%
\pgfsetfillcolor{currentfill}%
\pgfsetlinewidth{0.000000pt}%
\definecolor{currentstroke}{rgb}{1.000000,1.000000,1.000000}%
\pgfsetstrokecolor{currentstroke}%
\pgfsetdash{}{0pt}%
\pgfpathmoveto{\pgfqpoint{0.000000in}{0.000000in}}%
\pgfpathlineto{\pgfqpoint{6.000000in}{0.000000in}}%
\pgfpathlineto{\pgfqpoint{6.000000in}{3.000000in}}%
\pgfpathlineto{\pgfqpoint{0.000000in}{3.000000in}}%
\pgfpathlineto{\pgfqpoint{0.000000in}{0.000000in}}%
\pgfpathclose%
\pgfusepath{fill}%
\end{pgfscope}%
\begin{pgfscope}%
\pgfsetbuttcap%
\pgfsetmiterjoin%
\definecolor{currentfill}{rgb}{1.000000,1.000000,1.000000}%
\pgfsetfillcolor{currentfill}%
\pgfsetlinewidth{0.000000pt}%
\definecolor{currentstroke}{rgb}{0.000000,0.000000,0.000000}%
\pgfsetstrokecolor{currentstroke}%
\pgfsetstrokeopacity{0.000000}%
\pgfsetdash{}{0pt}%
\pgfpathmoveto{\pgfqpoint{0.490023in}{0.579475in}}%
\pgfpathlineto{\pgfqpoint{6.000000in}{0.579475in}}%
\pgfpathlineto{\pgfqpoint{6.000000in}{3.000000in}}%
\pgfpathlineto{\pgfqpoint{0.490023in}{3.000000in}}%
\pgfpathlineto{\pgfqpoint{0.490023in}{0.579475in}}%
\pgfpathclose%
\pgfusepath{fill}%
\end{pgfscope}%
\begin{pgfscope}%
\pgfsetbuttcap%
\pgfsetroundjoin%
\definecolor{currentfill}{rgb}{0.000000,0.000000,0.000000}%
\pgfsetfillcolor{currentfill}%
\pgfsetlinewidth{0.803000pt}%
\definecolor{currentstroke}{rgb}{0.000000,0.000000,0.000000}%
\pgfsetstrokecolor{currentstroke}%
\pgfsetdash{}{0pt}%
\pgfsys@defobject{currentmarker}{\pgfqpoint{0.000000in}{-0.048611in}}{\pgfqpoint{0.000000in}{0.000000in}}{%
\pgfpathmoveto{\pgfqpoint{0.000000in}{0.000000in}}%
\pgfpathlineto{\pgfqpoint{0.000000in}{-0.048611in}}%
\pgfusepath{stroke,fill}%
}%
\begin{pgfscope}%
\pgfsys@transformshift{0.490023in}{0.579475in}%
\pgfsys@useobject{currentmarker}{}%
\end{pgfscope}%
\end{pgfscope}%
\begin{pgfscope}%
\definecolor{textcolor}{rgb}{0.000000,0.000000,0.000000}%
\pgfsetstrokecolor{textcolor}%
\pgfsetfillcolor{textcolor}%
\pgftext[x=0.490023in,y=0.482253in,,top]{\color{textcolor}\rmfamily\fontsize{16.000000}{19.200000}\selectfont \(\displaystyle {0}\)}%
\end{pgfscope}%
\begin{pgfscope}%
\pgfsetbuttcap%
\pgfsetroundjoin%
\definecolor{currentfill}{rgb}{0.000000,0.000000,0.000000}%
\pgfsetfillcolor{currentfill}%
\pgfsetlinewidth{0.803000pt}%
\definecolor{currentstroke}{rgb}{0.000000,0.000000,0.000000}%
\pgfsetstrokecolor{currentstroke}%
\pgfsetdash{}{0pt}%
\pgfsys@defobject{currentmarker}{\pgfqpoint{0.000000in}{-0.048611in}}{\pgfqpoint{0.000000in}{0.000000in}}{%
\pgfpathmoveto{\pgfqpoint{0.000000in}{0.000000in}}%
\pgfpathlineto{\pgfqpoint{0.000000in}{-0.048611in}}%
\pgfusepath{stroke,fill}%
}%
\begin{pgfscope}%
\pgfsys@transformshift{1.742291in}{0.579475in}%
\pgfsys@useobject{currentmarker}{}%
\end{pgfscope}%
\end{pgfscope}%
\begin{pgfscope}%
\definecolor{textcolor}{rgb}{0.000000,0.000000,0.000000}%
\pgfsetstrokecolor{textcolor}%
\pgfsetfillcolor{textcolor}%
\pgftext[x=1.742291in,y=0.482253in,,top]{\color{textcolor}\rmfamily\fontsize{16.000000}{19.200000}\selectfont \(\displaystyle {5}\)}%
\end{pgfscope}%
\begin{pgfscope}%
\pgfsetbuttcap%
\pgfsetroundjoin%
\definecolor{currentfill}{rgb}{0.000000,0.000000,0.000000}%
\pgfsetfillcolor{currentfill}%
\pgfsetlinewidth{0.803000pt}%
\definecolor{currentstroke}{rgb}{0.000000,0.000000,0.000000}%
\pgfsetstrokecolor{currentstroke}%
\pgfsetdash{}{0pt}%
\pgfsys@defobject{currentmarker}{\pgfqpoint{0.000000in}{-0.048611in}}{\pgfqpoint{0.000000in}{0.000000in}}{%
\pgfpathmoveto{\pgfqpoint{0.000000in}{0.000000in}}%
\pgfpathlineto{\pgfqpoint{0.000000in}{-0.048611in}}%
\pgfusepath{stroke,fill}%
}%
\begin{pgfscope}%
\pgfsys@transformshift{2.994558in}{0.579475in}%
\pgfsys@useobject{currentmarker}{}%
\end{pgfscope}%
\end{pgfscope}%
\begin{pgfscope}%
\definecolor{textcolor}{rgb}{0.000000,0.000000,0.000000}%
\pgfsetstrokecolor{textcolor}%
\pgfsetfillcolor{textcolor}%
\pgftext[x=2.994558in,y=0.482253in,,top]{\color{textcolor}\rmfamily\fontsize{16.000000}{19.200000}\selectfont \(\displaystyle {10}\)}%
\end{pgfscope}%
\begin{pgfscope}%
\pgfsetbuttcap%
\pgfsetroundjoin%
\definecolor{currentfill}{rgb}{0.000000,0.000000,0.000000}%
\pgfsetfillcolor{currentfill}%
\pgfsetlinewidth{0.803000pt}%
\definecolor{currentstroke}{rgb}{0.000000,0.000000,0.000000}%
\pgfsetstrokecolor{currentstroke}%
\pgfsetdash{}{0pt}%
\pgfsys@defobject{currentmarker}{\pgfqpoint{0.000000in}{-0.048611in}}{\pgfqpoint{0.000000in}{0.000000in}}{%
\pgfpathmoveto{\pgfqpoint{0.000000in}{0.000000in}}%
\pgfpathlineto{\pgfqpoint{0.000000in}{-0.048611in}}%
\pgfusepath{stroke,fill}%
}%
\begin{pgfscope}%
\pgfsys@transformshift{4.246826in}{0.579475in}%
\pgfsys@useobject{currentmarker}{}%
\end{pgfscope}%
\end{pgfscope}%
\begin{pgfscope}%
\definecolor{textcolor}{rgb}{0.000000,0.000000,0.000000}%
\pgfsetstrokecolor{textcolor}%
\pgfsetfillcolor{textcolor}%
\pgftext[x=4.246826in,y=0.482253in,,top]{\color{textcolor}\rmfamily\fontsize{16.000000}{19.200000}\selectfont \(\displaystyle {15}\)}%
\end{pgfscope}%
\begin{pgfscope}%
\pgfsetbuttcap%
\pgfsetroundjoin%
\definecolor{currentfill}{rgb}{0.000000,0.000000,0.000000}%
\pgfsetfillcolor{currentfill}%
\pgfsetlinewidth{0.803000pt}%
\definecolor{currentstroke}{rgb}{0.000000,0.000000,0.000000}%
\pgfsetstrokecolor{currentstroke}%
\pgfsetdash{}{0pt}%
\pgfsys@defobject{currentmarker}{\pgfqpoint{0.000000in}{-0.048611in}}{\pgfqpoint{0.000000in}{0.000000in}}{%
\pgfpathmoveto{\pgfqpoint{0.000000in}{0.000000in}}%
\pgfpathlineto{\pgfqpoint{0.000000in}{-0.048611in}}%
\pgfusepath{stroke,fill}%
}%
\begin{pgfscope}%
\pgfsys@transformshift{5.499093in}{0.579475in}%
\pgfsys@useobject{currentmarker}{}%
\end{pgfscope}%
\end{pgfscope}%
\begin{pgfscope}%
\definecolor{textcolor}{rgb}{0.000000,0.000000,0.000000}%
\pgfsetstrokecolor{textcolor}%
\pgfsetfillcolor{textcolor}%
\pgftext[x=5.499093in,y=0.482253in,,top]{\color{textcolor}\rmfamily\fontsize{16.000000}{19.200000}\selectfont \(\displaystyle {20}\)}%
\end{pgfscope}%
\begin{pgfscope}%
\definecolor{textcolor}{rgb}{0.000000,0.000000,0.000000}%
\pgfsetstrokecolor{textcolor}%
\pgfsetfillcolor{textcolor}%
\pgftext[x=3.245012in,y=0.213349in,,top]{\color{textcolor}\rmfamily\fontsize{16.000000}{19.200000}\selectfont iterations}%
\end{pgfscope}%
\begin{pgfscope}%
\pgfsetbuttcap%
\pgfsetroundjoin%
\definecolor{currentfill}{rgb}{0.000000,0.000000,0.000000}%
\pgfsetfillcolor{currentfill}%
\pgfsetlinewidth{0.803000pt}%
\definecolor{currentstroke}{rgb}{0.000000,0.000000,0.000000}%
\pgfsetstrokecolor{currentstroke}%
\pgfsetdash{}{0pt}%
\pgfsys@defobject{currentmarker}{\pgfqpoint{-0.048611in}{0.000000in}}{\pgfqpoint{-0.000000in}{0.000000in}}{%
\pgfpathmoveto{\pgfqpoint{-0.000000in}{0.000000in}}%
\pgfpathlineto{\pgfqpoint{-0.048611in}{0.000000in}}%
\pgfusepath{stroke,fill}%
}%
\begin{pgfscope}%
\pgfsys@transformshift{0.490023in}{0.689499in}%
\pgfsys@useobject{currentmarker}{}%
\end{pgfscope}%
\end{pgfscope}%
\begin{pgfscope}%
\definecolor{textcolor}{rgb}{0.000000,0.000000,0.000000}%
\pgfsetstrokecolor{textcolor}%
\pgfsetfillcolor{textcolor}%
\pgftext[x=0.110068in, y=0.606166in, left, base]{\color{textcolor}\rmfamily\fontsize{16.000000}{19.200000}\selectfont 0K}%
\end{pgfscope}%
\begin{pgfscope}%
\pgfsetbuttcap%
\pgfsetroundjoin%
\definecolor{currentfill}{rgb}{0.000000,0.000000,0.000000}%
\pgfsetfillcolor{currentfill}%
\pgfsetlinewidth{0.803000pt}%
\definecolor{currentstroke}{rgb}{0.000000,0.000000,0.000000}%
\pgfsetstrokecolor{currentstroke}%
\pgfsetdash{}{0pt}%
\pgfsys@defobject{currentmarker}{\pgfqpoint{-0.048611in}{0.000000in}}{\pgfqpoint{-0.000000in}{0.000000in}}{%
\pgfpathmoveto{\pgfqpoint{-0.000000in}{0.000000in}}%
\pgfpathlineto{\pgfqpoint{-0.048611in}{0.000000in}}%
\pgfusepath{stroke,fill}%
}%
\begin{pgfscope}%
\pgfsys@transformshift{0.490023in}{1.239618in}%
\pgfsys@useobject{currentmarker}{}%
\end{pgfscope}%
\end{pgfscope}%
\begin{pgfscope}%
\definecolor{textcolor}{rgb}{0.000000,0.000000,0.000000}%
\pgfsetstrokecolor{textcolor}%
\pgfsetfillcolor{textcolor}%
\pgftext[x=0.110068in, y=1.156285in, left, base]{\color{textcolor}\rmfamily\fontsize{16.000000}{19.200000}\selectfont 2K}%
\end{pgfscope}%
\begin{pgfscope}%
\pgfsetbuttcap%
\pgfsetroundjoin%
\definecolor{currentfill}{rgb}{0.000000,0.000000,0.000000}%
\pgfsetfillcolor{currentfill}%
\pgfsetlinewidth{0.803000pt}%
\definecolor{currentstroke}{rgb}{0.000000,0.000000,0.000000}%
\pgfsetstrokecolor{currentstroke}%
\pgfsetdash{}{0pt}%
\pgfsys@defobject{currentmarker}{\pgfqpoint{-0.048611in}{0.000000in}}{\pgfqpoint{-0.000000in}{0.000000in}}{%
\pgfpathmoveto{\pgfqpoint{-0.000000in}{0.000000in}}%
\pgfpathlineto{\pgfqpoint{-0.048611in}{0.000000in}}%
\pgfusepath{stroke,fill}%
}%
\begin{pgfscope}%
\pgfsys@transformshift{0.490023in}{1.789737in}%
\pgfsys@useobject{currentmarker}{}%
\end{pgfscope}%
\end{pgfscope}%
\begin{pgfscope}%
\definecolor{textcolor}{rgb}{0.000000,0.000000,0.000000}%
\pgfsetstrokecolor{textcolor}%
\pgfsetfillcolor{textcolor}%
\pgftext[x=0.110068in, y=1.706404in, left, base]{\color{textcolor}\rmfamily\fontsize{16.000000}{19.200000}\selectfont 5K}%
\end{pgfscope}%
\begin{pgfscope}%
\pgfsetbuttcap%
\pgfsetroundjoin%
\definecolor{currentfill}{rgb}{0.000000,0.000000,0.000000}%
\pgfsetfillcolor{currentfill}%
\pgfsetlinewidth{0.803000pt}%
\definecolor{currentstroke}{rgb}{0.000000,0.000000,0.000000}%
\pgfsetstrokecolor{currentstroke}%
\pgfsetdash{}{0pt}%
\pgfsys@defobject{currentmarker}{\pgfqpoint{-0.048611in}{0.000000in}}{\pgfqpoint{-0.000000in}{0.000000in}}{%
\pgfpathmoveto{\pgfqpoint{-0.000000in}{0.000000in}}%
\pgfpathlineto{\pgfqpoint{-0.048611in}{0.000000in}}%
\pgfusepath{stroke,fill}%
}%
\begin{pgfscope}%
\pgfsys@transformshift{0.490023in}{2.339857in}%
\pgfsys@useobject{currentmarker}{}%
\end{pgfscope}%
\end{pgfscope}%
\begin{pgfscope}%
\definecolor{textcolor}{rgb}{0.000000,0.000000,0.000000}%
\pgfsetstrokecolor{textcolor}%
\pgfsetfillcolor{textcolor}%
\pgftext[x=0.110068in, y=2.256523in, left, base]{\color{textcolor}\rmfamily\fontsize{16.000000}{19.200000}\selectfont 7K}%
\end{pgfscope}%
\begin{pgfscope}%
\pgfsetbuttcap%
\pgfsetroundjoin%
\definecolor{currentfill}{rgb}{0.000000,0.000000,0.000000}%
\pgfsetfillcolor{currentfill}%
\pgfsetlinewidth{0.803000pt}%
\definecolor{currentstroke}{rgb}{0.000000,0.000000,0.000000}%
\pgfsetstrokecolor{currentstroke}%
\pgfsetdash{}{0pt}%
\pgfsys@defobject{currentmarker}{\pgfqpoint{-0.048611in}{0.000000in}}{\pgfqpoint{-0.000000in}{0.000000in}}{%
\pgfpathmoveto{\pgfqpoint{-0.000000in}{0.000000in}}%
\pgfpathlineto{\pgfqpoint{-0.048611in}{0.000000in}}%
\pgfusepath{stroke,fill}%
}%
\begin{pgfscope}%
\pgfsys@transformshift{0.490023in}{2.889976in}%
\pgfsys@useobject{currentmarker}{}%
\end{pgfscope}%
\end{pgfscope}%
\begin{pgfscope}%
\definecolor{textcolor}{rgb}{0.000000,0.000000,0.000000}%
\pgfsetstrokecolor{textcolor}%
\pgfsetfillcolor{textcolor}%
\pgftext[x=0.000000in, y=2.806643in, left, base]{\color{textcolor}\rmfamily\fontsize{16.000000}{19.200000}\selectfont 10K}%
\end{pgfscope}%
\begin{pgfscope}%
\pgfpathrectangle{\pgfqpoint{0.490023in}{0.579475in}}{\pgfqpoint{5.509977in}{2.420525in}}%
\pgfusepath{clip}%
\pgfsetrectcap%
\pgfsetroundjoin%
\pgfsetlinewidth{1.505625pt}%
\definecolor{currentstroke}{rgb}{0.117647,0.533333,0.898039}%
\pgfsetstrokecolor{currentstroke}%
\pgfsetdash{}{0pt}%
\pgfpathmoveto{\pgfqpoint{0.490023in}{0.698301in}}%
\pgfpathlineto{\pgfqpoint{0.740477in}{0.711504in}}%
\pgfpathlineto{\pgfqpoint{0.990930in}{0.739010in}}%
\pgfpathlineto{\pgfqpoint{1.241384in}{0.812726in}}%
\pgfpathlineto{\pgfqpoint{1.491837in}{1.005267in}}%
\pgfpathlineto{\pgfqpoint{1.742291in}{1.521719in}}%
\pgfpathlineto{\pgfqpoint{1.992744in}{2.774891in}}%
\pgfpathmoveto{\pgfqpoint{1.992744in}{0.704022in}}%
\pgfpathlineto{\pgfqpoint{2.243198in}{0.713704in}}%
\pgfpathlineto{\pgfqpoint{2.493651in}{0.733949in}}%
\pgfpathlineto{\pgfqpoint{2.744105in}{0.764315in}}%
\pgfpathlineto{\pgfqpoint{2.994558in}{0.809425in}}%
\pgfpathlineto{\pgfqpoint{3.245012in}{0.893703in}}%
\pgfpathlineto{\pgfqpoint{3.495465in}{1.058959in}}%
\pgfpathlineto{\pgfqpoint{3.745919in}{1.348102in}}%
\pgfpathlineto{\pgfqpoint{3.996372in}{1.865214in}}%
\pgfpathmoveto{\pgfqpoint{3.996372in}{0.718105in}}%
\pgfpathlineto{\pgfqpoint{4.246826in}{0.829229in}}%
\pgfpathlineto{\pgfqpoint{4.497279in}{1.015170in}}%
\pgfpathlineto{\pgfqpoint{4.747733in}{1.313774in}}%
\pgfpathlineto{\pgfqpoint{4.998186in}{1.913404in}}%
\pgfpathmoveto{\pgfqpoint{4.998186in}{0.724046in}}%
\pgfpathlineto{\pgfqpoint{5.248640in}{0.761234in}}%
\pgfpathlineto{\pgfqpoint{5.499093in}{0.796662in}}%
\pgfpathlineto{\pgfqpoint{5.749547in}{0.827909in}}%
\pgfusepath{stroke}%
\end{pgfscope}%
\begin{pgfscope}%
\pgfpathrectangle{\pgfqpoint{0.490023in}{0.579475in}}{\pgfqpoint{5.509977in}{2.420525in}}%
\pgfusepath{clip}%
\pgfsetbuttcap%
\pgfsetroundjoin%
\definecolor{currentfill}{rgb}{0.117647,0.533333,0.898039}%
\pgfsetfillcolor{currentfill}%
\pgfsetlinewidth{1.003750pt}%
\definecolor{currentstroke}{rgb}{0.117647,0.533333,0.898039}%
\pgfsetstrokecolor{currentstroke}%
\pgfsetdash{}{0pt}%
\pgfsys@defobject{currentmarker}{\pgfqpoint{-0.020833in}{-0.020833in}}{\pgfqpoint{0.020833in}{0.020833in}}{%
\pgfpathmoveto{\pgfqpoint{0.000000in}{-0.020833in}}%
\pgfpathcurveto{\pgfqpoint{0.005525in}{-0.020833in}}{\pgfqpoint{0.010825in}{-0.018638in}}{\pgfqpoint{0.014731in}{-0.014731in}}%
\pgfpathcurveto{\pgfqpoint{0.018638in}{-0.010825in}}{\pgfqpoint{0.020833in}{-0.005525in}}{\pgfqpoint{0.020833in}{0.000000in}}%
\pgfpathcurveto{\pgfqpoint{0.020833in}{0.005525in}}{\pgfqpoint{0.018638in}{0.010825in}}{\pgfqpoint{0.014731in}{0.014731in}}%
\pgfpathcurveto{\pgfqpoint{0.010825in}{0.018638in}}{\pgfqpoint{0.005525in}{0.020833in}}{\pgfqpoint{0.000000in}{0.020833in}}%
\pgfpathcurveto{\pgfqpoint{-0.005525in}{0.020833in}}{\pgfqpoint{-0.010825in}{0.018638in}}{\pgfqpoint{-0.014731in}{0.014731in}}%
\pgfpathcurveto{\pgfqpoint{-0.018638in}{0.010825in}}{\pgfqpoint{-0.020833in}{0.005525in}}{\pgfqpoint{-0.020833in}{0.000000in}}%
\pgfpathcurveto{\pgfqpoint{-0.020833in}{-0.005525in}}{\pgfqpoint{-0.018638in}{-0.010825in}}{\pgfqpoint{-0.014731in}{-0.014731in}}%
\pgfpathcurveto{\pgfqpoint{-0.010825in}{-0.018638in}}{\pgfqpoint{-0.005525in}{-0.020833in}}{\pgfqpoint{0.000000in}{-0.020833in}}%
\pgfpathlineto{\pgfqpoint{0.000000in}{-0.020833in}}%
\pgfpathclose%
\pgfusepath{stroke,fill}%
}%
\begin{pgfscope}%
\pgfsys@transformshift{0.490023in}{0.698301in}%
\pgfsys@useobject{currentmarker}{}%
\end{pgfscope}%
\begin{pgfscope}%
\pgfsys@transformshift{0.740477in}{0.711504in}%
\pgfsys@useobject{currentmarker}{}%
\end{pgfscope}%
\begin{pgfscope}%
\pgfsys@transformshift{0.990930in}{0.739010in}%
\pgfsys@useobject{currentmarker}{}%
\end{pgfscope}%
\begin{pgfscope}%
\pgfsys@transformshift{1.241384in}{0.812726in}%
\pgfsys@useobject{currentmarker}{}%
\end{pgfscope}%
\begin{pgfscope}%
\pgfsys@transformshift{1.491837in}{1.005267in}%
\pgfsys@useobject{currentmarker}{}%
\end{pgfscope}%
\begin{pgfscope}%
\pgfsys@transformshift{1.742291in}{1.521719in}%
\pgfsys@useobject{currentmarker}{}%
\end{pgfscope}%
\begin{pgfscope}%
\pgfsys@transformshift{1.992744in}{2.774891in}%
\pgfsys@useobject{currentmarker}{}%
\end{pgfscope}%
\begin{pgfscope}%
\pgfsys@transformshift{1.992744in}{0.704022in}%
\pgfsys@useobject{currentmarker}{}%
\end{pgfscope}%
\begin{pgfscope}%
\pgfsys@transformshift{2.243198in}{0.713704in}%
\pgfsys@useobject{currentmarker}{}%
\end{pgfscope}%
\begin{pgfscope}%
\pgfsys@transformshift{2.493651in}{0.733949in}%
\pgfsys@useobject{currentmarker}{}%
\end{pgfscope}%
\begin{pgfscope}%
\pgfsys@transformshift{2.744105in}{0.764315in}%
\pgfsys@useobject{currentmarker}{}%
\end{pgfscope}%
\begin{pgfscope}%
\pgfsys@transformshift{2.994558in}{0.809425in}%
\pgfsys@useobject{currentmarker}{}%
\end{pgfscope}%
\begin{pgfscope}%
\pgfsys@transformshift{3.245012in}{0.893703in}%
\pgfsys@useobject{currentmarker}{}%
\end{pgfscope}%
\begin{pgfscope}%
\pgfsys@transformshift{3.495465in}{1.058959in}%
\pgfsys@useobject{currentmarker}{}%
\end{pgfscope}%
\begin{pgfscope}%
\pgfsys@transformshift{3.745919in}{1.348102in}%
\pgfsys@useobject{currentmarker}{}%
\end{pgfscope}%
\begin{pgfscope}%
\pgfsys@transformshift{3.996372in}{1.865214in}%
\pgfsys@useobject{currentmarker}{}%
\end{pgfscope}%
\begin{pgfscope}%
\pgfsys@transformshift{3.996372in}{0.718105in}%
\pgfsys@useobject{currentmarker}{}%
\end{pgfscope}%
\begin{pgfscope}%
\pgfsys@transformshift{4.246826in}{0.829229in}%
\pgfsys@useobject{currentmarker}{}%
\end{pgfscope}%
\begin{pgfscope}%
\pgfsys@transformshift{4.497279in}{1.015170in}%
\pgfsys@useobject{currentmarker}{}%
\end{pgfscope}%
\begin{pgfscope}%
\pgfsys@transformshift{4.747733in}{1.313774in}%
\pgfsys@useobject{currentmarker}{}%
\end{pgfscope}%
\begin{pgfscope}%
\pgfsys@transformshift{4.998186in}{1.913404in}%
\pgfsys@useobject{currentmarker}{}%
\end{pgfscope}%
\begin{pgfscope}%
\pgfsys@transformshift{4.998186in}{0.724046in}%
\pgfsys@useobject{currentmarker}{}%
\end{pgfscope}%
\begin{pgfscope}%
\pgfsys@transformshift{5.248640in}{0.761234in}%
\pgfsys@useobject{currentmarker}{}%
\end{pgfscope}%
\begin{pgfscope}%
\pgfsys@transformshift{5.499093in}{0.796662in}%
\pgfsys@useobject{currentmarker}{}%
\end{pgfscope}%
\begin{pgfscope}%
\pgfsys@transformshift{5.749547in}{0.827909in}%
\pgfsys@useobject{currentmarker}{}%
\end{pgfscope}%
\end{pgfscope}%
\begin{pgfscope}%
\pgfpathrectangle{\pgfqpoint{0.490023in}{0.579475in}}{\pgfqpoint{5.509977in}{2.420525in}}%
\pgfusepath{clip}%
\pgfsetrectcap%
\pgfsetroundjoin%
\pgfsetlinewidth{1.505625pt}%
\definecolor{currentstroke}{rgb}{1.000000,0.756863,0.027451}%
\pgfsetstrokecolor{currentstroke}%
\pgfsetdash{}{0pt}%
\pgfpathmoveto{\pgfqpoint{0.490023in}{0.698301in}}%
\pgfpathlineto{\pgfqpoint{0.740477in}{0.710183in}}%
\pgfpathlineto{\pgfqpoint{0.990930in}{0.733949in}}%
\pgfpathlineto{\pgfqpoint{1.241384in}{0.796442in}}%
\pgfpathlineto{\pgfqpoint{1.491837in}{0.950475in}}%
\pgfpathlineto{\pgfqpoint{1.742291in}{1.340180in}}%
\pgfpathlineto{\pgfqpoint{1.992744in}{2.242156in}}%
\pgfpathmoveto{\pgfqpoint{1.992744in}{0.704022in}}%
\pgfpathlineto{\pgfqpoint{2.243198in}{0.712824in}}%
\pgfpathlineto{\pgfqpoint{2.493651in}{0.730868in}}%
\pgfpathlineto{\pgfqpoint{2.744105in}{0.755293in}}%
\pgfpathlineto{\pgfqpoint{2.994558in}{0.787640in}}%
\pgfpathlineto{\pgfqpoint{3.245012in}{0.849033in}}%
\pgfpathlineto{\pgfqpoint{3.495465in}{0.974241in}}%
\pgfpathlineto{\pgfqpoint{3.745919in}{1.175364in}}%
\pgfpathlineto{\pgfqpoint{3.996372in}{1.519959in}}%
\pgfpathmoveto{\pgfqpoint{3.996372in}{0.718105in}}%
\pgfpathlineto{\pgfqpoint{4.246826in}{0.803264in}}%
\pgfpathlineto{\pgfqpoint{4.497279in}{0.923630in}}%
\pgfpathlineto{\pgfqpoint{4.747733in}{1.090646in}}%
\pgfpathlineto{\pgfqpoint{4.998186in}{1.437441in}}%
\pgfpathmoveto{\pgfqpoint{4.998186in}{0.724046in}}%
\pgfpathlineto{\pgfqpoint{5.248640in}{0.748912in}}%
\pgfpathlineto{\pgfqpoint{5.499093in}{0.776418in}}%
\pgfpathlineto{\pgfqpoint{5.749547in}{0.796882in}}%
\pgfusepath{stroke}%
\end{pgfscope}%
\begin{pgfscope}%
\pgfpathrectangle{\pgfqpoint{0.490023in}{0.579475in}}{\pgfqpoint{5.509977in}{2.420525in}}%
\pgfusepath{clip}%
\pgfsetbuttcap%
\pgfsetroundjoin%
\definecolor{currentfill}{rgb}{1.000000,0.756863,0.027451}%
\pgfsetfillcolor{currentfill}%
\pgfsetlinewidth{1.003750pt}%
\definecolor{currentstroke}{rgb}{1.000000,0.756863,0.027451}%
\pgfsetstrokecolor{currentstroke}%
\pgfsetdash{}{0pt}%
\pgfsys@defobject{currentmarker}{\pgfqpoint{-0.041667in}{-0.041667in}}{\pgfqpoint{0.041667in}{0.041667in}}{%
\pgfpathmoveto{\pgfqpoint{-0.041667in}{0.000000in}}%
\pgfpathlineto{\pgfqpoint{0.041667in}{0.000000in}}%
\pgfpathmoveto{\pgfqpoint{0.000000in}{-0.041667in}}%
\pgfpathlineto{\pgfqpoint{0.000000in}{0.041667in}}%
\pgfusepath{stroke,fill}%
}%
\begin{pgfscope}%
\pgfsys@transformshift{0.490023in}{0.698301in}%
\pgfsys@useobject{currentmarker}{}%
\end{pgfscope}%
\begin{pgfscope}%
\pgfsys@transformshift{0.740477in}{0.710183in}%
\pgfsys@useobject{currentmarker}{}%
\end{pgfscope}%
\begin{pgfscope}%
\pgfsys@transformshift{0.990930in}{0.733949in}%
\pgfsys@useobject{currentmarker}{}%
\end{pgfscope}%
\begin{pgfscope}%
\pgfsys@transformshift{1.241384in}{0.796442in}%
\pgfsys@useobject{currentmarker}{}%
\end{pgfscope}%
\begin{pgfscope}%
\pgfsys@transformshift{1.491837in}{0.950475in}%
\pgfsys@useobject{currentmarker}{}%
\end{pgfscope}%
\begin{pgfscope}%
\pgfsys@transformshift{1.742291in}{1.340180in}%
\pgfsys@useobject{currentmarker}{}%
\end{pgfscope}%
\begin{pgfscope}%
\pgfsys@transformshift{1.992744in}{2.242156in}%
\pgfsys@useobject{currentmarker}{}%
\end{pgfscope}%
\begin{pgfscope}%
\pgfsys@transformshift{1.992744in}{0.704022in}%
\pgfsys@useobject{currentmarker}{}%
\end{pgfscope}%
\begin{pgfscope}%
\pgfsys@transformshift{2.243198in}{0.712824in}%
\pgfsys@useobject{currentmarker}{}%
\end{pgfscope}%
\begin{pgfscope}%
\pgfsys@transformshift{2.493651in}{0.730868in}%
\pgfsys@useobject{currentmarker}{}%
\end{pgfscope}%
\begin{pgfscope}%
\pgfsys@transformshift{2.744105in}{0.755293in}%
\pgfsys@useobject{currentmarker}{}%
\end{pgfscope}%
\begin{pgfscope}%
\pgfsys@transformshift{2.994558in}{0.787640in}%
\pgfsys@useobject{currentmarker}{}%
\end{pgfscope}%
\begin{pgfscope}%
\pgfsys@transformshift{3.245012in}{0.849033in}%
\pgfsys@useobject{currentmarker}{}%
\end{pgfscope}%
\begin{pgfscope}%
\pgfsys@transformshift{3.495465in}{0.974241in}%
\pgfsys@useobject{currentmarker}{}%
\end{pgfscope}%
\begin{pgfscope}%
\pgfsys@transformshift{3.745919in}{1.175364in}%
\pgfsys@useobject{currentmarker}{}%
\end{pgfscope}%
\begin{pgfscope}%
\pgfsys@transformshift{3.996372in}{1.519959in}%
\pgfsys@useobject{currentmarker}{}%
\end{pgfscope}%
\begin{pgfscope}%
\pgfsys@transformshift{3.996372in}{0.718105in}%
\pgfsys@useobject{currentmarker}{}%
\end{pgfscope}%
\begin{pgfscope}%
\pgfsys@transformshift{4.246826in}{0.803264in}%
\pgfsys@useobject{currentmarker}{}%
\end{pgfscope}%
\begin{pgfscope}%
\pgfsys@transformshift{4.497279in}{0.923630in}%
\pgfsys@useobject{currentmarker}{}%
\end{pgfscope}%
\begin{pgfscope}%
\pgfsys@transformshift{4.747733in}{1.090646in}%
\pgfsys@useobject{currentmarker}{}%
\end{pgfscope}%
\begin{pgfscope}%
\pgfsys@transformshift{4.998186in}{1.437441in}%
\pgfsys@useobject{currentmarker}{}%
\end{pgfscope}%
\begin{pgfscope}%
\pgfsys@transformshift{4.998186in}{0.724046in}%
\pgfsys@useobject{currentmarker}{}%
\end{pgfscope}%
\begin{pgfscope}%
\pgfsys@transformshift{5.248640in}{0.748912in}%
\pgfsys@useobject{currentmarker}{}%
\end{pgfscope}%
\begin{pgfscope}%
\pgfsys@transformshift{5.499093in}{0.776418in}%
\pgfsys@useobject{currentmarker}{}%
\end{pgfscope}%
\begin{pgfscope}%
\pgfsys@transformshift{5.749547in}{0.796882in}%
\pgfsys@useobject{currentmarker}{}%
\end{pgfscope}%
\end{pgfscope}%
\begin{pgfscope}%
\pgfpathrectangle{\pgfqpoint{0.490023in}{0.579475in}}{\pgfqpoint{5.509977in}{2.420525in}}%
\pgfusepath{clip}%
\pgfsetrectcap%
\pgfsetroundjoin%
\pgfsetlinewidth{1.505625pt}%
\definecolor{currentstroke}{rgb}{0.000000,0.301961,0.250980}%
\pgfsetstrokecolor{currentstroke}%
\pgfsetdash{}{0pt}%
\pgfpathmoveto{\pgfqpoint{0.490023in}{0.689499in}}%
\pgfpathlineto{\pgfqpoint{0.740477in}{0.690819in}}%
\pgfpathlineto{\pgfqpoint{0.990930in}{0.693240in}}%
\pgfpathlineto{\pgfqpoint{1.241384in}{0.700721in}}%
\pgfpathlineto{\pgfqpoint{1.491837in}{0.732188in}}%
\pgfpathlineto{\pgfqpoint{1.742291in}{0.847273in}}%
\pgfpathlineto{\pgfqpoint{1.992744in}{1.244459in}}%
\pgfpathmoveto{\pgfqpoint{1.992744in}{0.689499in}}%
\pgfpathlineto{\pgfqpoint{2.243198in}{0.690379in}}%
\pgfpathlineto{\pgfqpoint{2.493651in}{0.691919in}}%
\pgfpathlineto{\pgfqpoint{2.744105in}{0.696540in}}%
\pgfpathlineto{\pgfqpoint{2.994558in}{0.706883in}}%
\pgfpathlineto{\pgfqpoint{3.245012in}{0.721846in}}%
\pgfpathlineto{\pgfqpoint{3.495465in}{0.746931in}}%
\pgfpathlineto{\pgfqpoint{3.745919in}{0.822408in}}%
\pgfpathlineto{\pgfqpoint{3.996372in}{0.961918in}}%
\pgfpathmoveto{\pgfqpoint{3.996372in}{0.689499in}}%
\pgfpathlineto{\pgfqpoint{4.246826in}{0.716125in}}%
\pgfpathlineto{\pgfqpoint{4.497279in}{0.761234in}}%
\pgfpathlineto{\pgfqpoint{4.747733in}{0.842432in}}%
\pgfpathlineto{\pgfqpoint{4.998186in}{0.994485in}}%
\pgfpathmoveto{\pgfqpoint{4.998186in}{0.689499in}}%
\pgfpathlineto{\pgfqpoint{5.248640in}{0.701822in}}%
\pgfpathlineto{\pgfqpoint{5.499093in}{0.697421in}}%
\pgfpathlineto{\pgfqpoint{5.749547in}{0.703582in}}%
\pgfusepath{stroke}%
\end{pgfscope}%
\begin{pgfscope}%
\pgfpathrectangle{\pgfqpoint{0.490023in}{0.579475in}}{\pgfqpoint{5.509977in}{2.420525in}}%
\pgfusepath{clip}%
\pgfsetbuttcap%
\pgfsetroundjoin%
\definecolor{currentfill}{rgb}{0.000000,0.301961,0.250980}%
\pgfsetfillcolor{currentfill}%
\pgfsetlinewidth{1.003750pt}%
\definecolor{currentstroke}{rgb}{0.000000,0.301961,0.250980}%
\pgfsetstrokecolor{currentstroke}%
\pgfsetdash{}{0pt}%
\pgfsys@defobject{currentmarker}{\pgfqpoint{-0.041667in}{-0.041667in}}{\pgfqpoint{0.041667in}{0.041667in}}{%
\pgfpathmoveto{\pgfqpoint{-0.041667in}{-0.041667in}}%
\pgfpathlineto{\pgfqpoint{0.041667in}{0.041667in}}%
\pgfpathmoveto{\pgfqpoint{-0.041667in}{0.041667in}}%
\pgfpathlineto{\pgfqpoint{0.041667in}{-0.041667in}}%
\pgfusepath{stroke,fill}%
}%
\begin{pgfscope}%
\pgfsys@transformshift{0.490023in}{0.689499in}%
\pgfsys@useobject{currentmarker}{}%
\end{pgfscope}%
\begin{pgfscope}%
\pgfsys@transformshift{0.740477in}{0.690819in}%
\pgfsys@useobject{currentmarker}{}%
\end{pgfscope}%
\begin{pgfscope}%
\pgfsys@transformshift{0.990930in}{0.693240in}%
\pgfsys@useobject{currentmarker}{}%
\end{pgfscope}%
\begin{pgfscope}%
\pgfsys@transformshift{1.241384in}{0.700721in}%
\pgfsys@useobject{currentmarker}{}%
\end{pgfscope}%
\begin{pgfscope}%
\pgfsys@transformshift{1.491837in}{0.732188in}%
\pgfsys@useobject{currentmarker}{}%
\end{pgfscope}%
\begin{pgfscope}%
\pgfsys@transformshift{1.742291in}{0.847273in}%
\pgfsys@useobject{currentmarker}{}%
\end{pgfscope}%
\begin{pgfscope}%
\pgfsys@transformshift{1.992744in}{1.244459in}%
\pgfsys@useobject{currentmarker}{}%
\end{pgfscope}%
\begin{pgfscope}%
\pgfsys@transformshift{1.992744in}{0.689499in}%
\pgfsys@useobject{currentmarker}{}%
\end{pgfscope}%
\begin{pgfscope}%
\pgfsys@transformshift{2.243198in}{0.690379in}%
\pgfsys@useobject{currentmarker}{}%
\end{pgfscope}%
\begin{pgfscope}%
\pgfsys@transformshift{2.493651in}{0.691919in}%
\pgfsys@useobject{currentmarker}{}%
\end{pgfscope}%
\begin{pgfscope}%
\pgfsys@transformshift{2.744105in}{0.696540in}%
\pgfsys@useobject{currentmarker}{}%
\end{pgfscope}%
\begin{pgfscope}%
\pgfsys@transformshift{2.994558in}{0.706883in}%
\pgfsys@useobject{currentmarker}{}%
\end{pgfscope}%
\begin{pgfscope}%
\pgfsys@transformshift{3.245012in}{0.721846in}%
\pgfsys@useobject{currentmarker}{}%
\end{pgfscope}%
\begin{pgfscope}%
\pgfsys@transformshift{3.495465in}{0.746931in}%
\pgfsys@useobject{currentmarker}{}%
\end{pgfscope}%
\begin{pgfscope}%
\pgfsys@transformshift{3.745919in}{0.822408in}%
\pgfsys@useobject{currentmarker}{}%
\end{pgfscope}%
\begin{pgfscope}%
\pgfsys@transformshift{3.996372in}{0.961918in}%
\pgfsys@useobject{currentmarker}{}%
\end{pgfscope}%
\begin{pgfscope}%
\pgfsys@transformshift{3.996372in}{0.689499in}%
\pgfsys@useobject{currentmarker}{}%
\end{pgfscope}%
\begin{pgfscope}%
\pgfsys@transformshift{4.246826in}{0.716125in}%
\pgfsys@useobject{currentmarker}{}%
\end{pgfscope}%
\begin{pgfscope}%
\pgfsys@transformshift{4.497279in}{0.761234in}%
\pgfsys@useobject{currentmarker}{}%
\end{pgfscope}%
\begin{pgfscope}%
\pgfsys@transformshift{4.747733in}{0.842432in}%
\pgfsys@useobject{currentmarker}{}%
\end{pgfscope}%
\begin{pgfscope}%
\pgfsys@transformshift{4.998186in}{0.994485in}%
\pgfsys@useobject{currentmarker}{}%
\end{pgfscope}%
\begin{pgfscope}%
\pgfsys@transformshift{4.998186in}{0.689499in}%
\pgfsys@useobject{currentmarker}{}%
\end{pgfscope}%
\begin{pgfscope}%
\pgfsys@transformshift{5.248640in}{0.701822in}%
\pgfsys@useobject{currentmarker}{}%
\end{pgfscope}%
\begin{pgfscope}%
\pgfsys@transformshift{5.499093in}{0.697421in}%
\pgfsys@useobject{currentmarker}{}%
\end{pgfscope}%
\begin{pgfscope}%
\pgfsys@transformshift{5.749547in}{0.703582in}%
\pgfsys@useobject{currentmarker}{}%
\end{pgfscope}%
\end{pgfscope}%
\begin{pgfscope}%
\pgfpathrectangle{\pgfqpoint{0.490023in}{0.579475in}}{\pgfqpoint{5.509977in}{2.420525in}}%
\pgfusepath{clip}%
\pgfsetbuttcap%
\pgfsetroundjoin%
\pgfsetlinewidth{1.505625pt}%
\definecolor{currentstroke}{rgb}{0.501961,0.501961,0.501961}%
\pgfsetstrokecolor{currentstroke}%
\pgfsetdash{{5.550000pt}{2.400000pt}}{0.000000pt}%
\pgfpathmoveto{\pgfqpoint{0.490023in}{2.889976in}}%
\pgfpathlineto{\pgfqpoint{6.000000in}{2.889976in}}%
\pgfusepath{stroke}%
\end{pgfscope}%
\begin{pgfscope}%
\pgfpathrectangle{\pgfqpoint{0.490023in}{0.579475in}}{\pgfqpoint{5.509977in}{2.420525in}}%
\pgfusepath{clip}%
\pgfsetbuttcap%
\pgfsetroundjoin%
\pgfsetlinewidth{1.505625pt}%
\definecolor{currentstroke}{rgb}{0.333333,0.000000,0.831373}%
\pgfsetstrokecolor{currentstroke}%
\pgfsetdash{{5.550000pt}{2.400000pt}}{0.000000pt}%
\pgfpathmoveto{\pgfqpoint{1.992744in}{0.689499in}}%
\pgfpathlineto{\pgfqpoint{1.992744in}{2.889976in}}%
\pgfusepath{stroke}%
\end{pgfscope}%
\begin{pgfscope}%
\pgfpathrectangle{\pgfqpoint{0.490023in}{0.579475in}}{\pgfqpoint{5.509977in}{2.420525in}}%
\pgfusepath{clip}%
\pgfsetbuttcap%
\pgfsetroundjoin%
\pgfsetlinewidth{1.505625pt}%
\definecolor{currentstroke}{rgb}{0.333333,0.000000,0.831373}%
\pgfsetstrokecolor{currentstroke}%
\pgfsetdash{{5.550000pt}{2.400000pt}}{0.000000pt}%
\pgfpathmoveto{\pgfqpoint{3.996372in}{0.689499in}}%
\pgfpathlineto{\pgfqpoint{3.996372in}{2.889976in}}%
\pgfusepath{stroke}%
\end{pgfscope}%
\begin{pgfscope}%
\pgfpathrectangle{\pgfqpoint{0.490023in}{0.579475in}}{\pgfqpoint{5.509977in}{2.420525in}}%
\pgfusepath{clip}%
\pgfsetbuttcap%
\pgfsetroundjoin%
\pgfsetlinewidth{1.505625pt}%
\definecolor{currentstroke}{rgb}{0.333333,0.000000,0.831373}%
\pgfsetstrokecolor{currentstroke}%
\pgfsetdash{{5.550000pt}{2.400000pt}}{0.000000pt}%
\pgfpathmoveto{\pgfqpoint{4.998186in}{0.689499in}}%
\pgfpathlineto{\pgfqpoint{4.998186in}{2.889976in}}%
\pgfusepath{stroke}%
\end{pgfscope}%
\begin{pgfscope}%
\pgfsetrectcap%
\pgfsetmiterjoin%
\pgfsetlinewidth{0.803000pt}%
\definecolor{currentstroke}{rgb}{0.000000,0.000000,0.000000}%
\pgfsetstrokecolor{currentstroke}%
\pgfsetdash{}{0pt}%
\pgfpathmoveto{\pgfqpoint{0.490023in}{0.579475in}}%
\pgfpathlineto{\pgfqpoint{0.490023in}{3.000000in}}%
\pgfusepath{stroke}%
\end{pgfscope}%
\begin{pgfscope}%
\pgfsetrectcap%
\pgfsetmiterjoin%
\pgfsetlinewidth{0.803000pt}%
\definecolor{currentstroke}{rgb}{0.000000,0.000000,0.000000}%
\pgfsetstrokecolor{currentstroke}%
\pgfsetdash{}{0pt}%
\pgfpathmoveto{\pgfqpoint{6.000000in}{0.579475in}}%
\pgfpathlineto{\pgfqpoint{6.000000in}{3.000000in}}%
\pgfusepath{stroke}%
\end{pgfscope}%
\begin{pgfscope}%
\pgfsetrectcap%
\pgfsetmiterjoin%
\pgfsetlinewidth{0.803000pt}%
\definecolor{currentstroke}{rgb}{0.000000,0.000000,0.000000}%
\pgfsetstrokecolor{currentstroke}%
\pgfsetdash{}{0pt}%
\pgfpathmoveto{\pgfqpoint{0.490023in}{0.579475in}}%
\pgfpathlineto{\pgfqpoint{6.000000in}{0.579475in}}%
\pgfusepath{stroke}%
\end{pgfscope}%
\begin{pgfscope}%
\pgfsetrectcap%
\pgfsetmiterjoin%
\pgfsetlinewidth{0.803000pt}%
\definecolor{currentstroke}{rgb}{0.000000,0.000000,0.000000}%
\pgfsetstrokecolor{currentstroke}%
\pgfsetdash{}{0pt}%
\pgfpathmoveto{\pgfqpoint{0.490023in}{3.000000in}}%
\pgfpathlineto{\pgfqpoint{6.000000in}{3.000000in}}%
\pgfusepath{stroke}%
\end{pgfscope}%
\begin{pgfscope}%
\definecolor{textcolor}{rgb}{0.333333,0.000000,0.831373}%
\pgfsetstrokecolor{textcolor}%
\pgfsetfillcolor{textcolor}%
\pgftext[x=2.117971in,y=2.669928in,left,base]{\color{textcolor}\rmfamily\fontsize{16.000000}{19.200000}\selectfont sketch}%
\end{pgfscope}%
\begin{pgfscope}%
\definecolor{textcolor}{rgb}{0.333333,0.000000,0.831373}%
\pgfsetstrokecolor{textcolor}%
\pgfsetfillcolor{textcolor}%
\pgftext[x=2.117971in,y=2.449881in,left,base]{\color{textcolor}\rmfamily\fontsize{16.000000}{19.200000}\selectfont guide}%
\end{pgfscope}%
\begin{pgfscope}%
\definecolor{textcolor}{rgb}{0.333333,0.000000,0.831373}%
\pgfsetstrokecolor{textcolor}%
\pgfsetfillcolor{textcolor}%
\pgftext[x=2.117971in,y=2.229833in,left,base]{\color{textcolor}\rmfamily\fontsize{16.000000}{19.200000}\selectfont n°1}%
\end{pgfscope}%
\begin{pgfscope}%
\definecolor{textcolor}{rgb}{0.333333,0.000000,0.831373}%
\pgfsetstrokecolor{textcolor}%
\pgfsetfillcolor{textcolor}%
\pgftext[x=4.121599in,y=2.669928in,left,base]{\color{textcolor}\rmfamily\fontsize{16.000000}{19.200000}\selectfont sketch}%
\end{pgfscope}%
\begin{pgfscope}%
\definecolor{textcolor}{rgb}{0.333333,0.000000,0.831373}%
\pgfsetstrokecolor{textcolor}%
\pgfsetfillcolor{textcolor}%
\pgftext[x=4.121599in,y=2.449881in,left,base]{\color{textcolor}\rmfamily\fontsize{16.000000}{19.200000}\selectfont guide}%
\end{pgfscope}%
\begin{pgfscope}%
\definecolor{textcolor}{rgb}{0.333333,0.000000,0.831373}%
\pgfsetstrokecolor{textcolor}%
\pgfsetfillcolor{textcolor}%
\pgftext[x=4.121599in,y=2.229833in,left,base]{\color{textcolor}\rmfamily\fontsize{16.000000}{19.200000}\selectfont n°2}%
\end{pgfscope}%
\begin{pgfscope}%
\definecolor{textcolor}{rgb}{0.333333,0.000000,0.831373}%
\pgfsetstrokecolor{textcolor}%
\pgfsetfillcolor{textcolor}%
\pgftext[x=5.123413in,y=2.669928in,left,base]{\color{textcolor}\rmfamily\fontsize{16.000000}{19.200000}\selectfont sketch}%
\end{pgfscope}%
\begin{pgfscope}%
\definecolor{textcolor}{rgb}{0.333333,0.000000,0.831373}%
\pgfsetstrokecolor{textcolor}%
\pgfsetfillcolor{textcolor}%
\pgftext[x=5.123413in,y=2.449881in,left,base]{\color{textcolor}\rmfamily\fontsize{16.000000}{19.200000}\selectfont guide}%
\end{pgfscope}%
\begin{pgfscope}%
\definecolor{textcolor}{rgb}{0.333333,0.000000,0.831373}%
\pgfsetstrokecolor{textcolor}%
\pgfsetfillcolor{textcolor}%
\pgftext[x=5.123413in,y=2.229833in,left,base]{\color{textcolor}\rmfamily\fontsize{16.000000}{19.200000}\selectfont n°3}%
\end{pgfscope}%
\end{pgfpicture}%
\makeatother%
\endgroup%

%% file: section/related-work.tex
\section{Related Work}

\paragraph{\textbf{Controlling optimizations}}
Historically, programmers had either to explicitly write optimized code, e.g. explicit vectorization or loop ordering, or to entrust optimization to a black box compiler.

\emph{Automating optimization.}
Some compiler optimizations are fully automated via equality saturation \cite{tate2009-equality-saturation, yang2021-eqsat-tensor} or heuristic searches \cite{lift-rewrite-2015, polymage-2015}.
Although this approach can automatically yield high performance, it is not always feasible or even desirable, as it may result in poor performance or may be too time-consuming \cite{maleki2011-vectorizing-compilers, parello2004-pragmatic-opt}.
When automatic optimization is unsatisfactory, programmers often fall back to manual optimization in order to achieve their performance goals \cite{niittylahti2002-high-perf, hlt-lacassagne-2014, lemaitre2017-cholesky}.

\emph{Rewriting strategies and schedules.}
More recently, compiler optimizations can be precisely controlled by the programmer specifying rewriting strategies \cite{visser1998-strategies, hagedorn2020-elevate, koehler2021-elevate-imgproc} or schedules \cite{halide-2012, chen2018-tvm}.
However, these are challenging to write as the phase ordering problem is passed on to the programmer~\cite{ikarashi2021-guided}, as discussed in~\cref{elevaterewrite}.

\paragraph{\textbf{Guidance used outside of optimizations}}
Although equality saturation exploits e-graphs for program optimization~\cite{tate2009-equality-saturation}, e-graphs were originally designed for efficient congruence closure in theorem provers \cite{nelson1980-techniques, de2008-z3}, and are useful in other settings such as program synthesis, or semantic code search \cite{premtoon2020-yogo}.

\emph{Guidance in theorem proving.}
In the realm of theorem proving, guidance is typically required because automation is both theoretically undecidable and difficult in practice.
Just as rewriting strategies specify how to transform a program step-by-step, proof tactics \cite{gordon1979-edinburgh} specify how to transform a proof state step-by-step in procedural proof languages.
In declarative proof languages, proof sketches specify partial proofs and can guide theorem proving \cite{wiedijk2003-formal-proof-sketches, corbineau2007-declarative-coq}. Proof sketches are analogous to program sketches that specify partial programs to guide optimization.


\emph{Sketching for synthesizing programs.}
The idea of sketching has been used for program synthesis \cite{solar2008-synthesis-sketching}, along with counterexample guided inductive synthesis that combines a synthesizer with a validation procedure.
Our approach differs as we use sketches for optimizations rather than program synthesis. 
We use sketches as program patterns to filter a set of equivalent programs generated via equality saturation, and as a result do not require a validation procedure.

\paragraph{\textbf{Other techniques for scaling equality saturation}}
There exists a number of prior techniques for scaling equality saturation in practice.

\emph{Languages with binding.}
We are not the first to attempt applying equality saturation to languages with binding.
\citet{willsey2021-egg} implements a partial evaluator for the lambda calculus using explicit substitution.
Although conceptually simple, this is too inefficient for complex optimizations, as demonstrated in \cref{lambda-eval}.
\citet{smith2021-access-patterns} proposes \emph{access patterns} to avoid the need for binding structures when representing tensor programs.
In \cref{eqsat-bindings}, we instead present lambda calculus encoding techniques that make equality saturation significantly more efficient.


\emph{Rewrite rule scheduling.}
To reduce e-graph growth, previous work proposes rewrite rule schedulers as a way to control which rewrite rules should be applied on a given equality saturation iteration \cite{willsey2021-egg}.
By default, the egg library uses a \texttt{BackoffScheduler} preventing specific rules from being applied too often, and reducing e-graph growth in the presence of ``explosive'' rules such as associativity and commutativity.
Our experience with \Rise{} optimization is that using the \texttt{BackoffScheduler} is often counterproductive as the desired optimization depends on some explosive rules.
Future work may explore better ways to schedule rewrite rules, but at present \kles{} does
not use a rewrite rule scheduler.

\emph{External solvers.}
External solvers may add equivalences to the e-graph~\cite{nandi2020-synthesizing-CAD}, but this requires the identification of sub-tasks that can benefit from being delegated.

%% file: section/conclusion.tex
\section{Conclusion}

This paper broadens the applicability of equality saturation by making it scale to complex optimizations in a functional language in two ways.

Firstly, sketch-guided equality saturation is proposed as a semi-automated technique for scaling to more complex optimizations by factoring a single equality saturation search into a sequence of smaller equality saturation searches (\cref{sketching}).
Programmers guide the process by describing how a program should evolve using a sequence of sketches.
We demonstrate that sketch-guided equality saturation enables seven complex optimizations of matrix multiplication to be applied within seconds in the \Rise{} functional language, using under 1 GB of RAM, using no more than three sketch guides (\cref{fig:s3}).
By contrast, traditional unguided equality saturation cannot discover the five more complex optimizations even with an hour of runtime and 60 GB of RAM (\cref{fig:s2}).
For each optimization, the generated code is identical to the high-performance code generated by manually ordering thousands of rewrites via rewriting strategies.
The performance of this code is comparable with code produced by the state-of-the-art TVM compiler~\cite{hagedorn2020-elevate}.

Secondly, to effectively apply equality saturation to functional languages, we propose new techniques to efficiently encode lambda calculi for equality saturation. 
The key innovations are extraction-based substitution and representing identifiers as De Bruijn indices (\cref{eqsat-bindings}). Specifically for our  \Rise{} use case, we apply the techniques to a polymorphically typed lambda calculus. Combining the techniques reduces the runtime and memory consumption of equality saturation over lambda terms by orders of magnitude, enabling equality saturation to scale to more complex optimization goals  (\cref{fig:lambda-res}).

\medskip

Future work may investigate how to identify appropriate sets of rules to be applied in each search.
We are interested in exploring how to design effective sketch guides for more diverse applications, and to explore ways to possibly even synthesize sketch guides automatically.
We believe that combining imperative, step-by-step approaches (e.g. rewriting strategies) with more declarative approaches (e.g. sketch guidance) deserves further research and can lead to the creation of practical, interactive optimization assistants.
Finally, we hope that the community will be inspired to apply and extend sketch-guided equality saturation in new domains.


\clearpage